\documentclass[]{aa} 

\usepackage{graphicx}
\usepackage{txfonts}
\usepackage{comment}
\usepackage{lscape}
\usepackage{longtable}

\usepackage{natbib,twoopt}
\bibpunct{(}{)}{;}{a}{}{,}             

\def\ojo{\fbox{\bf !$\odot$j$\odot$!}} 			

\usepackage[usenames]{color}                                   
\def\ATT{\color{red}}                				
\def\ATC{\color{cyan}}                				

\newcommand\pycasso{{\sc p}y{\sc casso}}          	
\newcommand\starlight{{\sc starlight}}          	

\begin{document}

\title{The spatially-resolved star formation histories of CALIFA galaxies: } 
\subtitle{Implications for galaxy formation }

\authorrunning{IAA et al.}
\titlerunning{The 2D star formation history of CALIFA galaxies}

\author{
R. M. Gonz\'alez Delgado\inst{1},
E. P\'erez\inst{1},
R. Cid Fernandes\inst{2},
R. Garc\'{\i}a-Benito\inst{1}, 
R. L\'opez Fern\'andez\inst{1},
N. Vale Asari\inst{2},
C. Cortijo-Ferrero\inst{1},
A. L.\ de Amorim\inst{2},
E. A. D. Lacerda\inst{2},
S. F. S\'anchez\inst{3},
M.D. Lehnert\inst{4},
C. J. Walcher\inst{5}
}

\institute{
Instituto de Astrof\'{\i}sica de Andaluc\'{\i}a (CSIC), P.O. Box 3004, 18080 Granada, Spain. (\email{rosa@iaa.es})
\and
Departamento de F\'{\i}sica, Universidade Federal de Santa Catarina, P.O. Box 476, 88040-900, Florian\'opolis, SC, Brazil
\and
Instituto de Astronom\'\i a,Universidad Nacional Auton\'oma de M\'exico, A.P. 70-264, 04510 M\'exico D.F., Mexico
\and
Institut d'Astrophysique de Paris, UMR 7095, CNRS, Universit\'e Pierre et Marie Curie, 98bis boulevard Arago, 75014 Paris, France
\and
Leibniz-Institut f\"ur Astrophysik Potsdam (AIP), An der Sternwarte 16, D-14482 Potsdam, Germany
}

\date{22/March/2017}


\abstract{
This paper presents the spatially resolved star formation history (SFH) of nearby galaxies with the aim of furthering our understanding of the different processes involved in the formation and evolution of galaxies.
To this end, we apply the fossil record method of stellar population synthesis to a rich and diverse data set of 436 galaxies observed with integral field spectroscopy in the CALIFA survey. The sample covers a wide range of Hubble types, 
with  stellar masses ranging from $M_\star \sim 10^9$ to $7 \times 10^{11} M_\odot$. Spectral synthesis techniques are applied to the datacubes to retrieve the spatially resolved time evolution of the star formation rate (SFR), its intensity ($\Sigma_{\rm SFR}$), and other descriptors of the 2D-SFH in seven bins of galaxy morphology (E, S0, Sa, Sb, Sbc, Sc, and Sd), and five bins of stellar mass. 
Our main results are: 
{\em (a)} Galaxies form very fast independently of their current stellar mass, with the peak of star formation  at high redshift ($z > 2$). Subsequent star formation is driven by $M_\star$ and morphology, with less massive and later type spirals showing more prolonged periods of star formation. 
{\em (b)}  At any epoch in the past the SFR is proportional to $M_\star$, with most massive galaxies having the highest absolute (but lowest specific) SFRs.  
{\em (c)}  While nowadays $\Sigma_{\rm SFR}$ is similar for all spirals, and significantly lower in early type galaxies (ETG),  in the past $\Sigma_{\rm SFR}$ scales well with  morphology.
The central regions of today's ETGs are where $\Sigma_{\rm SFR}$ reached the highest values ($>  10^3 \,M_\odot\,$Gyr$^{-1}\,$pc$^{-2}$), similar to those measured in high redshift star forming galaxies. 
{\em (d)}  The evolution of  $\Sigma_{\rm SFR}$ in Sbc systems matches that of models for Milky-Way-like galaxies, suggesting that the formation of a thick disk may be a common phase in spirals at early epochs. 
{\em (e)} The SFR and $\Sigma_{\rm SFR}$ in outer regions of E's and S0's show that they have undergone an extended phase of growth in mass between $z = 2$ and 0.4. The mass assembled in this phase is in agreement with the two-phase scenario  proposed for the formation of ETG. 
{\em (f)}
Evidence of an early and fast quenching is found only in the most massive ($M_\star > 2 \times 10^{11} M_\odot$) E galaxies of the sample, but not in spirals of similar mass, suggesting that  halo-quenching is not the main mechanism 
for the shut down of star formation in galaxies. Less massive E and disk galaxies show more extended SFHs and a slow quenching. 
{\em (g)}
Evidence of fast quenching is also found in the nuclei of ETG and early spirals, with SFR and $\Sigma_{\rm SFR}$ indicating that they can be the relic of the ``red nuggets'' detected at high redshift.
}



\maketitle

\section{Introduction}
\label{sec:Introduction}

Our understanding of how galaxies form has undergone significant progress in recent years due to improvements in observational facilities and the diversity of techniques to estimate galaxy properties. A large number of surveys, performed at a range of redshifts that sample the age of the Universe, have allowed to establish relations of the distribution and the properties of galaxies with the complex distribution of dark matter halos, filaments, and voids that form the cosmic web. However, it is not yet known which is the main driver for galaxy assembly, with accretion and merging the two most important mechanisms proposed. 

On the theoretical front, the $\Lambda$ cold dark matter paradigm poses that galaxies grow their mass by merging of dark matter halos, progressively assembling more massive systems by bringing stars from subsystems with different histories to what eventually becomes a single massive galaxy. Simulations suggest that major and minor mergers make up as much as  50\% of the outer envelopes of massive galaxies \citep{naab09}, but observations indicate that equal mass mergers are rare and relatively unimportant for the cosmic star formation budget \citep{man12, williams11}, and there are also difficulties in matching the number of thin disk galaxies in this scenario \citep[]{naab16}. 

Galaxies with stellar masses\footnote{The units of $M_\star$ are $M_\odot$ throughout; hereinafter we will not specify them for the sake of clarity.} $M_\star\sim10^{11}$  mark the transition between those dominated by in-situ star formation growth (at low mass) and those domninated by merger growth (at high mass; \citealt{behroozi13}). 
Milky Way-like galaxies and those of lower mass seem to have assembled their mass through streams of cold gas from the cosmic web \citep{jsa14}. In this context the galaxy's gas accretion and star formation rates (SFR) are expected to be associated with the cosmological dark matter specific accretion rate \citep{neistein06,birnboim07, neistein08, dutton10}.  

Observationally, several fundamental results related to how galaxies grow their stellar mass are now well established: 
{\em(1)} The star formation rate density in the Universe peaked $\sim 10$ Gyr ago, at redshift $z \sim 2$, and declined thereafter \citep{lilly96, madau98, hopkinsbeacom06, fardal07, madau14}. 
 {\em(2)}  At any $z$,  star forming galaxies show a correlation between SFR and  $M_\star$, known as the main sequence of star formation (MSSF;  \citealt{brinchmann04, noeske07,daddi07, elbaz07, wuyts11, whitaker12, renzinipeng15, catalan15, canodiaz16}).
 {\em(3)}  The specific star formation rate, sSFR = SFR/$M_\star$, declines weakly with increasing galaxy mass \citep{salim07, schiminovich07}, evolving rapidly at $z < 2$ \citep{rodighiero10, oliver10, karim11, elbaz11, speagle14}, and increasing  slowly at $z>2$ \citep{magdis10,stark13}. 
 {\em(4)}  Galaxy colors, morphology, chemical composition, and  spectral type
all show bimodal distributions, with blue star-forming galaxies separated from red quiescent ones \citep{blanton03, kauffmann03, baldry04, gallazzi05, mateus06, blanton09}. This bimodality is also intrinsic in the Hubble classification of galaxies (e.g., \citealt{roberts94, kennicutt98}).
 {\em(5)}  Galaxies in the blue cloud must evolve to the red sequence, since the population of local quiescent galaxies is not consistent with a simple passive evolution of the population of quiescent galaxies at $z \sim 1$ \citep[]{Faber07, bell07, gallazzi14}.

These  results suggest that the growth of galaxies is not driven solely by gas supply; 
other physical processes, such as feedback from supernovae,  are required to decouple the star formation from the gas accretion \citep{dekelbirnboim09, lilly13}.  
For example,  \citet{lehnert14} proposes a two-phase  scenario to explain the evolution of the sSFR: 
{\em (1)} At $z > 2$  the star formation is self-regulated  by starburst driven outflows. {\em (2)} At $z < 2$ the decline of the sSFR is driven not only by declining gas fractions, but also by the evolution in the angular momentum to support the accreting gas to the disk, and by gas density and stellar mass surface density relations through a generalized Schmidt-Kennicut law \citep{dopita94, shi11}. 
Furthermore, feedback from AGN can cause the suppression of gas accretion onto galaxies by heating the surrounding gas \citep{diMatteoT05}. 

Alternative mechanisms  to  feedback (from stars or AGN) have been proposed to explain the complete shut-down
of the star formation and the stopping of the growth of massive galaxies. 
In the halo-quenching scenario, there is a critical halo mass of $\sim 10^{12}$ above which the circumgalactic gas is shock heated and stops cooling \citep{dekel06, croton06}.
The morphological transformation by the formation of a spheroidal component, known as morphological quenching, can also stop the star formation by stabilizing the gas disk against fragmentation \citep{martig09, genzel14, gonzalezdelgado15, belfiore17}. Recently, thanks to  spatial resolution information on the sSFR, it has been found that this quenching progresses inside-out \citep{tacchella15, gonzalezdelgado15, tacchella16, gonzalezdelgado16}.

In these scenarios the growth of galaxies is a uniform phenomenon that is interrupted by internal or external quenching processes. However,  alternative views are emerging in which galaxy growth is a more  heterogeneous phenomenon, as suggested by the diversity of star formation histories (SFH hereafter) obtained in several surveys \citep{gladders13, oemler13, dressler16, abramson16}.  These works show that the SFH 
may  be extended or compressed in time  \citep{asari07,  pacifici16}.

Simple analytic models for the SFH have been explored for decades to infer the SFR (and sSFR) of 
galaxies at different redshifts by fitting their spectral energy distributions.  However, parametrizations like an exponential declining function or the so called delayed $\tau$ models \citep{searle73,bruzual80} are not able to predict the rising epoch of the SFH of the Universe. Models that assume a rising SFH 
have been more successful to explain the evolution of the SFR at $z >2$ (e.g., \citealt{papovich11, maraston13, behroozi13}). 
A lognormal function has been proposed by \citet{gladders13} to represent both the cosmic evolution of the SFR of the Universe and the diversity of SFHs observed in individual galaxies.
Furthermore, the impact that the SFH of galaxies have on the MSSF relation \citep{cassara16} or how galaxies stop their mass growth has just started to be explored \citep{pacifici16}.

Alternatively to redshift studies, the SFH of galaxies can be inferred using the fossil record encoded in their present day stellar populations. This technique has changed significantly since the seminal works by \citet{tinsley68, tinsley72}, \citet{searle73}, \citet{gallagher84}, and \citet{sandage86}, who first used  optical colors to study how the SFH of galaxies vary along the Hubble sequence.  It has gained significantly with the development of full spectral synthesis and fitting codes where no a priori assumption is made about 
the functional form of the SFH 
\citep{panter03, heavens04, cidfernandes05, ocvirk06, asari07, panter08, tojeiro11, koleva11, mcdermid15, citro16}.

Regardless of the methodology, most of the observational work reviewed above is based on  spatially integrated data, where the different morphological components (bulge and disk) are not separated or the data only cover partially the extent of a galaxy.

Recently, a new generation of Integral Field Spectroscopy (IFS) surveys has emerged to overcome these limitations, such as ATLAS3D \citep{cappellari11}, CALIFA \citep{sanchez12, husemann13, garciabenito15, sanchez16}, SAMI \citep{bryant15}, and MaNGA \citep{bundy15, law15}. These surveys are important to understand how the spheroidal and disk components in a galaxy form and evolve, and what are their relative contributions to the SFH of the Universe.
CALIFA is particularly well suited for such a study because: 
{\em (1)} its large field of view covers the galaxies fully; 
{\em (2)} its spectral range and resolution allow us to apply full spectral fitting to retrieve the SFH;
{\em (3)} it includes  galaxies of all  Hubble types (E, S0, and spirals from Sa to Sd) and stellar masses ($\sim 10^9$ to $10^{12}$);
{\em (4)} its well defined selection function allows for reliable volume-corrections, and thus for extrapolation of results to a cosmic context \citep{walcher14,bekeraite16,gonzalezdelgado16}.

\subsection{Previous work}
\label{sec:Introduction_PreviousWork}

In previous papers in this series we have employed fossil record tools to CALIFA datacubes, obtaining:

\begin{enumerate}

\item The mass assembly history of $\sim100$ galaxies \citep{perez13}. We find that galaxies grow inside-out, and that the signal of downsizing is spatially preserved, with both inner and outer regions growing faster for massive galaxies. This inside-out scenario has been recently confirmed with MaNGA data by \citet{ibarra16}. 

\item 
The radial distribution of stellar populations properties (age, metallicity, extinction) and their gradients as a function of Hubble type and $M_\star$ \citep{gonzalezdelgado14a, gonzalezdelgado15}. We find that more massive galaxies are more compact, older, more metal rich, and less dusty, and that these trends are preserved with radial distance to the nucleus. The age gradients also confirm an inside-out formation for both early and late type galaxies. These results are also sustained with CALIFA data by \citet{sanchez-blazquez14}, and by the analysis of \citet{zheng17} of MaNGA data, though not by \citet{goddard16}, who find positive age radial gradients in their analysis of MaNGA data for early type galaxies (ETG).

\item 
The local relations between the stellar mass surface density ($\mu_\star$) and age \citep{gonzalezdelgado14a}, and  stellar metallicity \citep{gonzalezdelgado14b}.
 These indicate that  $\mu_\star$  regulates the star formation and chemical enrichment in disks, while in  spheroids  $M_\star$ is a more important driver. The bimodal behavior of the stellar ages and $\mu_\star$ has been confirmed by \citet{zibetti17} in an independent analysis of CALIFA data.

\item 
The existence of a tight relation between $\mu_\star$ and the intensity of the star formation $\Sigma_{\rm SFR}$ (defined as the SFR per unit area), defining a local MSSF relation of slope similar to the global one between total SFR and $M_\star$ \citep{gonzalezdelgado16}.
This suggests that local processes are important in determining the star formation in disks, probably through a density dependence of the SFR law.  A similar $\Sigma_{\rm SFR}$-$\mu_\star$ relation was  
derived by \citet{canodiaz16} using the  spatial distribution of H$\alpha$ in CALIFA disk galaxies. 

\item 
The radial profile of the recent sSFR as a function of Hubble type \citep{gonzalezdelgado16}. These profiles 
increase  outwards, with a steeper slope in the inner $R < 1$ half light radius (HLR). This behavior suggests that galaxies are quenched inside-out and that this process is faster in the  bulge-dominated  regions than in the disks. A similar result is found by \citet{tacchella16} analyzing a sample of star forming galaxies at $z \sim 2.2$. These results are interpreted as evidence of the morphological quenching that  these galaxies experience in the growth of their central bulges. 

\end{enumerate}

\subsection{This work}

The list above illustrates the potential of the combination of spatially resolved spectroscopy with stellar population analysis tools as a means to gather clues on the processes leading up to the present day galaxy population. 
Even though these studies resolve galaxies in time ($t$) and space ($R$), so far in this series the radial axis has been explored in far greater detail than the temporal one. Indeed, while radial profiles of physical properties were mapped in the full resolution allowed by the data, the temporal information was mostly condensed to mean stellar ages or to an assessment of the star formation activity in the very recent past. This approach is justified because these low order moments of the SFH are naturally the most robust descriptors obtained from any stellar population analysis. Yet, despite the uncertainties involved, there is evidently  more information encoded in the age distribution inferred from our fossil record method. 

In this paper we explore this information by examining the temporally and spatially resolved SFH of CALIFA galaxies
as a function of  Hubble type and $M_\star$. Different representations of the SFH are used, including the SFR, sSFR, and $\Sigma_{\rm SFR}$ as a function of lookback time and for different radial regions. These are then averaged over bins in stellar mass and morphology to obtain a panoramic view of the growth of galaxies. 
The aim is that this approach leads to useful insights into questions like:

\begin{enumerate}

\item What are the main epochs of star formation in galaxies and how do they change in time, place, and intensity as a function of current mass and morphology?

\item Can we identify phases like spheroid, thick and thin disk formation envisaged in current galaxy formation scenarios?

\item Do merger or accretion of stars from smaller systems leave detectable imprints in the spatially resolved SFHs of galaxies?

\item How do SFHs derived from a fossil record analysis compare to those inferred from the snapshots of galaxy evolution obtained by studies at different redshifts?

\end{enumerate}

An ulterior goal of this paper is to make theorists aware of the power and limitations of these new tools to study galaxy evolution, and so help establishing meaningful and enlightening ways of testing their models.

This paper is organized as follows. Section \ref{sec:Data} describes the observations and  summarizes the properties of the galaxies analyzed here. In Sections \ref{sec:Method} and  \ref{sec:Results} we summarize our method for extracting the SFH and present the spatially resolved results. 
Section \ref{sec:SFH} deals with the cosmic evolution of the mass fraction, SFR,  sSFR, and $\Sigma_{\rm SFR}$ of galaxies as a function of Hubble type and $M_\star$ for three spatial regions: within $R < 0.5$ HLR, outside 1.5 HLR, and for the whole galaxy.
In Section \ref{sec:Discussion}, we discuss the results in relation to different paradigms for the growth of spirals and early type galaxies. 
Section \ref{sec:Summary} reviews our main findings.

\section{Data and sample}
\label{sec:Data}

\subsection{Observations and data reduction}
\label{sec:Observations}

The observations were carried out at Calar Alto observatory (CAHA) with the 3.5m telescope and the Potsdam Multi-Aperture Spectrometer PMAS \citep{Roth05} in the PPaK mode \citep{verheijen04}. PPaK is an integral field spectrograph with a field of view of $74{\tt''} \times 64{\tt''}$  and 382 fibers of $2.7{\tt''}$  diameter each \citep{kelz06}. The galaxies analyzed here were observed with two spectral setups, using the gratings V500 and V1200, with spectral resolutions  $\sim 6$  and 2.3 \AA\ (FWHM), respectively. To reduce the effects of vignetting on the data, we combine the observations in the two setups (COMBO data, in the jargon of CALIFA), 
covering the 3700--7300 \AA\ range with the same resolution as the V500 and a spatial sampling of 1 arcsec/spaxel. The data were calibrated  with version V2.2 of the reduction pipeline \citet{sanchez16}. We refer to \citet{sanchez12}, \citet{husemann13}, \citet{garciabenito15}, and \citet{sanchez16} for  details on the observational strategy and data processing.


\subsection{Sample and morphological classification}
\label{sec:Sample}

The galaxies are from the main and extended CALIFA  samples published in the third and final COMBO data release (DR3) by  \citet{sanchez16}. The galaxies targeted are a random subset of the mother sample plus some additional galaxies from  the  extended sample (a heterogeneous set of 38 sources observed in different ancillary projects).  
We further add 35 galaxies  from the 2$^{\rm nd}$ data release previously analyzed by us in \citet{gonzalezdelgado16}. Type 1 Seyferts and galaxies that show merger or interaction features are excluded. 
This leaves a final sample of 436 galaxies,  398  of which belong to the mother sample. 
A full description and characterization of the mother sample is given by \citet{walcher14}, and by
\citet{sanchez16} for the extended sample. Briefly, they have: {\em (a)} angular isophotal diameter between 
$45{\tt''}$  and $79{\tt''}$;  {\em (b)} redshift range $0.005 \leq z \leq 0.03 $;  {\em (c)}  colors ($u-r < 5$) and magnitudes ($-24 < M_r < -17$)  covering the whole color-magnitude diagram.

\begin{figure}
\includegraphics[width=0.48\textwidth]{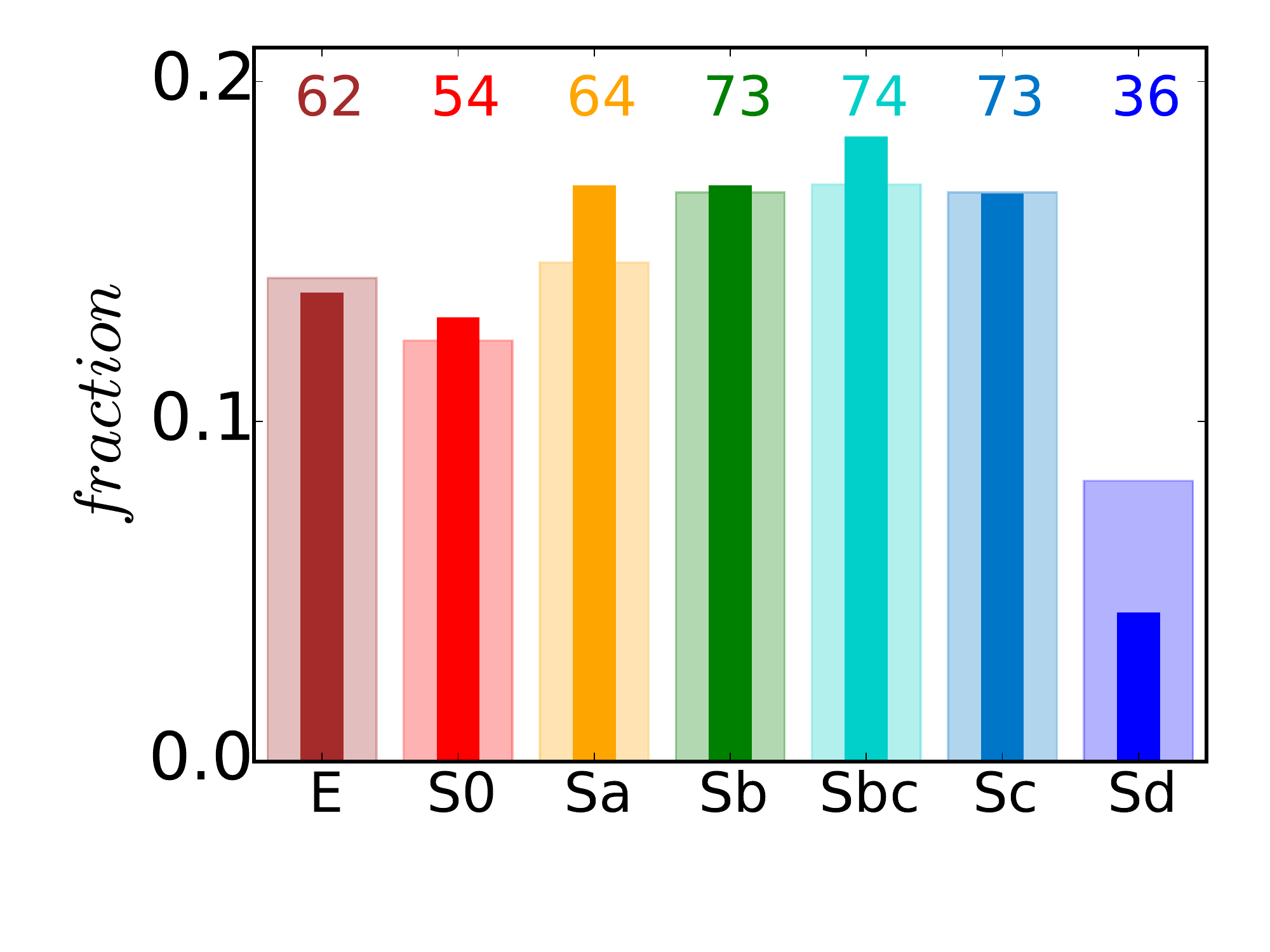}
\caption{ Comparison of Hubble type distributions in the CALIFA mother sample (939 galaxies, broad bars) and the 436 galaxies analyzed here (narrow, darker bars). Both histograms are normalized to unit sum. The number of galaxies in each morphology bin is labeled in color. 
 }
\label{fig:hist-type}
\end{figure}

All galaxies were  morphologically classified by members of the collaboration through visual inspection of the SDSS $r$-band images. As in previous works \citep{gonzalezdelgado15}, we group the galaxies into seven morphology bins: E (62 galaxies), S0 (54, including S0 and S0a), Sa (64, including Sa and Sab), Sb (73), Sbc (74), Sc (73, including Sc and Scd), and Sd (36, including 1 Sm and 1 Irr). Fig.\ \ref{fig:hist-type} shows the morphological distribution of our 436 galaxies (filled bars), as well as that of the mother sample (wider bars). This distribution shows that this working sample is consistent with a random subset of the mother sample except for galaxies in the Sd-bin, which is under-sampled in comparison with the others.

\section{Star formation history analysis}
\label{sec:Method}

\subsection{Method of analysis: \starlight\ and \pycasso}
\label{sec:Base}

The spatially resolved SFHs and related stellar population properties are extracted from the datacubes following the  method  originally presented in \citet{perez13} and used in a series of works reviewed in Section \ref{sec:Introduction_PreviousWork}. The method includes basic pre-processing steps, such as spatial masking of foreground and background sources, rest-framing, spectral resampling, extraction of individual spectra and their stacking into Voronoi zones \citep{cappellari03} whenever necessary to reach a signal-to-noise ratio\footnote{Measured in a 90 \AA\ window centered at 5635 \AA\ rest-frame.} ${\rm S/N} \geq 20$. Each spectrum in then fitted with  \starlight\ \citep{cidfernandes05},  decomposing it in terms of stellar population  with ages $t_j$ and metallicities $Z_j$. The code outputs a population vector $\vec{x}$ whose components $x_j$ express the fractional contribution of base component $j$ to the observed continuum at a reference wavelength of 5635 \AA. The corresponding mass fractions ($m_j$) are also given. 

The results are then processed through \pycasso\ (the Python CALIFA \starlight\ Synthesis Organizer; \citealt{cidfernandes13}; de Amorim et al 2017, in preparation) to produce a suite of spatially resolved stellar population properties. \pycasso\ organizes the information into datacubes with spatial as well as age and metallicity dimensions. Two of the main properties products used in this paper are the images of luminosity (at 5635 \AA) per unit area, ${\cal L}_{\lambda5635}$ (in units of $L_\odot\,$\AA$^{-1}\,$pc$^{-2}$), and stellar mass surface density\footnote{The surface densities $\mu_\star$ and mass fractions $m_j$ reported in this paper are {\em not} corrected for the mass lost by stars during their evolution. This is justified because we also analyze SFRs, which count all mass turned into stars, even that which is eventually returned to the interstellar medium. The $M_\star$ values, however, do correct for this effect. The relation between the total mass turned into stars, $M_\star^\prime$, and the current mass in stars is $M^\prime_\star \sim 1.4 M_\star$  for the IMF used in this work.}, $\mu_\star$ (in $M_\odot\,$pc$^{-2}$).

SFHs can be expressed as the age distribution in either ${\cal L}_{\lambda5635}$ or $\mu_\star$, obtained with the corresponding $\vec{x}$ and $\vec{m}$ arrays. In fact, these light and mass fractions themselves are representations of the SFH routinely used in fossil record studies (e.g., \citealt{tojeiro07,tojeiro09,koleva09}). Star formation rates  can also be defined. We will work with SFRs both in absolute ($M_\odot\,$yr$^{-1}$) and specific (yr$^{-1}$) units, as well as with surface densities, $\Sigma_{\rm SFR}$ (in $M_\odot\,$yr$^{-1}\,$pc$^{-2}$, sometimes called star formation intensities; e.g.,  \citealt{lehnert15}). These different ways of representing SFHs are complementary to each other, in the sense that they are useful to highlight different aspects of the results.

Besides the organization of the \starlight\ output in a format suitable for 2D work, \pycasso\ performs a number of other  tasks which will be useful in our analysis. In particular, radial distances ($R$) to the nucleus are defined along elliptical rings as explained in \citet{cidfernandes13}. As in previous papers, these distances are expressed in units of the half light radius (HLR), defined as the semi-major axis length of the elliptical aperture which contains half of the total light of the galaxy at the rest-frame wavelength 5635 \AA.

Finally, the galaxy masses ($M_\star$) used in our analysis are obtained by adding the masses of each spatial zone, thus taking into account spatial variations on stellar populations and extinction. Masked spaxels due to foreground stars or other artifacts are corrected for in \pycasso\ using the  radial profile of $\mu_\star$, as explained in \citet{gonzalezdelgado14a}.

\subsection{The base of stellar population models}
\label{sec:CSP_base}

This study introduces some novelties related to the base used in \starlight\ to perform the spectral decomposition. Whereas previous works used a base of simple stellar populations (SSP, also known as instantaneous bursts), here we build a base of composite stellar populations (CSP) consisting of ``square bursts'', i.e., episodes of constant SFR during a certain period of time and zero elsewhere. The SFR in each base component is scaled to form $1 M_\odot$ over the corresponding age interval. 
 
We define 18 such CSPs, spanning ages between 1 Myr and  14 Gyr \footnote{In the \citet{Vazdekis15} models, the time interval between SSPs older than 10 Gyr is 0.5 Gyr. To include in the CSP contributions from 13.7 Gyr, the age of the Universe, we include SSPs up to 14 Gyr to generate the CSP with average age of 11.50 Gyr.}, centered at $t_j = 0.00245$,   0.00575,   0.011, 0.018, 0.028, 0.045, 0.072, 0.114, 0.180, 0.285, 0.455, 0.725, 1.14, 1.8, 2.85, 4.55, 7.25, and 11.50 Gyr. These ages are separated by $\Delta \log t = 0.2$ dex, except for the first two, which span 0.4 dex. A corollary of this logarithmic sampling is that $\Delta t$ grows with $t$. The last age bin, for instance, spans a $\Delta t = 5.2$ Gyr long interval from 8.9 to 14.1 Gyr, while the last but one  spans  $\Delta t = 3.3$ Gyr (from 5.6 to 8.9 Gyr), and the one before that covers the $\Delta t = 2.1$ Gyr from 3.5 to 5.6 Gyr. On average, $\Delta t /t= 0.46$, a property which must be kept in mind when evaluating the results of the synthesis. 

An advantage of using CSPs instead of SSPs is that it makes the spectral fits less dependent on the specific SSPs that form the base. This is because each CSP includes all SSPs available in the given time interval, averaging the contribution of each SSP in proportion to its mass-luminosity ratio and the time intervals between the SSPs. In this way, the interpretation of the results in terms of SFR is more straightforward since for each time interval the SFR is constant on average. Other important advantage of this method is that the computation time to fit each spectrum is reduced by a factor 3-4 with respect to using SSP templates. The reason is that now we have for each metallicity only 18 CSPs covering from 1 Myr to 14 Gyr, instead of $\sim$35 SSPs for the same age period, and the computational time required for \starlight\ goes as N$^2$, being N the number of components used for the spectral fit.

The CSP spectra were built using the set of SSP models from \citet{Vazdekis15} for populations older than $t = 63$ Myr \footnote{ Due to limitations in the number of O and B stars in the MILES library \citep{sanchez-blazquez06}, the \citet{Vazdekis15}  models are unsafe for $t \leq 63$ Myr, and in particular for metallicities below half solar; further, BaSTI isochrones are not computed for ages younger than 10 Myr. In contrast, \citet{gonzalezdelgado05} are well optimized for young ages.}, and from the GRANADA models of \citet{gonzalezdelgado05} for younger ages. 
The evolutionary tracks in the GRANADA models are those of \citet{girardi00}, except for the youngest ages (1 and 3 Myr), which are based on the Geneva tracks \citep{schaller92,schaerer93,charbonnel93}; while the Vazdekis models are based on the BaSTI isochrones \citep{pietrinferni04, pietrinferni06, pietrinferni09, pietrinferni13, cordier07}. 
From these new sets of SSPs, we use here the base models that match the Galactic abundance pattern imprinted in the MILES stars \citep{sanchez-blazquez06}. Eight metallicities are considered: $\log Z/Z_\odot = -2.28$, $-1.79$, $-1.26$, $-0.66$, $-0.35$, $-0.06$, $0.25$ and $+0.40$. Each  of the CSPs  has the same initial chemical composition and are always affected by the same amount of extinction. 

As in previous papers of this series, the initial mass function (IMF) is Salpeter, and dust is modeled as a foreground screen with a \citet{cardelli89} reddening law with $R_V = 3.1$. We emphasize that the novelties reported above are essentially routine updates of the ingredients used in our full spectral synthesis with \starlight. The results reported in this paper would remain essentially the same if computed with the bases used in our studies (e.g., \citealt{gonzalezdelgado16}). The Appendix includes a comparison of the stellar population properties derived by using CSPs or SSPs.

\subsubsection{A note on ages, lookback times and redshifts}
\label{sec:NoteOnAgesAndTimes}

The  information on galaxy evolution analyzed here comes exclusively from the distribution of stellar population ages derived from our spectral fits. Because our galaxies are all basically at $z = 0$, lookback times or the corresponding redshifts\footnote{The $t(z)$ relation used in this paper adopts a flat cosmology:  $\Omega_M$= 0.3, $\Omega_\Lambda$ = 0.7, and H$_0$ = 70 km s$^{-1}$ Mpc$^{-1}$} are equivalent to stellar ages, and indeed will be treated as such throughout this study. We shall say, for instance, that a stellar age of 4 Gyr corresponds to 4 Gyr in lookback time or $z \sim 0.4$.

Subtleties appear, however, when interpreting SFHs. A feature at a certain $t$, say, an increase in the SFR at $t = 4$ Gyr, 
does not necessarily imply something happened in the galaxy itself 4 Gyr ago. For all we know this population could have formed elsewhere (say, a satellite) and accreted at any time between now and 4 Gyr ago. 
Only in events producing in-situ star formation (like wet mergers) one expects the cosmic date $t_e$ of the event to leave a signature at $t = t_e$ in our SFHs. Dry mergers, on the other hand, will change the SFH at the typical age of the accreted stars.

Given that major and minor, dry and wet mergers are such conspicuous elements of the current paradigm for how galaxies assemble their stars over time, it is important to make this distinction.

 
\subsection{The case for a combined mass-morphology analysis} 
\label{sec:Mass_x_Morphology}

\begin{figure}
\includegraphics[width=0.48\textwidth]{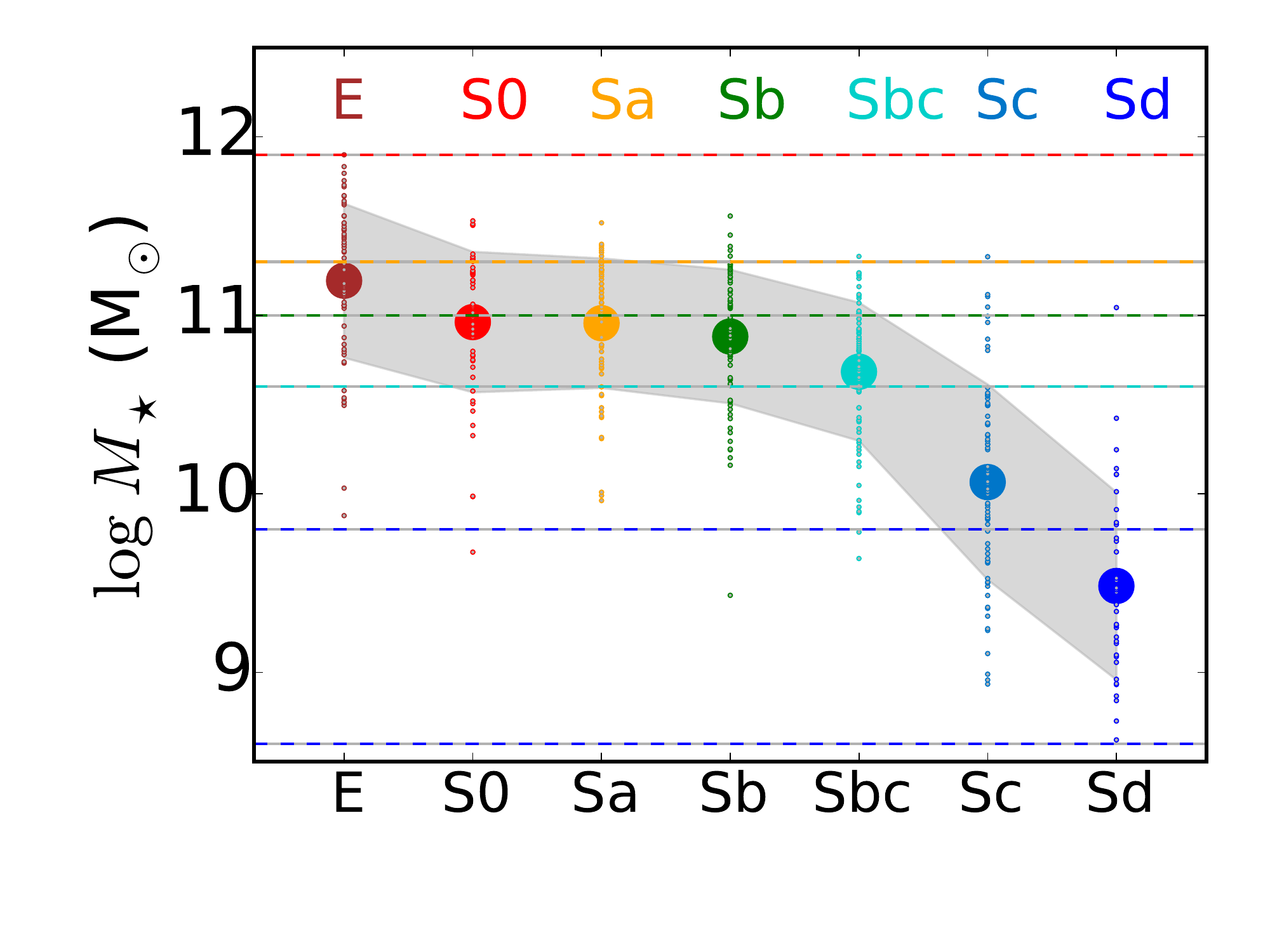}
\includegraphics[width=0.48\textwidth]{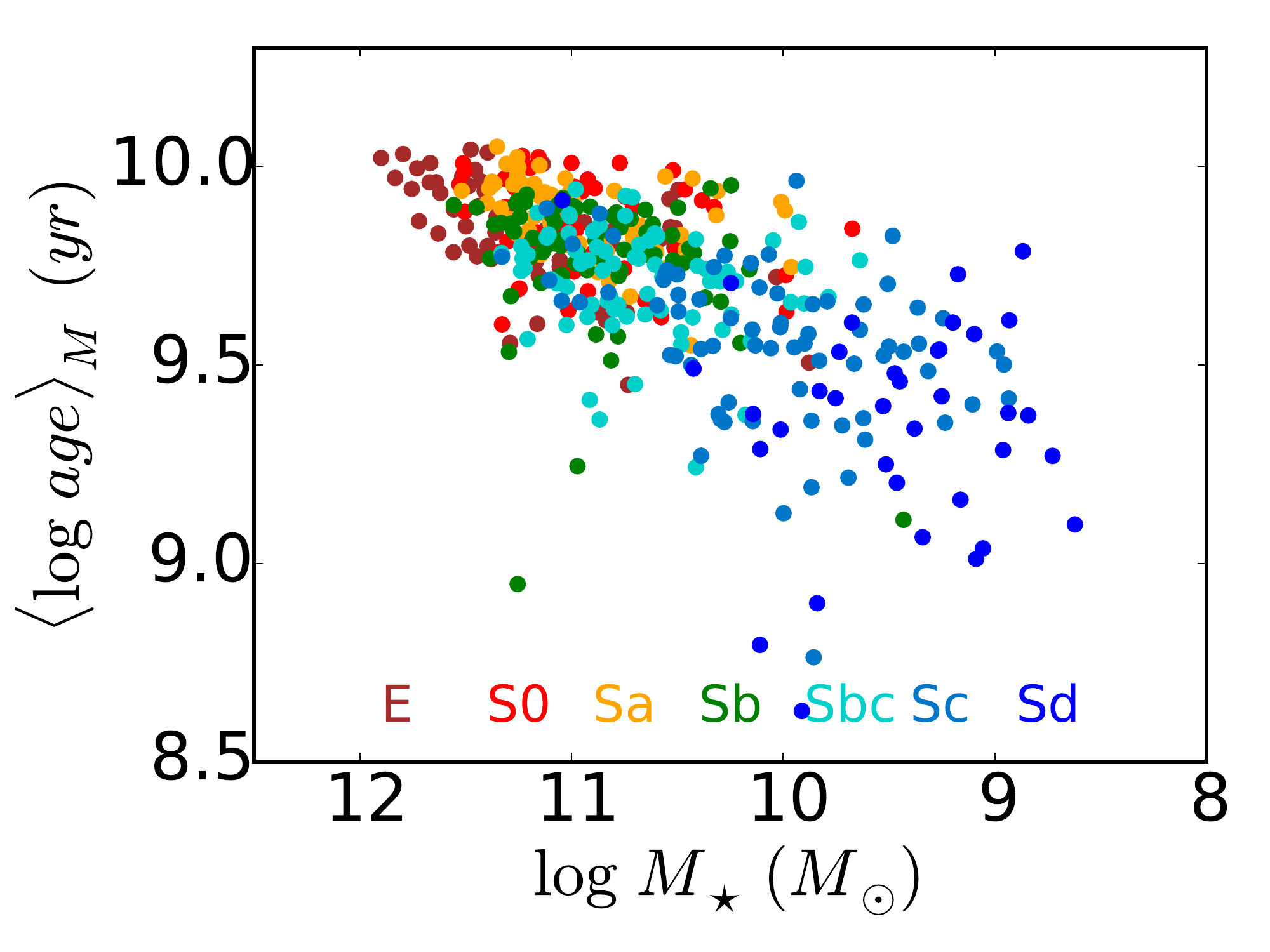}
\caption{Top: Distribution of stellar mass with Hubble type. Large circles are the mean $M_\star$ for each Hubble type, and  the grey area marks the one sigma dispersion in mass for each morphology. Horizontal lines mark the ranges of the five $M_\star$ bins defined in Table \ref{tab:Massdistribution} and used throughout the paper.
Bottom: Mass weighted mean log stellar age versus  $M_\star$ for the 436 galaxies in the sample, color coded by the morphological type.
}
\label{fig:mass}
\end{figure}

\begin{table}
\caption{Number of galaxies in each Hubble type and mass interval.}
\begin{tabular}{lccccccc}
\hline\hline
$\log  M_\star (M_\odot)$ &   E & S0 & Sa & Sb & Sbc & Sc & Sd    \\      
\hline
11.3--11.9  & 29 & 11 & 9   &  6    &  1  & 1  &  - \\
11.0--11.3  & 16 & 19 & 31 & 26  & 16 & 3  & 1  \\
10.6--11.0  & 8  & 15 & 12  & 26  & 33 &  6  & - \\
9.8--10.6    & 8  & 8   & 12  & 14  & 23 & 42 & 9  \\
8.6--9.8      &  -  & 1   & -    & 1    & 1    & 21 & 26 \\
\hline
 total (436)  & 61 & 54 & 64   & 73  & 74 & 73 & 36 \\
 \hline
 $\langle \log M_\star \rangle$ & 11.2& 11.0& 11.0 & 10.9 & 10.7 & 10.1 & 9.5\\
 $\sigma ( \log M_\star)$ & 0.4 & 0.4 & 0.4 & 0.4 & 0.4 & 0.5 & 0.5 \\
\hline\hline
\label{tab:Massdistribution}
\end{tabular}
\end{table}

Throughout the rest of the paper the SFHs derived with the methodology outlined above will be studied as a function of stellar mass and Hubble type. The five bins in $M_\star$ and seven morphological bins defined for this purpose are given in Table \ref{tab:Massdistribution}, which lists the number of galaxies in this mass-morphology space. This approach can be justified in two different ways.

First, although morphology and $M_\star$ are related, the relation is not even approximately univocal.
This is clearly seen in both Table \ref{tab:Massdistribution} and in the upper panel of Fig.\ \ref{fig:mass}, where the distribution of $M_\star$ with morphology is shown. As for the general galaxy population, stellar masses decrease from (on average) $\log M_\star = 11.2$ for ellipticals to 9.5 for the latest types, but the scatter is large. The dispersion within any Hubble type is 0.4--0.5 dex, as marked by the shaded band in  Fig.\ \ref{fig:mass}. Conversely, the dispersion is also large in morphology for a fixed mass. In particular, the bins covering $\log M_\star$ between 9.8 and 11.3 are populated by all Hubble types from E to Sc. Hence, despite the evident correlation between $M_\star$ and morphology, these two properties cannot be considered approximately equivalent, justifying our approach of investigating the effects of both upon our results.

A second motivation to study SFHs as a function of both mass and morphology is illustrated in the bottom panel of Fig.\ \ref{fig:mass}, where we plot the mean age vs.\ mass diagram for our 436 galaxies, color coded by morphology. 
In this plot the mean age of stars in a galaxy is represented by the mass-weighted mean log age, $\langle \log age \rangle_M$, measured at $R = 1$ HLR from the center. As shown by \cite{gonzalezdelgado14a}, properties at 1 HLR match very closely those defined for entire galaxies, so that the plot would hardly change if a galaxy-wide measure of $\langle \log age \rangle_M$ was used instead. Note also that this plot is the physical equivalent of the widely used color magnitude diagram, where absolute magnitude and color play the roles of proxies for $M_\star$ and mean age, respectively.

Fig.\ \ref{fig:mass} shows that the mean age increases with mass, reflecting the well documented pattern of galaxy ``downsizing'' (e.g. \citealt{gallazzi05}). The plot also shows that the spread in ages at any given $M_\star$ correlates with  morphology, with earlier types tending to be older than later types of the same mass  (see also \citealt{gonzalezdelgado15}).
Since the mean age of the stellar population is the first moment of a galaxy's SFH, this illustrates and justifies why our analysis needs to be done in terms of both $M_\star$ and  Hubble type.


\section{Spatially resolved star formation histories}
\label{sec:Results}

This section presents our results for the spatially resolved SFH  as a function of galaxy mass and morphological type. This involves mapping some SFH-function (like luminosity, mass, or SFR) in terms of two spatial coordinates and lookback time $t$, and its dependence on two further variables, $M_\star$ and Hubble type. Even after collapsing the 2D $(x,y)$ information into a single spatial dimension $R$, one is left with a complex manifold whose exploration poses challenges to visualization and scientific analysis.

We chose to first present our results in their most complete form by means of the $R \times t$ diagrams first introduced by \citet{cidfernandes13} as a means of compressing the spatially resolved SFH of galaxies in images (Section \ref{sec:Rxt_diagrams}). These diagrams contain essentially all the science discussed in this paper, but their full comprehension requires summarizing the information in simpler forms. 
One way of carrying out such a simplification is by discretizing the radial and temporal dimensions into a few bins, and this is also presented here (Section \ref{sec:LighAndMassFractions}). Easier to grasp projections of the SFH($t$,$R$,$M_\star$,morphology) manifold are presented in Section  \ref{sec:SFH}.

\subsection{$R \times t$ maps}
\label{sec:Rxt_diagrams}

\begin{figure*}
\includegraphics[width=0.98\textwidth]{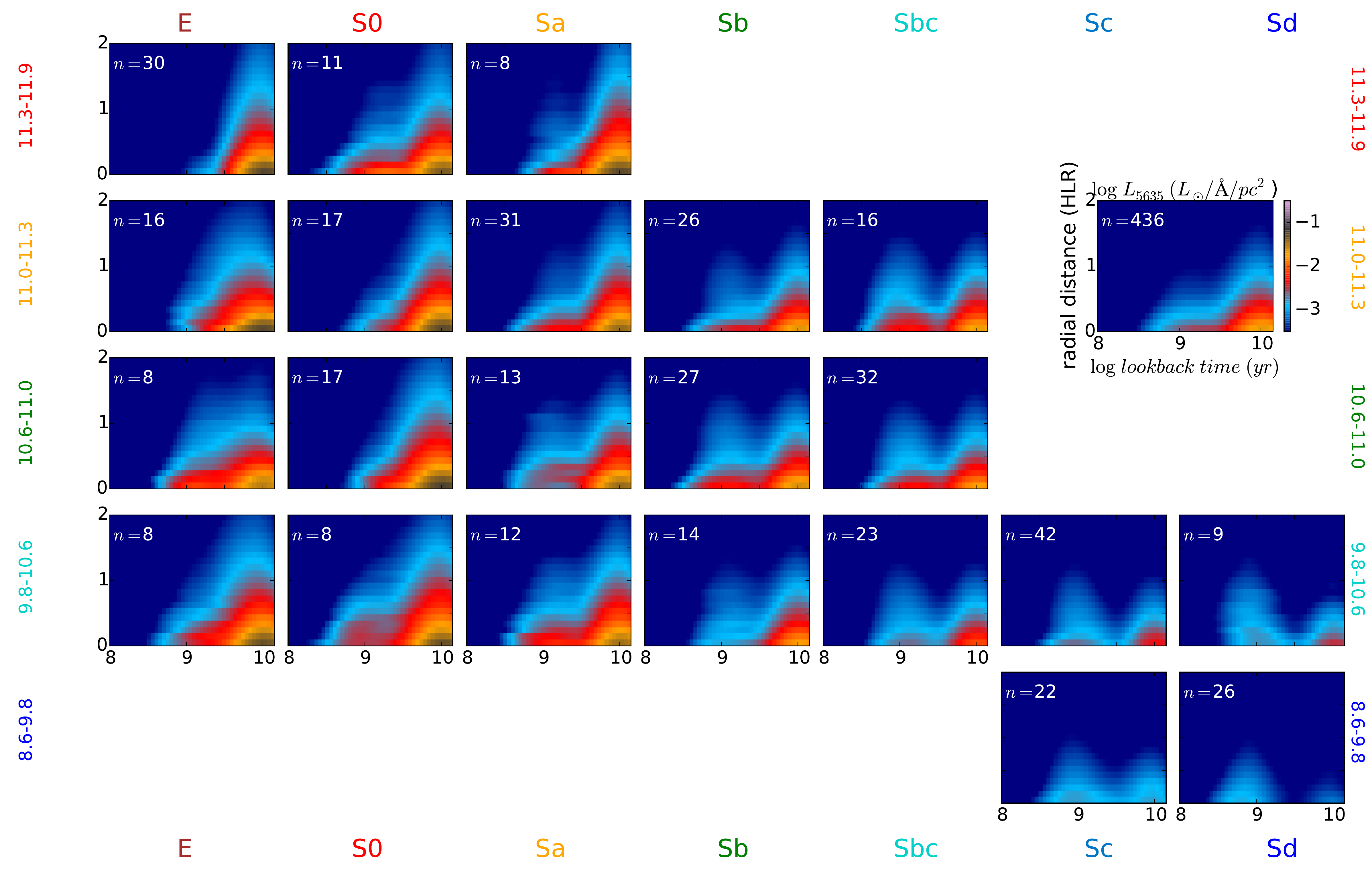}
\includegraphics[width=0.98\textwidth]{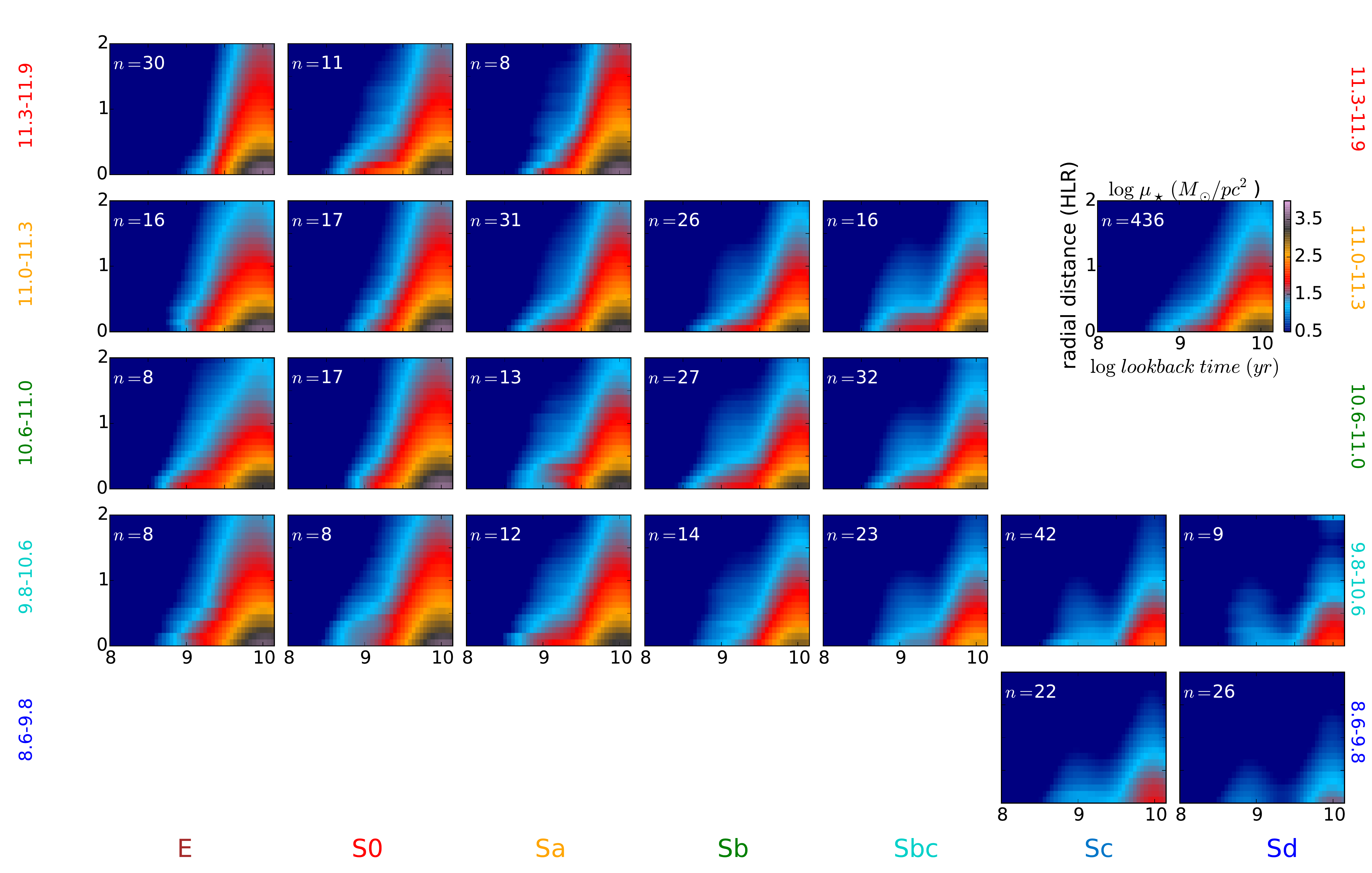}
\caption{  
$R \times t$ diagrams showing azimuthally averaged SFHs in terms of distributions of light and mass as a function of the  stellar ages ($=$ lookback time) and distance  to the nucleus (in HLR units).
The intensity of the map shows the luminosity (corrected for extinction) at 5635 \AA\ per unit area (${\cal L}_{\lambda5635}$, upper panels), and stellar mass formed per unit area ($\mu_\star$, bottom panels). 
Each frame represents the average results for bins in the $M_\star$-morphology plane, with stellar masses decreasing from the top down, and Hubble type running from E to Sd from left to right. The number of galaxies is indicated in each panel.
Average maps for the entire sample are shown on the top-right panels which also show the color bar.
%
}
\label{fig:2DSFH}
\end{figure*}

For a generic SFH-related function $F(x,y,t,Z)$, with age, metallicity, and spatial dimensions, the $R \times t$ map is obtained by first collapsing $F$ along the $Z$ axis, and then azimuthally averaging the $F(x,y,t)$ image along elliptical annulli of varying $R$. The result is a map in $(R,t)$ space. Cuts at fixed $t$ give radial profiles of $F$ at the given lookback time, while for a constant $R$ one obtains how $F$ is distributed among populations of different ages. For visualization purposes, and also to mitigate degeneracies intrinsic to stellar population synthesis (e.g,. \citealt{worthey94},  \citealt{cidfernandes05}, \citealt{asari07}), we apply a  gaussian smoothing of 0.5 dex in FWHM in $\log t$. As in our previous studies (e.g,. \citealt{perez13},  \citealt{cidfernandes13}, \citealt{gonzalezdelgado14a}), $R$ is expressed in units of the galaxy's HLR, a convenient metric when averaging radial information for different galaxies.

Fig.\ \ref{fig:2DSFH} presents two sets of $R \times t$ maps representing the mean SFH obtained for galaxies in bins in the mass-morphology space. 
Hubble type runs from E to Sd from left to right, while $M_\star$ grows upwards. The layout is exactly as that of Table \ref{tab:Massdistribution}, except that bins containing less than 8 galaxies are not shown.
Diagrams in the top half of Fig.\ \ref{fig:2DSFH}
 express the 2D SFH in terms of the continuum luminosity per unit area (${\cal L}_{\lambda5635}$)  associated with stars of a given age and radial position, as indicated in the reference panel in the top right, where the average for all 436 galaxies is shown. Diagrams in the  bottom half of the figure  show the corresponding images in terms of stellar mass surface density ($\mu_\star$).

Some clear differences in the SFHs along the Hubble sequence and galaxy mass are visible in these plots. For instance, for a fixed Hubble type one sees progressively more extended SFHs towards lower $M_\star$, the signature of downsizing. For galaxies in a same $M_\star$-bin, SFHs appear to be more extended in time among later Hubble types. This effect is more clearly visible in the outer regions, associated with the disks of spirals, than at small $R$, presumably dominated by the bulges or the thick disks. 

The richness of information in these maps is huge. Indeed, it can be somewhat overwhelming, in the sense that it can be hard to interpret and strongly dependent upon seemingly unimportant choices, like whether to plot absolute or specific quantities, to show light or mass-related properties, use logarithmic or linear scales, etc. This is why the next sections will present basically the same information contained in Fig.\ \ref{fig:2DSFH}, but expressed in ways that allow a fuller understanding of the results.


\subsection{Stellar population components}
\label{sec:LighAndMassFractions}

\begin{figure*}
\includegraphics[width=\textwidth]{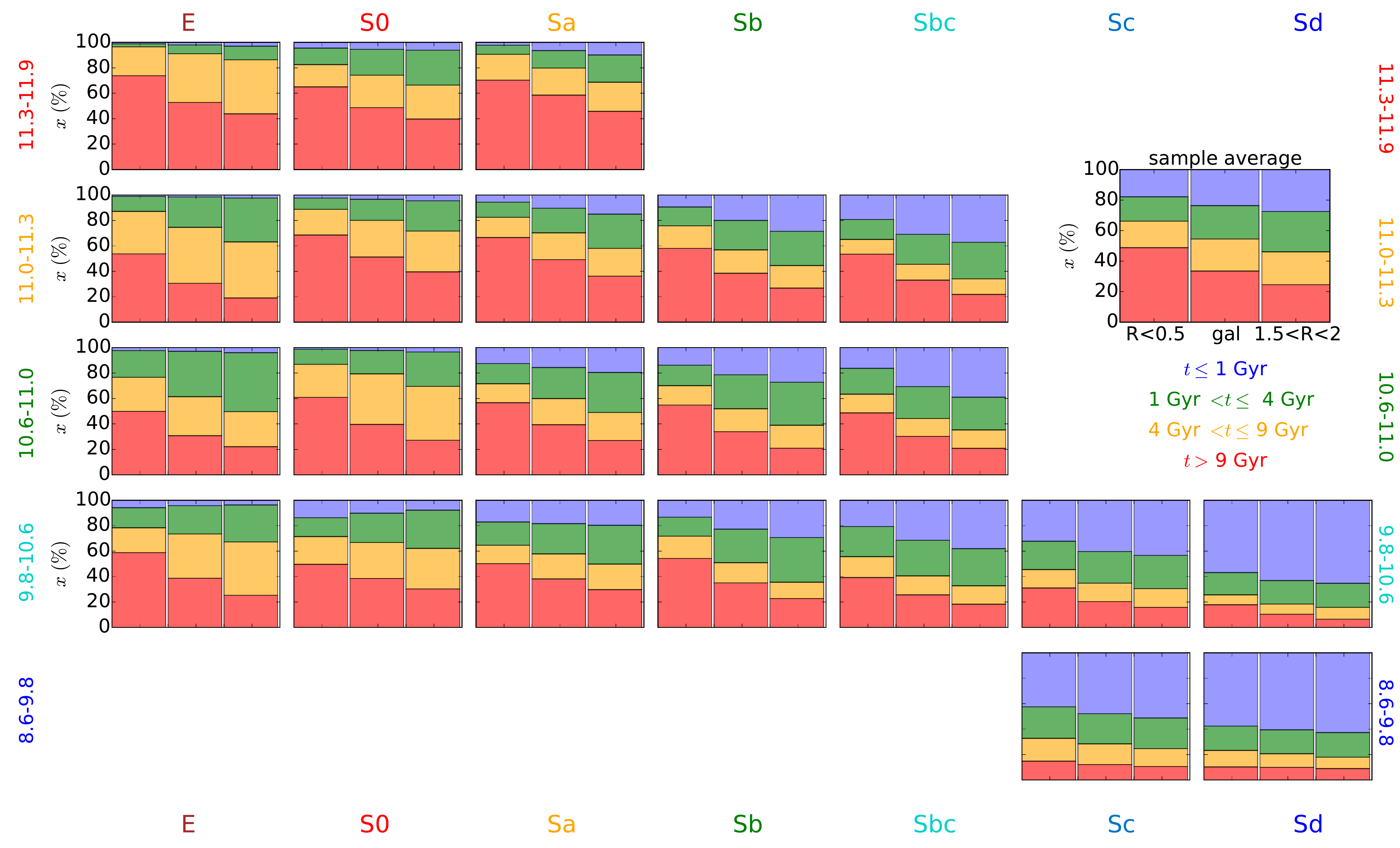}
\includegraphics[width=\textwidth]{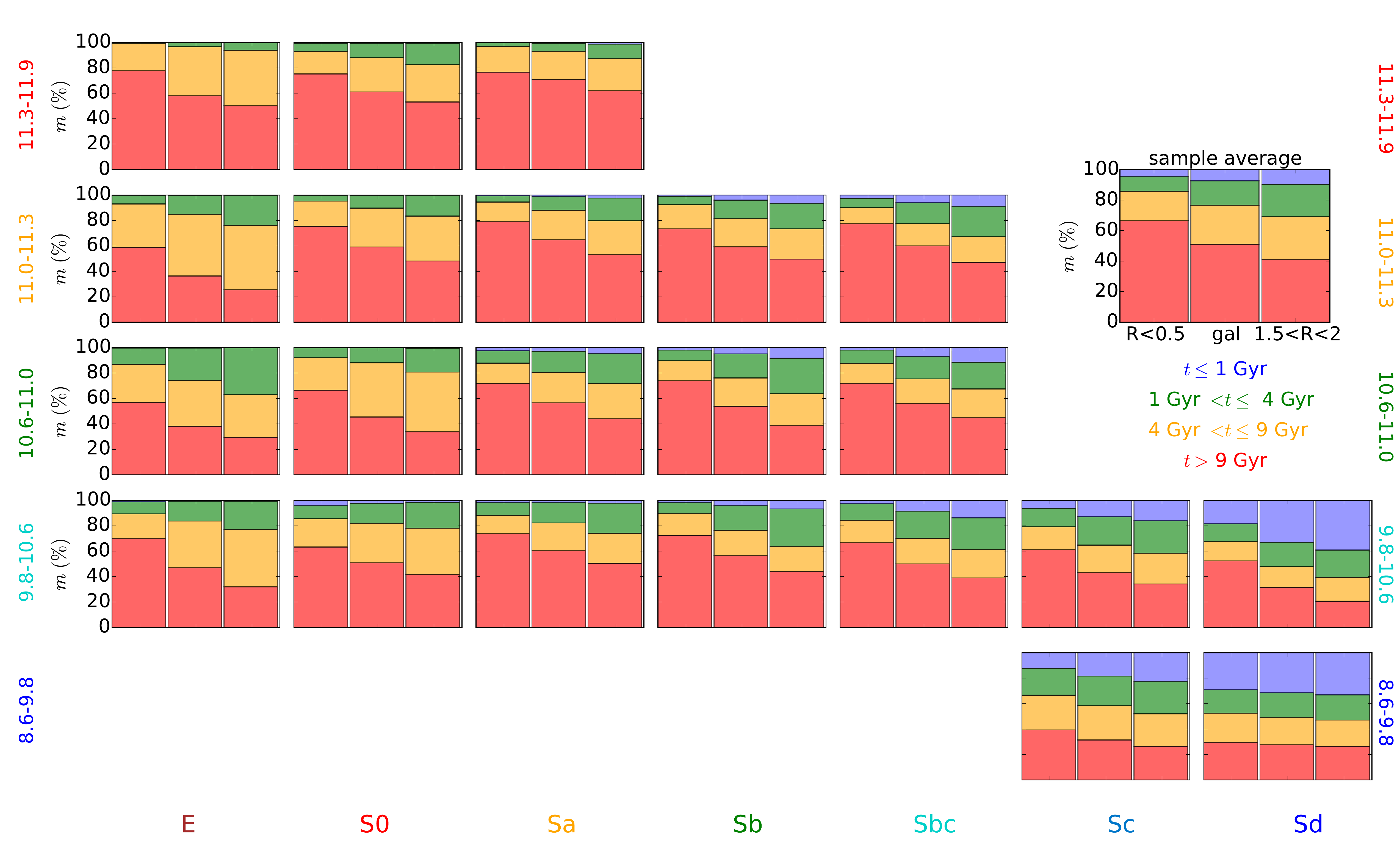}
\caption{
Average light (upper panels) and mass (bottom) fractions due to stars in different age  and radial ranges.
Each panel corresponds to a bin in the galaxy mass-morphology plane, exactly as in Fig.\ \ref{fig:2DSFH}.
The three bar chart histograms in each panel correspond to different galaxy regions: the inner region $R \leq 0.5$ HLR (left bar);  the whole galaxy $R \leq 2$ HLR (central), and outer regions $1.5<R<2$ HLR (right).
Colors represent the following age ranges (in Gyr): 
$< 1$ in blue, 1--4 in green,  4--9 in orange, and $ >9$ in red.
Panels on the top right show averages for the full sample. 
} 
\label{fig:pop}
\end{figure*}

The light ($x$)  and mass ($m$) fractions are two further examples of $F(x,y,t,Z)$ functions which can be represented in $R \times t$ diagrams like those in Fig.\ \ref{fig:2DSFH}. Instead, Fig.\ \ref{fig:pop} presents an alternative way of visualizing  results for these two functions by 
discretizing both the $R$ and $t$ axes  into a few relevant ranges. 
Each panel shows three rectangular ``bar charts'',  each corresponding to a spatial region: $R < 0.5$ HLR (left), $0 \le R < 2$ HLR (middle), and $R > 1.5$ HLR (right). The middle charts are meant to represent the galaxy as a whole, while those on the right represent distances to the nucleus similar to that of the solar neighborhood\footnote{1.75\,HLR is $\sim$8 kpc for the mean HLR of the CALIFA Sbc-Sc galaxies.}.
The age information is compressed into four color-coded representative ranges: 
$t\leq 1$ Gyr (blue), $1 < t \leq 4$ Gyr (green), $4 < t \leq 9$ Gyr (orange), and
$t > 9 $ Gyr (red). In terms of  redshift these ranges correspond to approximately $z \leq 0.1$, 
$0.1 < z \leq 0.4$, $0.4 < z \leq1.5$,  and $z > 1.5$, 
respectively. For reference, these cosmic epochs can be associated with 
(i) the present time, 
(ii) the growth of the thin disk in spirals,
(iii) the size growth epoch, and 
(iv) the early  formation of the galaxy halo and/or core.

The layout of Fig.\ \ref{fig:pop} is again as in Table \ref{tab:Massdistribution}, with Hubble type varying from column to column and $M_\star$ decreasing from top to bottom rows. Panels on the top half show the percentage contributions in light of populations in these four age ranges and three spatial regions. These are discussed first.

\subsubsection{Light fractions} 

The top panels of Fig.\ \ref{fig:pop} show a steady progression of young populations (blue, $t \leq 1$ Gyr) along the Hubble sequence, a variation that is stronger than with stellar mass. Breaking down the evolutionary information in terms of radial locations we find the following.

\begin{enumerate}

\item Galaxy-wide average ($R < 2$ HLR, middle charts): 
The contribution of $< 1$ Gyr populations varies from $\sim 1 \%$ in E to $\sim 60 \%$ in Sd systems, while
it hardly changes (23 to 21\%) from the least to the most massive Sb. 
Old populations show the opposite behavior, with  contributions of $> 4$ Gyr stars (orange + red) increasing from $20 \%$ for Sd to $90 \%$ for the most massive E. These  populations show a stronger dependence with $M_\star$ in Sa's, increasing from 58 to 80\% over the $M_\star$ range sampled, and similarly for ellipticals. S0's, however, do not show any systematic dependence with $M_\star$, as can also be noted in Fig.\ \ref{fig:2DSFH}.
  
\item Outer regions ($1.5 < R < 2$ HLR, right charts): 
$x(t < 1 {\mathrm Gyr})$ ranges from $\sim 3 \%$ in E to $\sim 65 \%$ in Sd, but only from 29 to 24\%  from the least to the most massive Sb. Stars older than 4 Gyr increase their contribution from 20\% for Sd to 86\% for the most massive E. 
These results are very similar to those for the $R = 0$--2 HLR region, indicating that in terms of light the ``disk'' plays a dominant role in SFHs derived for entire galaxies.

\item Central region ($R < 0.5$ HLR, left charts):
In this region, $x(t < 1 {\mathrm Gyr})$ is smaller and $x(t > 4 {\mathrm Gyr})$ is larger than in the ``disk regions'' for all morphological types and $M_\star$ bins. This is also the case for the mass fractions (lower panels). This is a clear indication of the inside-out growth process in CALIFA galaxies \citep{perez13, gonzalezdelgado15}, although this is not so evident in the two extreme morphology-$M_\star$ bins, namely, Sd's with $\log M_\star = 8.6$--9.8 and E's with $\log M_\star = 11.3$--11.9. Regarding the younger populations, $x(t < 1 {\mathrm Gyr})$ grows from $\sim 1 \%$ in E's to 13\% in Sb's and 57\% among Sd's. Stars older than 4 Gyr vary from 23\% in Sd's to 96\% in massive ellipticals. The dependence with $M_\star$ is very clear for ETGs, for which $x(t > 4 {\mathrm Gyr})$ changes from 64 to 91\% from the least to the most massive Sa's, and similarly for E's.

\end{enumerate}

\subsubsection{Mass fractions} 

Panels in the bottom half of Fig.\ \ref{fig:pop} show that most of the stellar mass formed very early on, with very little mass  in stars younger than 1 Gyr.  Sd's and the less massive Sc's are exceptions, with $m(t < 1 {\mathrm Gyr})$ fractions of up to 20--30\%, with a tendency to increase towards the outer regions. The main tendencies seen in these plots can be summarized as follows:

\begin{enumerate}

\item Galaxy-wide average ($R < 2$ HLR): 
Except for the latest types, the fractions in the $> 9$ Gyr bin (painted in red) are the largest amongst our four age-ranges, 
with $m(t > 9\, {\mathrm Gyr}) \sim 40$--70\%. This old component increases with $M_\star$.
The contribution of 4--9 Gyr stars (orange) shows a clear tendency to increase from late to early Hubble types, with little or no dependence on $M_\star$. 
Stars in the 1--4 Gyr range (green) generally account for $\lesssim 20\%$ of the mass, with a tendency to decrease in strength at the highest $M_\star$ bins. 
\item Outer regions ($1.5 < R < 2$ HLR): 
The mass fractions in these outer regions behave similarly to the previous case (0--2 HLR).
The oldest populations decrease their fractions slightly (signaling negative mean age gradients explicitly studied in  \citealt{gonzalezdelgado15}) but still account for much of the mass. Noticeably, the $m(t > 9\, {\mathrm Gyr})$ fractions for E and S0 are smaller than for Sa, while their 4--9 Gyr populations are larger.

\item Central region ($R < 0.5$ HLR): 
$t > 9$ Gyr stars dominate the mass in the central regions, accounting for over 60\% in most spirals and up to 80\% in the most massive galaxies. 
The exceptions are, again, Sd's  and low mass Sc's.
Except at the highest $M_\star$ bin, this old component contributes less in E's and S0's than in early type spirals. 

\end{enumerate}


\section{SFH as a function of mass and Hubble type}
\label{sec:SFH}

Having presented how our 2D SFHs vary with galaxy mass {\em and}  Hubble type, we now simplify the analysis by  investigating variations as a function of $M_\star$ {\em or} morphology separately. These projected views of SFHs in the mass-morphology space aid the interpretation of the results presented above. 

This section presents  SFHs in terms of mass fractions (Section \ref{sec:SFH_MassFractions}), absolute (Section \ref{sec:SFH_SFRs}) and specific (Section \ref{sec:SFH_sSFRs}) SFRs, and star formation intensities (Section \ref{sec:SFH_SFI}) as a function of lookback time. In order to emphasize evolutionary aspects, the spatial analysis is simplified to the same three radial regions used in the discussion of Fig.\ \ref{fig:pop}, namely, $R<0.5$ HLR, 0--2 HLR, and 1.5--2 HLR, corresponding to the nuclear regions, the whole-galaxy, and outer regions.


\subsection{SFHs: Mass fractions}
\label{sec:SFH_MassFractions}

\begin{figure*}
\includegraphics[width=\textwidth]{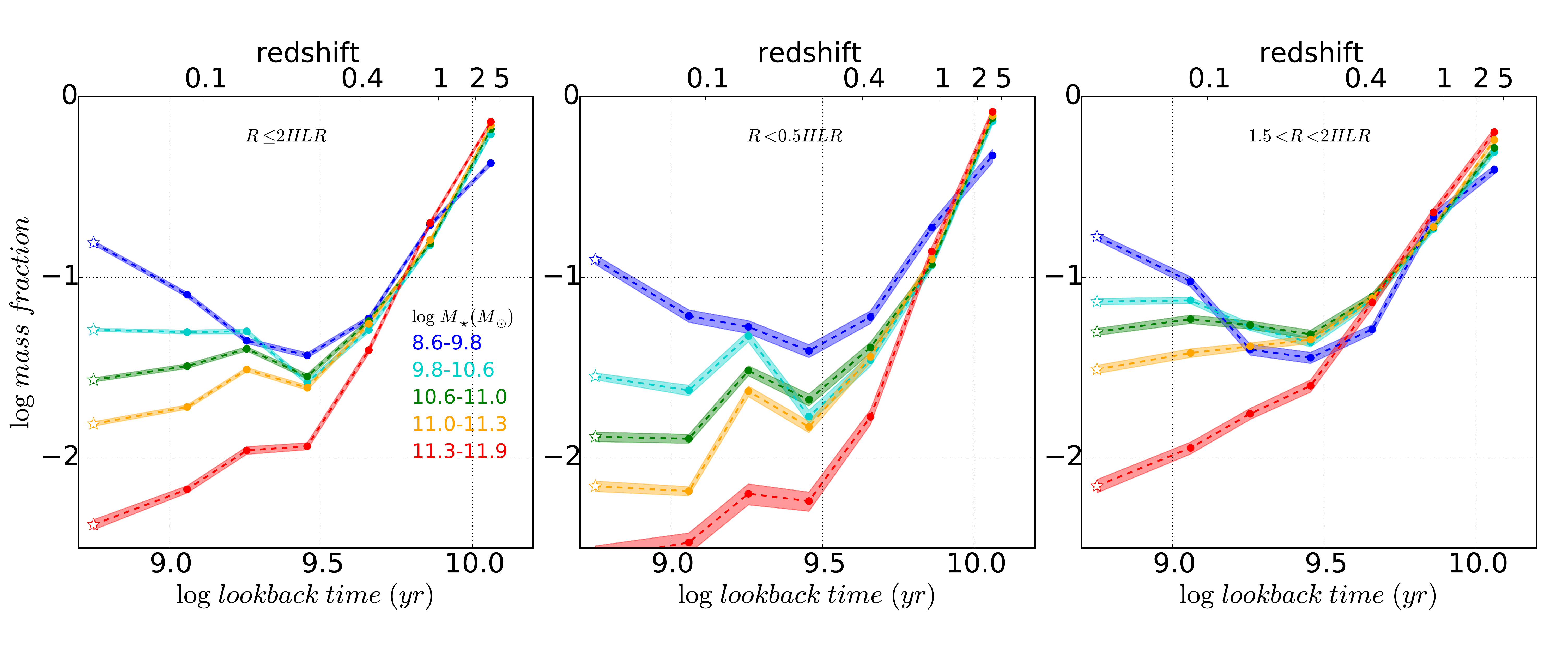}
\includegraphics[width=\textwidth]{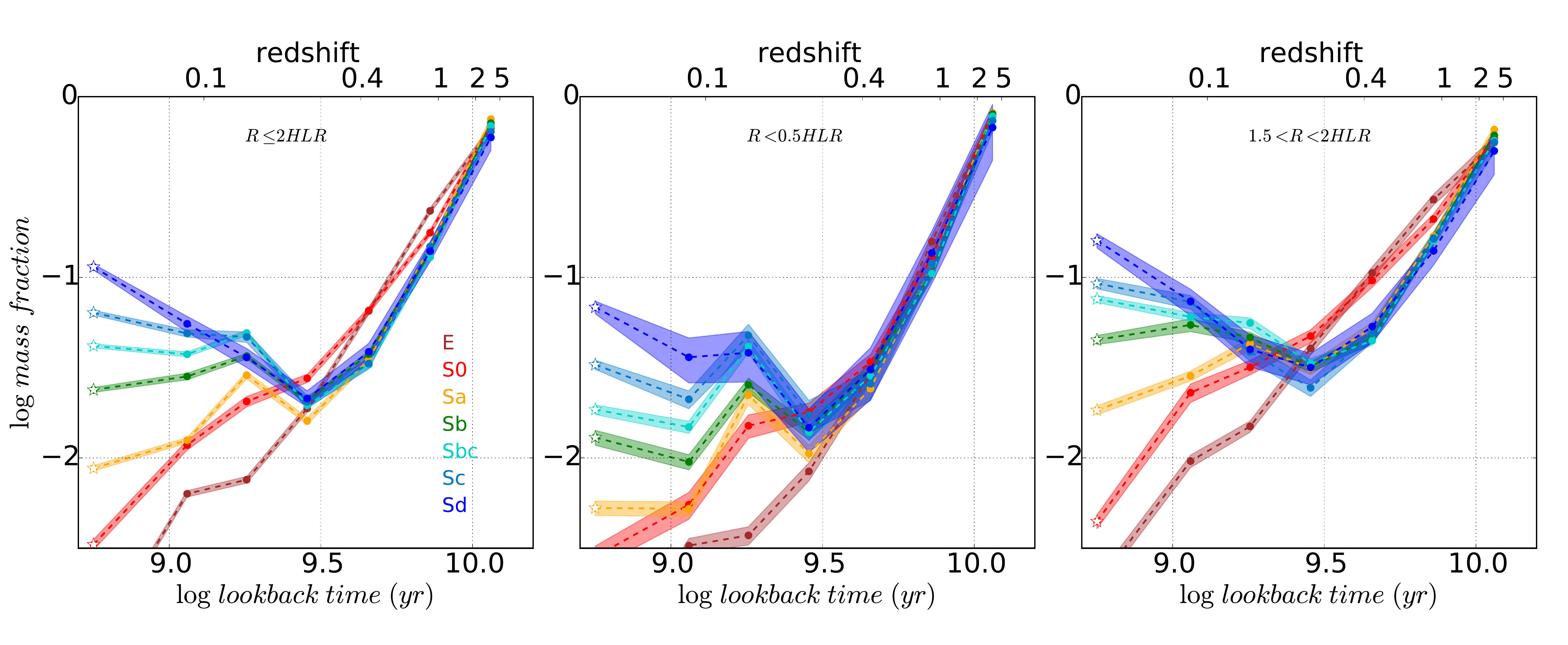}
\caption{ 
The star formation history is represented here by the fraction of stellar mass formed in each epoch, $m(t)$, averaged over  different $M_\star$ (upper panels) and Hubble type (bottom) bins.
Left, middle, and right panels show the average $m(t)$ curves for different regions in the galaxy: the whole galaxy (left), the inner regions (middle), and outer regions (right).
The shaded bands around the mean curves represent $\pm$ the error in the mean, computed as the r.m.s. dispersion of the corresponding $m(t)$ values divided by the square root of the number of galaxies in each bin.
} 
\label{fig:massfraction}
\end{figure*}

Fig.\ \ref{fig:massfraction} shows stellar mass fractions ($m$) as a function of lookback time. Left, middle, and right panels correspond to the whole galaxy, central and outer regions, respectively. Top panels show the average $m(t)$ curves for galaxies in our five $M_\star$ bins, while the ones in the bottom stack objects by Hubble type.
These curves are actually histograms, with each point representing a 0.2 dex wide bin in  $\log t$, but it is visually clearer to connect the points. Note also that, because of the logarithmic sampling in $t$ and because these are not masses but mass fractions, the results in Fig.\ \ref{fig:massfraction} should not be read as ${\rm SFR}(t)$ curves. SFRs are presented in the next section.

Inspection of the top panels show that the highest mass fractions invariably occur at the earliest times. Subsequent star formation varies systematically with $M_\star$,
 with the low $M_\star$ galaxies forming stars over extended periods of time,  and high $M_\star$ galaxies exhibiting the fastest decline in $m(t)$. This footprint of cosmic downsizing is more clearly observed in the inner regions 
(central panels), which (except for the latest Hubble types) are mainly associated with the spheroidal component. The decline in $m(t)$ among the most massive galaxies is visibly faster in the inner regions than away from the nucleus. The curves for the outer regions are similar to those obtained for the galaxy as a whole.

The bottom panels of Fig.\ \ref{fig:massfraction} show that the behaviour with morphology mimics that with $M_\star$, 
with $m(t)$ peaking at the earliest time and subsequent star formation increasing systematically from early to late types. However, there are at least two important differences: {\em (a)} In the inner regions, the $m(t)$ curves are very similar all the way from their peak at the oldest ages to $\sim 4$ Gyr ago.  After this epoch, $m(t)$ curves have the same qualitative behaviour whether binned by $M_\star$ or morphology, with ellipticals and massive galaxies declining rapidly, while spirals and low $M_\star$ systems stretch their star formation activity over a longer period of time. {\em (b)} In the outer regions, E's and  S0's have higher mass fractions than later types in the $t = 4$ and 7 Gyr age bins.
Our analysis does not say whether these stars were formed in-situ at these epochs or accreted from other systems, but it does show that they are now part of the envelopes of E and S0 galaxies.

In a first approximation, the decline of $m(t)$ in spirals and inner regions of E and S0 at early times can be fitted by a decaying function of cosmic time much like the so called $\tau$-models used in parametric modeling of SFHs, particularly in high-$z$ studies  \citep[]{maraston10,  maraston13}. However, the SFHs in Fig.\ \ref{fig:massfraction} clearly show that additional star formation has taken place at $t < 4$ Gyr. Evidence of such extended phases of star formation have already been reported in the literature. For instance, a global star forming episode in the last 2--4 Gyr has been detected in the disk of M31 \citep{williams15, bernard15}. Also, modeling the stellar populations of a sample of spiral galaxies with $M_\star > 10^{10}$, \citet{huang13} find that, in addition to an exponentially declining SFR, the observed $D_{4000}$ break and H$_\gamma$ indices require an extra burst of star formation in the last 2 Gyr to be adequately fitted. This burst is more significant in the outer disk of their galaxies, and accounts for a small fraction of the total mass formed. 
Such late bursts of star formation are also found in other contexts. For instance, in the photometric analysis of $z < 1.2$ galaxies by \citet{walcher08}.  Our results support these earlier findings, and show the advantages of obtaining the SFH with non-parametric methods.


\subsection{SFHs: Star formation rates}
\label{sec:SFH_SFRs}

\begin{figure*}
\includegraphics[width=\textwidth]{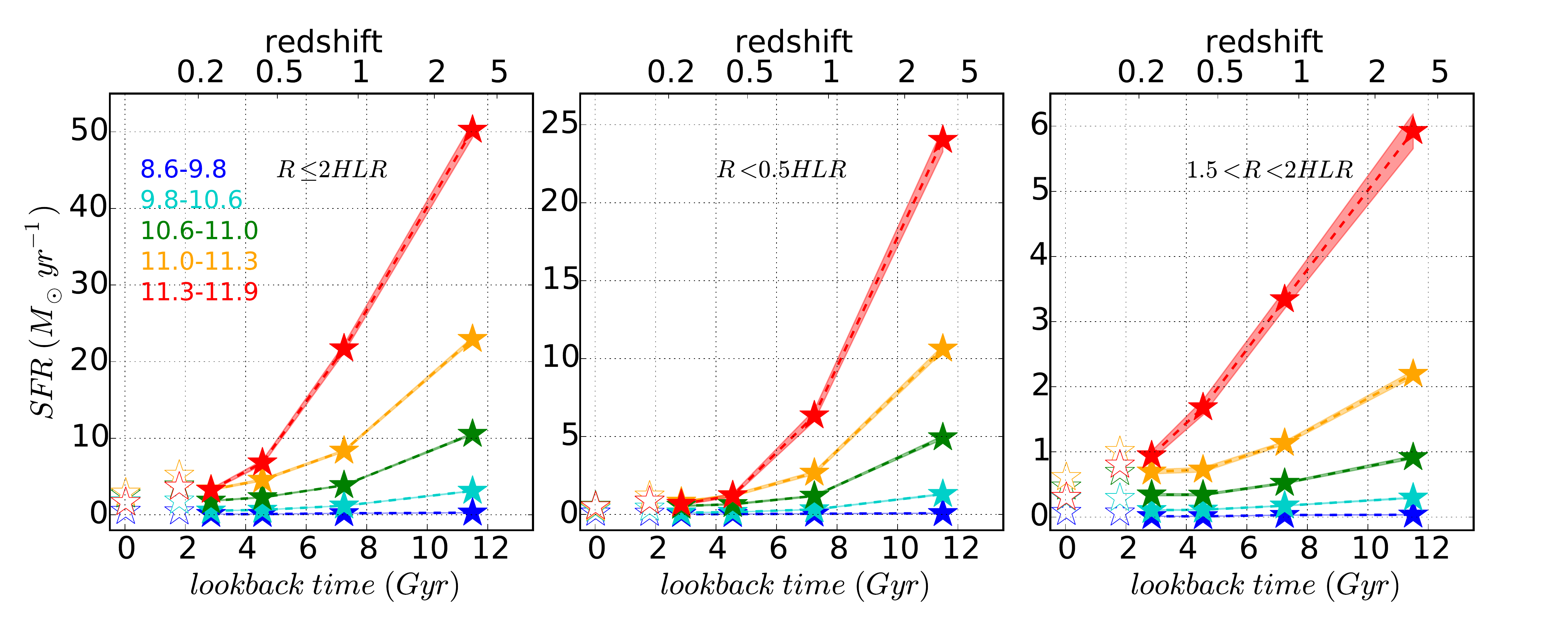}
\includegraphics[width=\textwidth]{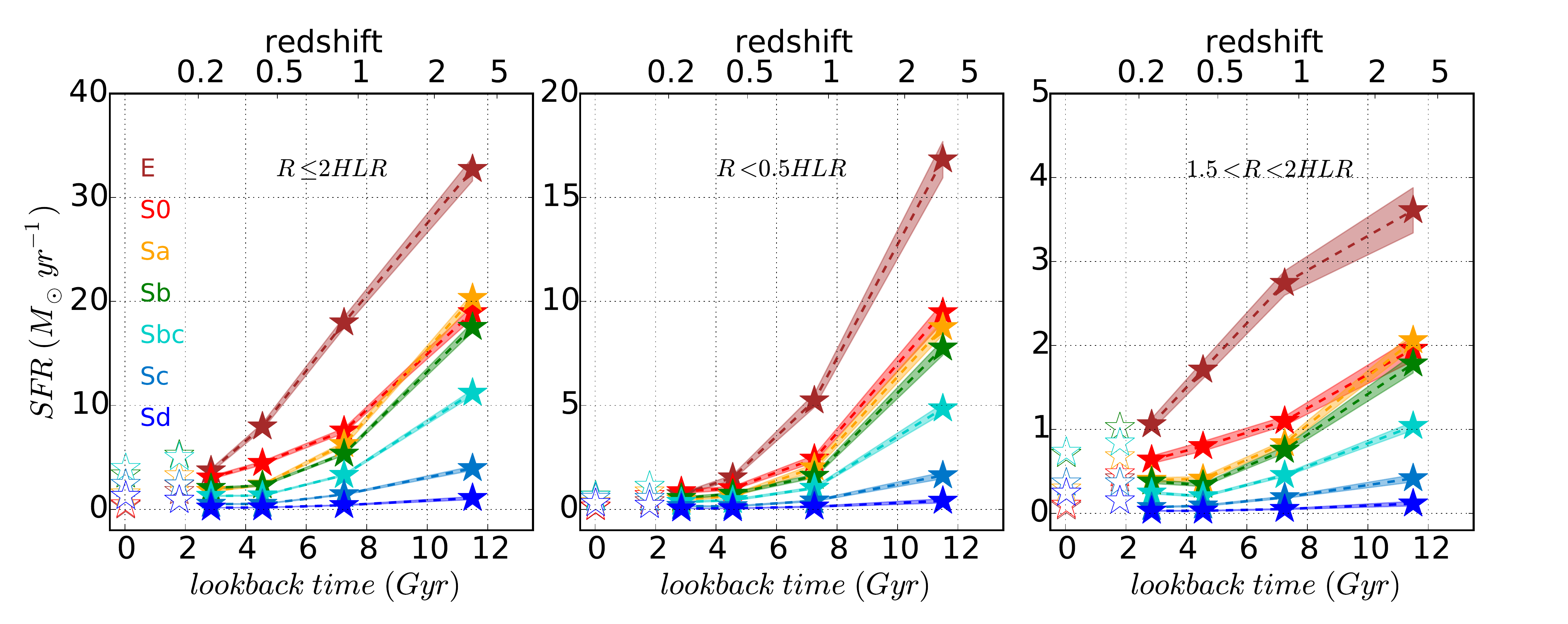}
\caption{The star formation rate (SFR) in each epoch averaged for different bins in stellar mass (upper panels) and Hubble type (bottom). 
As in Fig.\ \ref{fig:massfraction}, left, middle and right panels shows results for the whole galaxy, the inner $R < 0.5$ HLR, and radial distances between 1.5 and  2 HLR, respectively.
 Filled stars show SFR values at the original $t$-sampling of our base of composite stellar populations. The open star at $t = 1.8$ Gyr shows the SFR obtained by averaging over all populations younger than 2.2 Gyr. Similarly, the open star at the left-end of the ${\rm SFR}(t)$ curves show the value obtained averaging over components younger than 32 Myr. The shaded regions represent the statistical uncertainty on the mean curves.
} 
\label{fig:SFR}
\end{figure*}

The SFR is calculated at each epoch as the ratio of the stellar mass formed to the duration of the  ``square burst'' event represented by the corresponding CSP in our base. These are therefore time-averaged SFRs during each time interval. Though mathematically related to the mass fractions discussed above, ${\rm SFR}(t)$ curves are perhaps easier to interpret, besides being more directly connected to observables.
Recall that, as previously noted in Section \ref{sec:NoteOnAgesAndTimes}, we cannot know whether the stars within a given age bin $t$ were formed in-situ at lookback time $t$ or were accreted from other systems at any time since $t$. This caveat should be kept in mind when evaluating our ${\rm SFR}(t)$ curves and indeed all the SFH descriptors explored in this paper.

Fig.\ \ref{fig:SFR} shows the time evolution of the SFR for $M_\star$ (top panels) and morphology bins (bottom), again divided into global (left), central (middle), and outer (right) regions. The plot shows that SFRs decline rapidly as the universe evolves. The top panels show that, at any epoch, the SFR scales with $M_\star$. 
The most massive galaxies thus have the highest SFRs, reaching $\sim 60 M_\odot\,$yr$^{-1}$ at $z \ga 2$,  declining by a factor 2 at  $z=1$, and by a factor $>10$ by the time we get to $z = 0$.
In low mass galaxies the SFR at  early epochs is quite small, $\sim 0.3 M_\odot\,$yr$^{-1}$, rising by a factor of $\sim 2$ in recent times (hardly noticeable because of the scale).

The SFR in the inner regions (top central panels) shows an even faster decline  than the total (whole-galaxy) ${\rm SFR}(t)$. 
At early epochs this inner region contributes significantly ($\sim 40\%$) to the total SFR, except for the lowest $M_\star$ galaxies, for which the $R < 0.5$ HLR accounts for only a quarter of the total SFR. 
In contrast, SFR$(R < 0.5 {\mathrm{HLR}})$ presently  represents $< 20\%$ of the total. 
SFRs also decline with time in the outer regions (right panels), although this decline is only significant for our highest $M_\star$ bin. In galaxies with $M_\star < 10^{11}$ these outer regions can be approximated by a roughly constant SFR throughout the ages. Notice also that at early times these regions account for only $\sim 10 \%$ of the galaxy SFR.

Averaging the SFR$(t)$ functions by morphology (bottom panels in Fig.\ \ref{fig:SFR}) leads to similar results as averaging by $M_\star$, in particular for the inner regions. As a consequence of the relation between morphology and mass (Fig.\ \ref{fig:mass}), the scaling of SFRs with $M_\star$ is preserved with Hubble types, with ellipticals showing the highest SFRs at $z >0.5$. In fact, S0, Sa, and Sb galaxies, that span similar $M_\star$ values (see Table \ref{tab:Massdistribution}), have almost identical SFRs  at early times.

The SFR in the outer regions show interesting differences with respect  to the inner regions, and with respect to the stacking by $M_\star$. While in most cases ${\rm SFR}(t)$ can be approximated as exponentially decaying for lookback times between 12 and 4 Gyr, the SFR$(1.5<R<2 {\mathrm{HLR}})$ in ellipticals cannot be explained by these simple laws, as can be seen by the excess   SFR between $\sim 4$ and 7 Gyr ago.
 
This suggests that E galaxies are actively forming (or accreting) stars in their outskirts between $z = 1.5$ and 0.5. 
S0's also show an excess of SFR at these epochs with respect to galaxies of similar mass, like Sa's and Sb's. 
This same effect was  noted in the bottom-right panel of Fig.\ \ref{fig:massfraction}, where both E's and S0's show larger mass fractions in the 4 and 7 Gyr bins. 

Another interesting feature seen in Fig.\ \ref{fig:SFR} is the increase in  SFR$(1.5<R<2 {\mathrm{HLR}})$ in the last 3 Gyr of spirals. This effect is more evident in Sb-Sc galaxies, and  suggests that the disk of spirals has experienced a rejuvenation in the last 3 Gyr, achieving SFRs similar to those of earlier epochs.
Overall, however, the outer disk of late type spirals seem to have undergone relatively little variation in the levels of SFR throughout their lives.


\subsection{SFHs: Specific star formation rates}
\label{sec:SFH_sSFRs}

\begin{figure*}
\includegraphics[width=\textwidth]{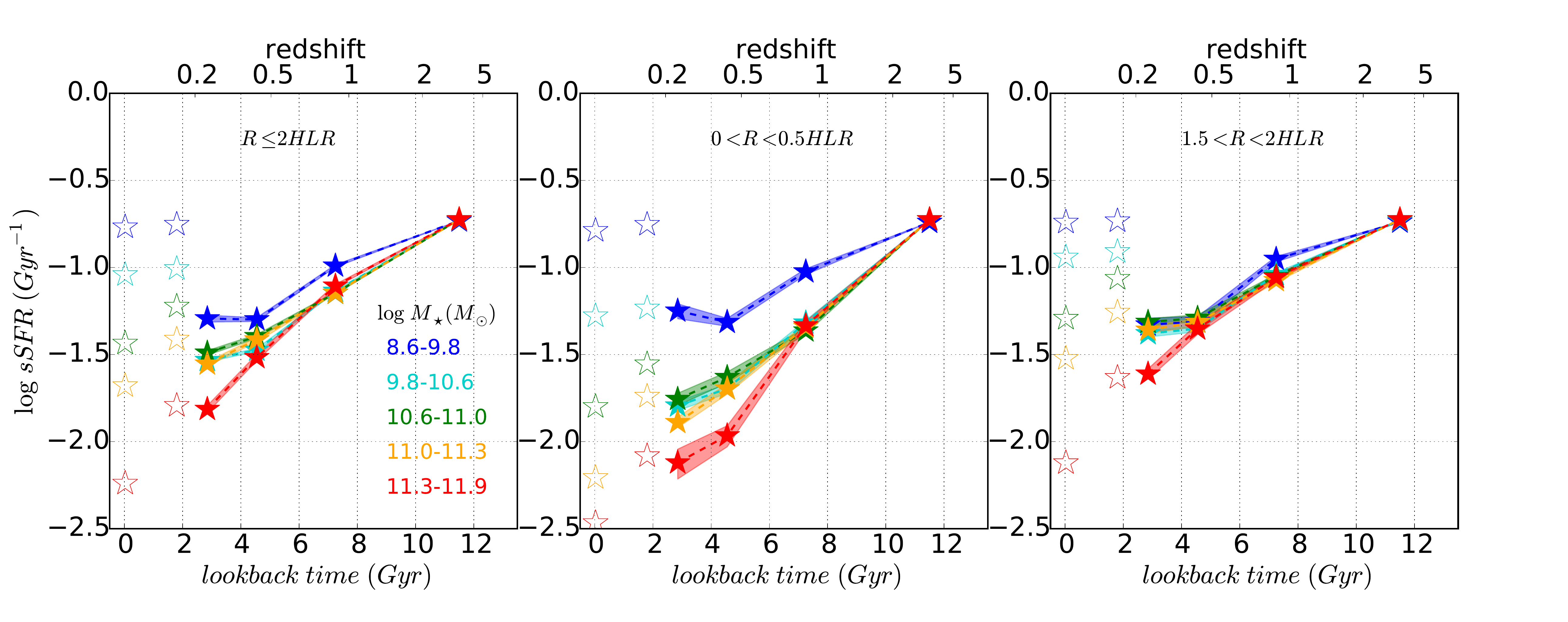}
\includegraphics[width=\textwidth]{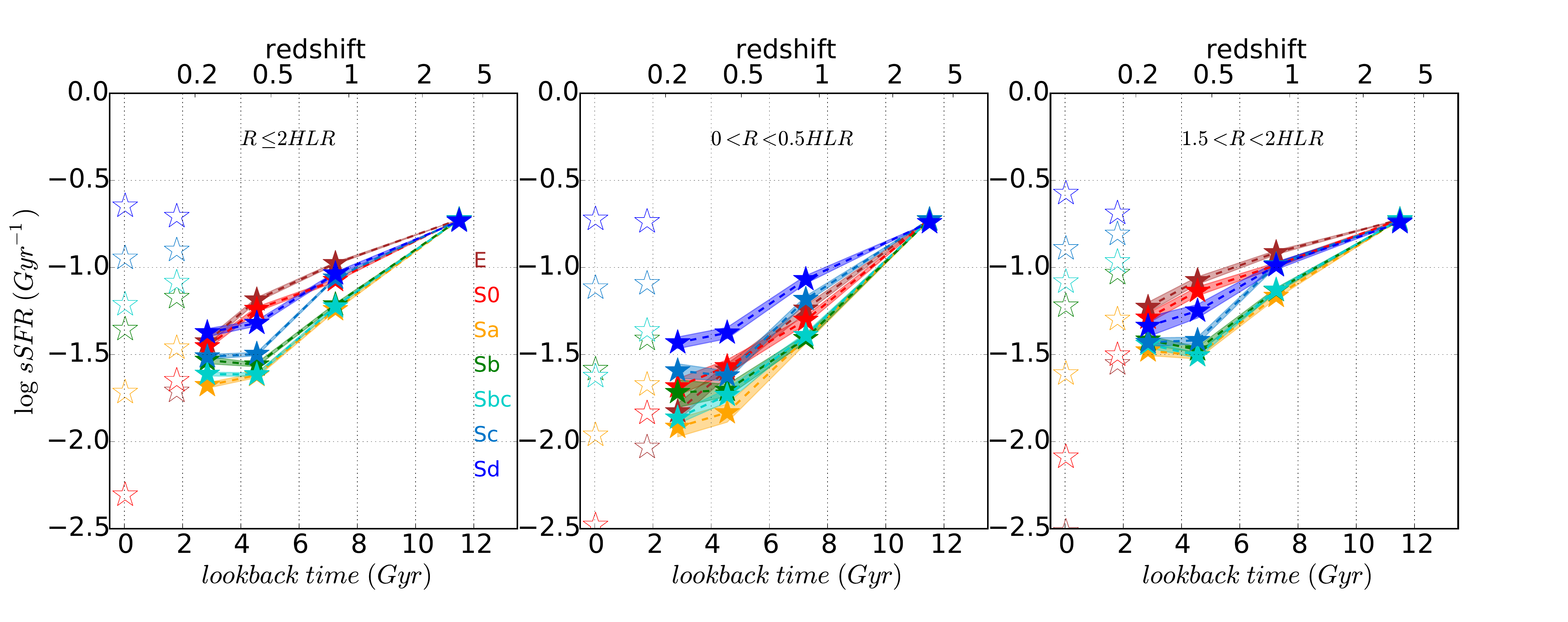}
\caption{As Fig.\ \ref{fig:SFR}, but for the specific SFR, defined as the SFR in a time-bin divided by the stellar mass formed up to that time.
} 
\label{fig:sSFR}
\end{figure*}

Yet another way to express SFHs is through the time evolution of specific SFRs. For a galaxy today,  the sSFR  is usually defined as  ${\rm SFR(today)}/M_\star$. Due to the slightly sub-linear relation between  SFR and $M_\star$ (the MSSF;  ${\rm SFR} \propto M_\star^{0.7-0.9}$), the present day sSFR declines slowly with mass, with  star forming galaxies occupying a tight sequence in the sSFR vs. $M_\star$ space. ETGs fall below this sequence and display a large spread of sSFRs at fixed $M_\star$ \citep{schiminovich07, salim07, karim11}. Spatially resolved data allow us to extend this concept and define a local sSFR, computed as the ratio between the surface densities of SFR and stellar mass, ${\rm sSFR} = \Sigma_{\rm SFR} / \mu_\star$. \citet{gonzalezdelgado16} used CALIFA data to show that this local sSFR increases with $R$, and that at the present time it grows faster with radius within $R < 1$ HLR than outwards, probably signaling the bulge-disk transition. 
Mapping how sSFRs vary as a function of radial location is thus useful for galaxy evolution studies.

Our fossil record analysis allows us to go one step further and study temporal variations of the sSFR. We do this by 
 computing ${\rm sSFR}(t) = {\rm SFR(t)} / M^\prime_\star(t)$, i.e., the SFR at lookback time $t$ (discussed in the previous section) divided by the stellar mass formed up to that epoch. Note that we prefer to use $M^\prime_\star$, which  includes all the stellar mass ever formed, instead of mass still in stars ($M_\star$) in the definition of the sSFR\footnote{
Defined in this way the inverse sSFR gives a doubling-mass time-scale, i.e., the time-span required to form as much mass as a galaxy has done to date at the current rate. Also, the product of the sSFR and the time since the start of star formation ($T$, in practice the age the Universe) gives the familiar birthrate parameter, $b = {\rm sSFR} \times T$  \citep{scalo86, cidfernandes13}}. 
In practice, $M^\prime_\star \sim 1.4 M_\star$ for a Salpeter IMF\footnote{It takes only $\sim 4$ Gyr for a stellar population to lose 1/1.4 of its original mass. Since most stars are older than this, a correction factor of 1.4 converts original to current stellar masses accurately.} 
so multiplication by 1.4 converts our values to the usual (but less natural) definition.

The results are shown in Fig.\ \ref{fig:sSFR}, where, as in the two previous figures, top and bottom panels average galaxies by current stellar mass and morphology, respectively, with galaxy-wide, central and outer regions presented in panels from left to right. 

First of all, let us clarify why, by construction, all curves start from the same point, ${\rm sSFR} = 0.19$ Gyr$^{-1}$ at lookback time 11.5 Gyr. This occurs because the mass formed in this first bin appears both in the numerator and denominator, such that the sSFR value obtained is simply the inverse of the $\Delta t = 5.2$ Gyr time-span of this first bin (see Section \ref{sec:CSP_base}). 
This mathematical triviality exposes a well known limitation of archeological studies. Because the spectral evolution of stellar populations follows a $\sim$ logarithmic clock, codes like \starlight\ are unable to distinguish populations of comparable ages \citep{ocvirk06, tojeiro09}. 
In particular, we cannot resolve SFHs to better than a few Gyr for ages $\ga 4$ Gyr. Our estimates of the sSFR at these large lookback times is therefore averaged for a period of time significantly longer than that sampled by high$-z$ studies based on H$\alpha$,  UV, or FIR emission (which sample time scales of 0.1 Gyr or less), a caveat which must be taken into account when comparing results.

Fig.\ \ref{fig:sSFR} shows that sSFRs decrease as the Universe evolves. This decline has been observed in archeological studies similar to ours (e.g,. \citealt{asari07}; \citealt{mcdermid15}) and in many redshift surveys \citep[]{speagle14}. These cosmological studies find that the sSFR increases with redshift as $(1+z)^3$  for $z<2$ \citep[e.g]{elbaz07, daddi07, oliver10, rodighiero10, elbaz11}. Our sSFRs declines more slowly with time, following roughly $(1+z)^2$ in the highest $M_\star$ bin, but this comparison is probably plagued by the resolution issues discussed above. 
For the same reason, we cannot possibly resolve the initial plateau or rising in sSFR between $z = 2$ and 6 \citep{feulner05, magdis10, stark13,lehnert15}.

Current sSFRs in Fig.\ \ref{fig:sSFR} were calculated considering all components younger than 2.2 Gyr in a single time-bin. 
This gives ${\rm sSFR} \sim 0.07$ Gyr$^{-1}$ for galaxies with $10^{10} \leq M_\star \leq 10^{11}$.
In terms of the more conventional definition (${\rm SFR}/M_\star$), this corresponds to 0.1 Gyr$^{-1}$, typical of galaxies in the MSSF \citep[]{brinchmann04, salim07}. These values are also in agreement with our estimations based on time scales of 32 Myr (for which our SFRs match those obtained from H$\alpha$; \citealt{gonzalezdelgado16}). In terms of time evolution, we find that there is an increase of the sSFR in last few Gyr in galaxies with $M_\star < 10^{11}$. This rejuvenation is also observed in Fig.\ \ref{fig:massfraction}.

In terms of the dependence  with galaxy mass,  the top-left panel in Fig.\ \ref{fig:sSFR}  shows that galaxies on the whole follow quite similar ${\rm sSFR}(t)$ curves down to $\sim 3$ Gyr ago, where the curves split into the well documented downsizing pattern, with sSFR and $M_\star$ going in opposite ways. This small dependence on $M_\star$ indicates that galaxies were on the MSSF on those epochs. This similarity is even more evident for regions located at distances similar to that of the solar neighborhood ($1.5 < R < 2$ HLR, top-right panel), where all the regions follow the same ${\rm sSFR}(t)$, independently of $M_\star$. By analogy with the result for entire galaxies, we interpret this as evidence  that these outer regions were in the past in the local main sequence relation between $\Sigma_{\rm SFR}$ and $\mu_\star$ \citep{gonzalezdelgado16, canodiaz16}. However, conditions are very different in the central regions of galaxies. The top-middle panel in Fig.\ \ref{fig:sSFR} shows that ${\rm sSFR}(t)$ declines more rapidly among the most massive galaxies. The inner $R < 0.5$ HLR of massive galaxies has  been below the local MSSF since $z \sim 0.7$. 
This may be interpreted as a consequence of an efficient, mass-dependent feedback mechanism that  quenches star formation more rapidly in galaxies with a larger potential well.

Let us now turn to the behaviour of ${\rm sSFR}(t)$ in terms of morphology, as shown in the bottom panels of Fig.\ \ref{fig:sSFR}. At the current epoch, E, S0, and Sa galaxies (all of which are off the MSSF) have sSFR $< 0.1$ Gyr$^{-1}$. In the inner regions, only late spirals have current sSFRs above 0.1 Gyr$^{-1}$ (although most  regions in the disks of spirals have sSFR above this value). At early epochs, the sSFR in these inner regions shows a similar declining dependence with cosmic time as that seen in the upper panels. Note, however, the remarkable behavior of the outer regions of E and S0 galaxies (bottom-right panel):  Over the $0.4 < z < 1$ period the ${\rm sSFR}(t)$ curves of these (current epoch) early type systems run above those of spirals.
We hypothesize that this may reflect the growth of the envelope of E and S0 through mergers.


\subsection{SFHs: Star formation rate intensities}
\label{sec:SFH_SFI}

\begin{figure*}
\includegraphics[width=\textwidth]{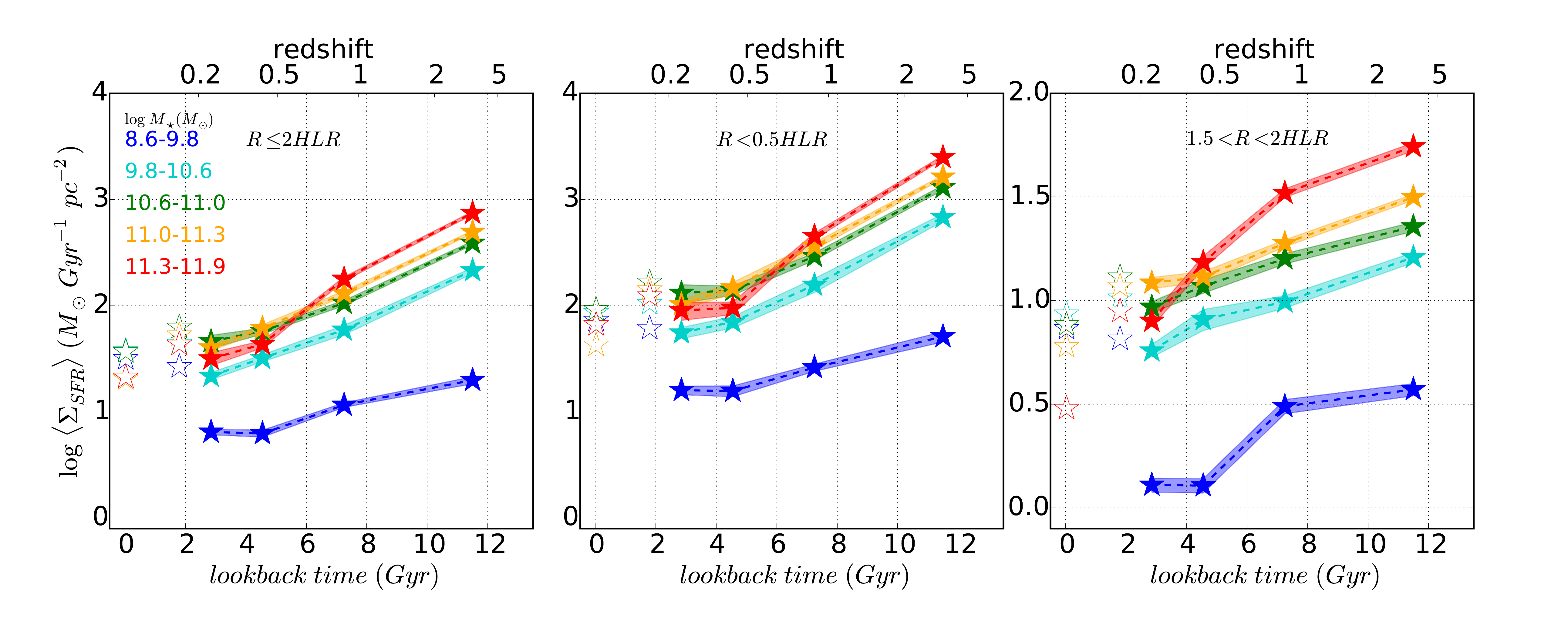}
\includegraphics[width=\textwidth]{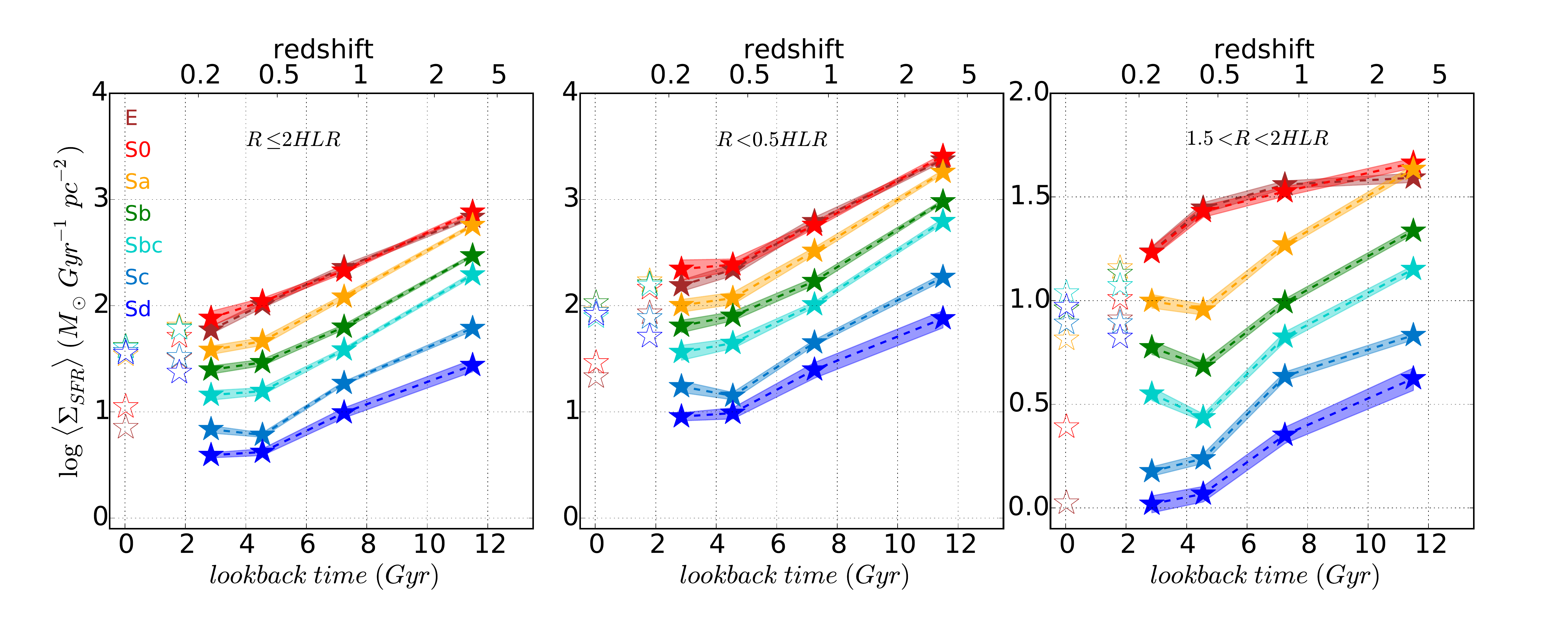}
\caption{As Fig.\ \ref{fig:SFR}, but for the star formation intensity, $\Sigma_{\rm SFR}$.
} 
\label{fig:SFRintensity}
\end{figure*}

As a final way of expressing our 2D-SFHs we show results in terms of the surface density of SFR, $\Sigma_{\rm SFR}$, also referred to as the star formation intensity (SFI). To investigate the time evolution of the SFI we divide $\mu_\star(t)$ (the stellar mass per unit area formed in each epoch, shown in the bottom panels of Fig.\ \ref{fig:2DSFH}) by the age-span of the corresponding CSP component in the base, obtaining $\Sigma_{\rm SFR}(t)$. As in the previous section, the information for the last $< 2.2$ Gyr is lumped into a single bin. We also show the SFI obtained for the last 32 Myr to facilitate comparison with \citet{gonzalezdelgado16}, where this shorter time scale was used as a measure of the current  $\Sigma_{\rm SFR}$. Obviously, our data give us no clue on how galaxy sizes and shapes change over time, just as they say nothing about where stars which are currently at some location within the galaxy were originally born, both of which affect the actual cosmic evolution of $\Sigma_{\rm SFR}$. Despite these limitations, SFIs bring valuable insight to this study.

Fig.\ \ref{fig:SFRintensity} shows the results in the same format as previous figures, with $M_\star$ and Hubble type averages shown in the top and bottom panels respectively, and right, middle, and left panels showing results for different spatial regions. In general terms the figure shows that the SFI increases with redshift. 
This is in line with simple expectations. Gas fractions and densities are both expected and observed to be larger at high $z$ \citep{tacconi13}, which, extrapolating from the Schmidt-Kennincutt relation, naturally leads to higher $\Sigma_{\rm SFR}$ \citep{barden05, yuma11, forsterscheiber11, mosleh12, carollo13}. On top of that, the smaller galaxy sizes in the past (e.g., \citealt{vandokkum13}) also lead to higher $\Sigma_{\rm SFR}$, but, as explained above, this effect is not included in our analysis. Higher SFIs are indeed observed in galaxies at high $z$. At $z\sim 2$--3, $\Sigma_{\rm SFR} \sim 1000\, M_\odot\,$Gyr$^{-1}\,$pc$^{-2}$ has been reported by \citet{lehnert13}, and Milky Way (MW) progenitor galaxies are proposed to have  $\Sigma_{\rm SFR} \sim 600\, M_\odot\,$Gyr$^{-1}\,$pc$^{-2}$ around $z \sim 2$ to explain the formation of a thick disk like the one in the Galaxy \citep{bovy12, lehnert14}. 

Except for low mass galaxies,  our $\Sigma_{\rm SFR}$ values increase with redshift up to  $\sim 1.5$ orders of magnitude with respect to 
values at $z = 0$. Our estimations of $\Sigma_{\rm SFR}$ at $z \geq 2$ averaged over the whole galaxy (left panels) range from $\sim 4000$ to $100\, M_\odot\,$Gyr$^{-1}\,$pc$^{-2}$ from the highest to the smallest $M_\star$ bin. The distribution of $\Sigma_{\rm SFR}$ with Hubble types ranges from $\sim 4000 \, M_\odot\,$Gyr$^{-1}\,$pc$^{-2}$ in ETGs  to $\sim 1000\, M_\odot\,$Gyr$^{-1}\,$pc$^{-2}$ in Sbc's, reaching a minimum of $150 \,M_\odot\,$Gyr$^{-1}\,$pc$^{-2}$ in Sd galaxies. With the exception of the less massive galaxies  ($M_\star < 10^{10}$) and late type spirals, these values are similar to the $\Sigma_{\rm SFR}$ observed in high $z$ galaxies.

The top panels in Fig.\ \ref{fig:SFRintensity} show that while at $z = 0$ the SFI is nearly independent of $M_\star$, in the past  $\Sigma_{\rm SFR}$ was higher for galaxies that today are more massive. Disregarding the lowest mass-bin and focusing on our oldest ages, $\Sigma_{\rm SFR}$ grows by $\sim 0.7$ dex for a $\sim 1.5$ dex increase in $M_\star$, an approximately $\Sigma_{\rm SFR} \propto M_\star^{0.4-0.5}$ scaling relation.
This may be understood in terms of the  dependence of galaxy size with mass. Recently, \citet{vanderwel14} have found that star forming galaxies with $M_\star >$ 3$\times$10$^9$  have effective radii that, for a given redshift, grow as $R_e \propto M_\star^{0.22}$. Similarly, \citet{vandokkum13} obtain a $R_e \propto M_\star^{0.27}$ size-mass relation (see also \citealt{newman12, mosleh12, buitrago13}). Combining this empirical scaling with a ${\rm SFR} \propto M_\star^{0.7-0.9}$ MSSF, and assuming its slope does not vary much at high $z$ \citep{whitaker14}, one finds $\Sigma_{\rm SFR} \propto {\rm SFR} / R_e^2 \propto M^{0.2-0.5}$, close to the relation obtained from Fig.\ \ref{fig:SFRintensity}.

A clear relation between the evolution of $\Sigma_{\rm SFR}$ and galaxy morphology is revealed in the bottom panels of Fig.\ \ref{fig:SFRintensity}. At present all spirals have similar $\Sigma_{\rm SFR}$ values, well above those of E's and S0's. This is particularly evident for the estimates over the last 32 Myr, but it remains true considering the last Gyr. (Note that this difference in current SFI is not as strong in the top panels because of the mixture of morphological types when binning galaxies by $M_\star$).
In the past, however, $\Sigma_{\rm SFR}(t)$ scales with  Hubble type, increasing systematically from late to early types.
A possible interpretation of this result is that it reflects a sequence in angular momentum, decreasing from early to late types. At high $z$, the combination of low angular momentum and abundant supply of gas favors high gas concentrations and thus, from the Schmidt-Kennicutt law, high SFI.
These higher $\Sigma_{\rm SFR}$ in the past would also imply higher $\mu_\star$ if a local MSSF analogous to the one seen nowadays \citep{wuyts13, gonzalezdelgado16, canodiaz16, maragkoudakis16} also existed in the past, so that galaxies that are denser today were also denser in the past.

It is interesting to note that the shape of the $\Sigma_{\rm SFR}(t)$ curves in the inner regions is similar to that of the curves obtained for the whole galaxy, both when binning by $M_\star$ (top panels) and by Hubble type (bottom). The same can be said about the outer regions of spirals. E and S0 galaxies, however, behave differently, showing a slower decay in their outer SFI than elsewhere in the galaxy, as can be seen comparing their $\Sigma_{\rm SFR}(t)$ curves (in brown and red) in the bottom-right panels to the corresponding ones in the left and central panels. We again speculate that this may be signaling the accretion of stars in the envelopes of these galaxies at  $z = 1$--0.5.


\section{Discussion: SFH constraints on galaxy formation scenarios}
\label{sec:Discussion}

The discovery of high redshift disk-like galaxies characterized by high density clumps of star formation \citep{elmegreen06, wuyts13, genzel14} have led to the suggestion that mergers at $z >2$ may not be the main mechanism for the formation of progenitors of today's massive galaxies.  A possible scenario is that these progenitors had a very rich gaseous disk-like component where high rates of star formation were possible by continuous fueling of cold and filamentary streams of gas from the cosmic web \citep{kere05, ocvirk08, dekel14}. These clumps can arise from gravitational instabilities driven by the continuous replenishment of the disk with high density gas. If the clumps survive for enough time, and are not destroyed by internal feedback,  dynamical friction could lead them to migrate inwards, where they can coalesce to form a compact central component (a ``bulge'';  \citealt{elmegreen08}).

In galaxies like the MW, where there is no classical bulge \citep{diMatteo14}, this early epoch of clump growth can lead to the formation of a turbulent thick disk. The accretion rate would then slow down, the star formation activity decrease, the angular momentum increase, and a thin disk would form.

In this section we try to shed light into some aspects of these scenarios from the perspective of our CALIFA-based 
2D-SFH analysis. Since the ideas discussed above  pertain to relatively massive galaxies, and also because of the incompleteness of our  
sample at low masses, we focus the discussion on  galaxies with $M_\star > 10^{10}$.


\subsection{The formation of disks in spirals: A comparison of CALIFA SFHs with models for the Milky Way}

\begin{figure*}
\includegraphics[width=0.48\textwidth]{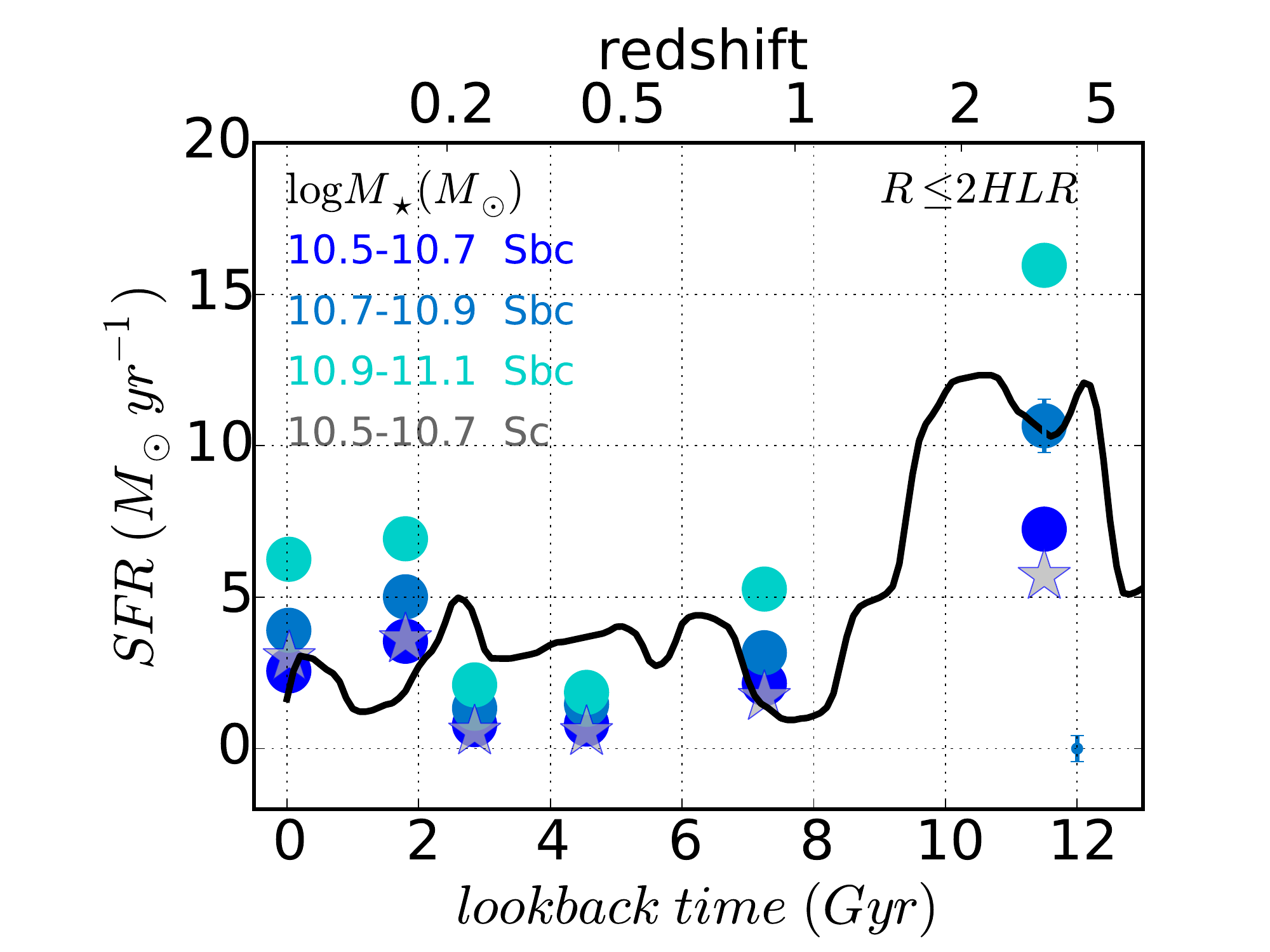}
\includegraphics[width=0.48\textwidth]{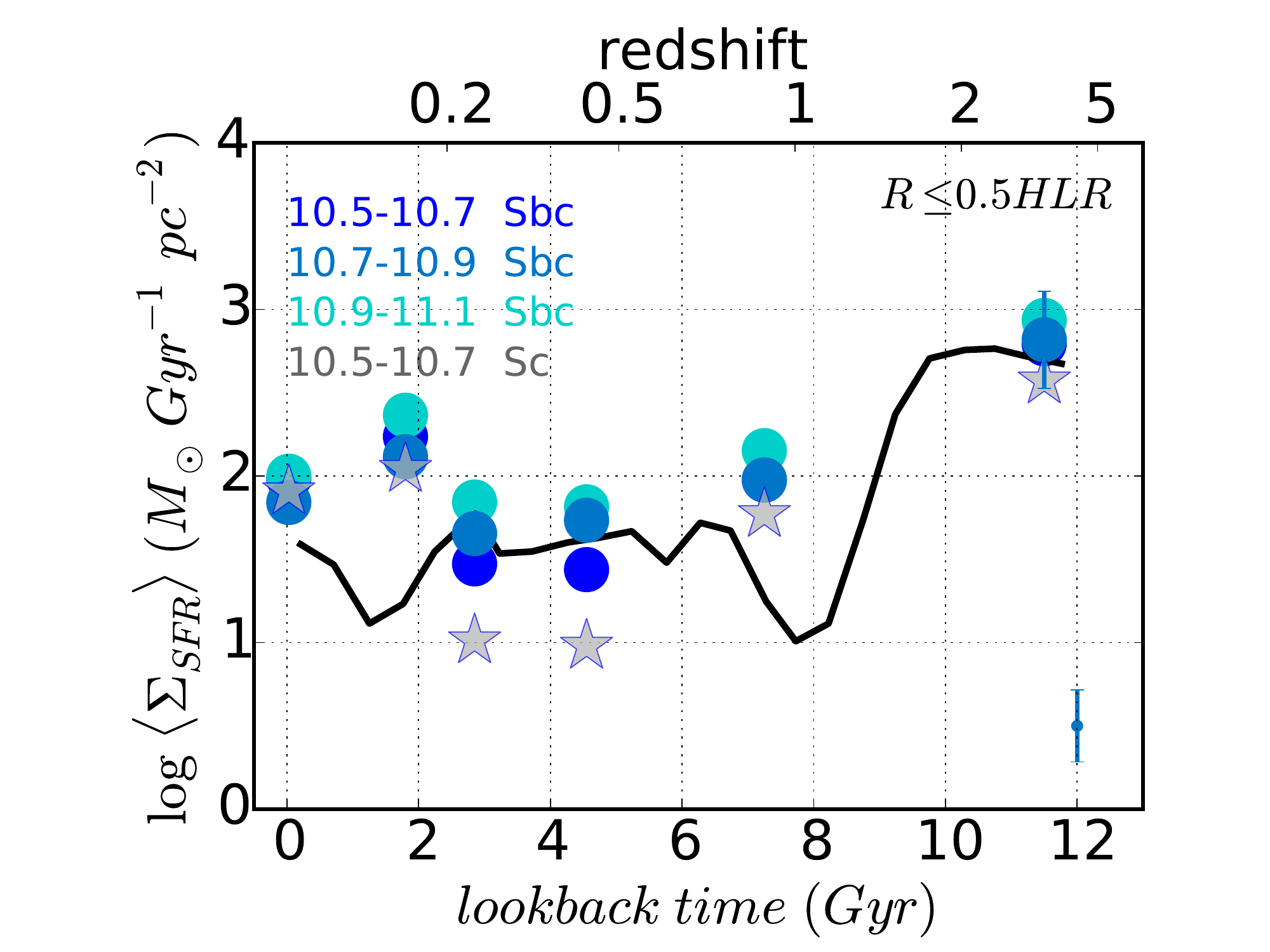}
\caption{The evolution of the SFR and SFI for Sbc galaxies (points) in three stellar mass bins (color coded as indicated in the label),  and Sc galaxies (stars) in our lowest $M_\star$ bin. The SFR is calculated adding the spaxels in the inner $R < 2$ HLR, while the $\Sigma_{\rm SFR}$-values are averages over the central 0.5 HLR. The black line shows the model proposed by \citet{haywood15} and \citet{lehnert14} for high redshift disk galaxies similar to the Milky Way ($\log M_\star \sim$ 10.9).
Bars located at the right side of the plots represent the standard deviations  for Sbc galaxies with $10.7 \leq \log M_\star \leq 10.9$ at $\sim$ 11.5 Gyr (upper bar)  and the standard deviations for these galaxies averaged for all the epochs (at the right lower corner).
}
\label{fig:thickdisk}
\end{figure*}

Recent modeling of the chemical abundances in the solar vicinity \citep{haywood13, snaith14, snaith15,haywood15} has shown that the inner disk ($R < 10$ kpc) of the Galaxy has gone through two phases of star formation:  {\em (i)} the formation of a thick disk, from $t = 13$ to 9 Gyr ago, and {\em (ii)} the formation of a thin disk, in the last 7 Gyr. \citet{lehnert14,lehnert15} and \citet{haywood16} extended this analysis, showing that this scenario is useful to explain the formation of MW-like galaxies at high $z$. They propose that there is a drop of the SFI in the thick disk phase to a more quiescent phase at $z<1$, a drop that is more significant than the decrease expected from the exhaustion of gas given by a Schmidt-Kennicutt relation. They argue that this cessation of the star formation marks the end of the growth of the thick disk.

Our sample contains several galaxies similar to the MW both in terms of mass ($M_\star = 8 \times10^{10}$,  \citealt{licquia15}\footnote{After converting to the IMF used in our analysis}), and morphology (Sbc to Sc). The spatially and temporally resolved SFRs and SFIs obtained with our analysis thus offer a powerful and independent way of testing the evolutionary scenarios for MW-like galaxies outlined above.

Fig.\ \ref{fig:thickdisk} shows the results of this comparison. The left panel  compares the evolution of the SFR of the inner disk of the MW as predicted by \citet{haywood16}, drawn as black solid lines, with the ${\rm SFR}(t)$ obtained with our analysis for Sbc (circles) and Sc (stars) galaxies in three $M_\star$ intervals covering stellar masses similar to that of the Galaxy. Our SFRs are computed for regions within $R < 2$ HLR, corresponding to $\sim 10$ kpc for Sbc galaxies, and thus matching the size of the MW inner disk \citep{haywood16}.  In the right panel of Fig.\ \ref{fig:thickdisk} we compare the SFI derived by \citet{lehnert15} with our estimations for the same Sbc and Sc galaxies averaging $\Sigma_{\rm SFR}$ over the central 0.5 HLR, that for these spirals corresponds to the size of the thick disk of the MW \citep{bland-hawthorn16}.

The similarity of our results for the SFH of Sbc galaxies with $\log M_\star = 10.7$--10.9 to the ${\rm SFR}(t)$ and $\Sigma_{\rm SFR}(t)$ proposed for the MW is remarkable. Both models and data in Fig.\ \ref{fig:thickdisk} show a significant drop of the SFR and its intensity from $z > 2$ to $z < 1$, with a slight rejuvenation in recent epochs that may be associated with star formation activity in the thin disk. This comparison suggests that the formation of a thick disk can be a common phase early in the life of MW-like galaxies. The similar shapes of the SFHs in the central regions of the late type spirals in CALIFA (as seen in Figs.\ \ref{fig:massfraction}--\ref{fig:SFRintensity}) further suggests that the formation of thick disks 10 Gyr ago is a generic feature in the build up of these systems.


\subsection{The growth of early type galaxies }

ATLAS3D has revealed a close link between ETG and spirals by showing that there is a critical dynamical  mass of $\sim 2 \times 10^{11}$ below which fast rotating ETGs form a parallel sequence to spirals in galaxy properties, while slow rotators dominate above this critical mass. Slow rotators assembled near the center of massive dark matter halos via intense star formation at high redshift, and remain slow rotators for the rest of their evolution via a channel dominated by gas poor mergers. Fast rotators, on the other hand,  start as star forming disks and evolve through a channel dominated by gas accretion, bulge growth, and quenching \citep{cappellari13, capellari13E, cappellari16}.

Using a non-parametric SFH analysis of integrated spectra\footnote{Up to $R = 0.5$--1 HLR for their data.}  of ATLAS3D galaxies, \citet{mcdermid15} found that ETGs more massive than $M_\star = 10^{10.5}$ form  90\% of their mass by $z \geq  2$. In contrast, lower mass ETGs have more extended SFHs, forming barely half of their mass before $z = 2$.

These results are similar to those obtained here. Fig.\ \ref{fig:SFRintensity} , for instance, indicates that: 
{\em (i)} E and S0 in our sample have on average equal SFHs at least during the first 10 Gyr. 
{\em (ii)} In the inner regions, the SFH of E and S0 are very similar in shape to the SFH of early type spirals, except in the last 3--4 Gyr. For instance, the mass fractions at $z > 2$ in the inner regions of E and S0 are in the 70--83\% range (depending on $M_\star$), indistinguishable from the 80\% found for Sa.
{\em (iii)} In the outer regions, not sampled by the ATLAS3D data, the SFH of E and S0 galaxies is quite different from that of Sa and later types.
Between $z = 2$ and 0.4, both $\Sigma_{\rm SFR}$ and sSFR  are significantly higher in ETG than in spirals.

These results indicate  that ETG have assembled their inner regions in a similar way to Sa galaxies, probably via gas accretion or mergers and bulge growth. However, their outer regions grow significantly slower than the inner ones, building 40--60\% of their stellar mass over the first few Gyr, compared to $\sim 80\%$ at $R < 0.5$ HLR. Thus, although the centers of ETG formed very fast and early on, their envelopes were assembled during a more extended period. This  epoch of active growth is roughly centered around $z$ of 1, but goes all way from $z \sim 2$ to 0.4.

It is now well established that massive galaxies ($M_\star > 10^{11}$) at $z \sim 2$ have significantly smaller sizes than their local E and S0 counterparts, and that they have grown significantly since then \citep{trujillo06, trujillo07, buitrago08, vandokkum10}. It has been suggested that dry mergers are the main driver for this late size evolution, expanding their envelope by means of small satellite accretion \citep[]{naab09, bell04}.  
Indeed, $z \sim 1$ has been identified as an epoch of galaxy merging \citep{hammer05, kaviraj15}, in which the progenitors of ETG increase in size and mass in proportion to one another, following approximately $\Delta \log  R_e  \sim 2 \times \Delta \log  M_\star$ \citep{vandokkum10, vandokkum15, huertas15}. 
This relation predicts that from $z = 2$ to 0.4 ETGs with a present-day mass of $3\times10^{11}$ increase their effective radii from $R_e \sim 2.5$ to 6 kpc, and their stellar mass by a factor of $\sim 1.5$. 
Although we cannot test the size evolution with our data, we find that from $z =  2$ to 0.4 our E and S0 galaxies have grown 
their mass by a factor of 1.5 on average\footnote{
This is a global estimate, but our mass growth factors vary with the radial location. Typically, inner regions grow by 20\% while outer ones grow by 60\% in mass over this same period.},
in excellent agreement with this estimate.

It thus seems that our results for the SFHs of ETGs are in agreement with the two phase formation scenario for ETG, where the central part builds most of its mass at high $z$, probably through highly dissipative processes involving gas accretion, while the outer envelope grows over a more extended period (down to $z \sim 0.4$), possibly through dry mergers.


\subsection{The shut down of star formation}

\begin{figure*}
\includegraphics[width=\textwidth]{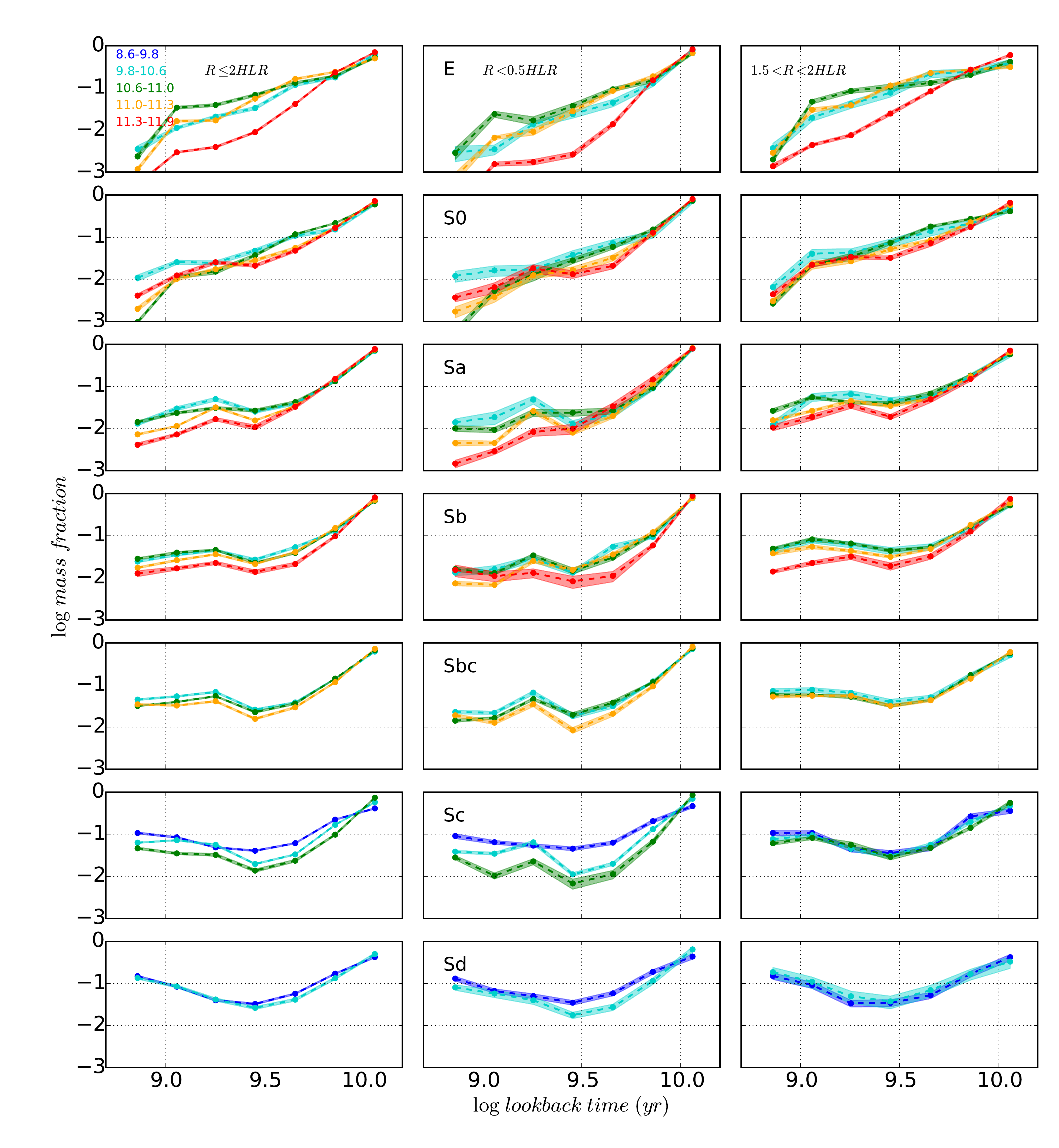}
\caption{As Fig.\ \ref{fig:massfraction}, but breaking the mass fraction SFHs in bins of Hubble type (E galaxies in the top and Sd in the bottom) and stellar mass (coded by colours, with $M_\star$ decreasing from red to blue). 
}
\label{fig:massfractionES}
\end{figure*}

High redshift studies have shown that massive galaxies ($M_\star>10^{11}$) form and quench fast \citep{cimatti04, mcCarthy04,  whitaker13}, a  scenario that is also supported by studies of local ETGs \citep{mcdermid15, citro16, pacifici16}. We too obtain that most galaxies formed most of their mass early on, but, in contrast with these other studies, we find that it took them a long time to complete the shut down of star formation. In other words, we seem to obtain a slower quenching than other studies. There is, however, one case for which we do find evidence for a fast quenching: the most massive ellipticals.

This result appears in several of the previous figures, perhaps more clearly so in Fig.\ \ref{fig:massfraction}, where the steepest SFHs around $z \sim 1$ occur for our highest-$M_\star$ bin (top panels), or ellipticals (bottom). A close inspection of Fig.\ \ref{fig:2DSFH} shows this fast quenching occurs not for massive or elliptical galaxies in general, but for galaxies which are both very massive {\em and} elliptical, something which cannot be fully appreciated in Fig.\ \ref{fig:massfraction} because its panels collapse over either the $M_\star$ or the morphology dimension. To highlight this point and also to disentangle the effects of $M_\star$ and morphology upon SFHs in a visually simpler way, Fig.\ \ref{fig:massfractionES} shows the average mass fraction curves, $m(t)$, for galaxies along the Hubble sequence (E to Sd running from top to bottom) for the same $M_\star$ bins used throughout the paper (coded by the colours of the curves). Left, middle and right panels separate radial regions, in the same order as in Figs.\ \ref{fig:massfraction}--\ \ref{fig:SFRintensity}.

The red curves in the top row of Fig.\ \ref{fig:massfractionES} stand out from all others in being the fastest declining SFHs in our sample. As pointed out above, this fast quenching is only seen among our most massive ($\log M_\star = 11.3$--11.9) ellipticals. S0 and Sa galaxies of the same mass do not have as steep SFHs in their first few Gyr of evolution, neither have ellipticals of lower masses. This is specially clear in the inner regions (central panels), but it also holds for the galaxy as a whole (left). In all other cases the decline in star formation activity is slower.

Assuming that $M_\star$ is a good tracer of the halo mass \citep[]{behroozi13}, this result suggests that the halo mass  is not the main mechanism to quench  these massive ellipticals.
There are other studies that suggest that the bias of the dark-matter halos depends on something other than their mass, where the initial conditions (e.g. voids, filaments, geometry of the environment) and halo assembly history (e.g. halo formation time) are relevant to 
set the differences in the SFH of galaxies \citep{dressler16} and their galaxy properties \citep{tojeiro16}.


\subsection{Red nuggets as galaxy nuclei}

The SFH of many of the nuclei\footnote{Central $< 0.2$ HLR, equivalent to $\sim 1$ kpc in our sample} of the galaxies in our sample show that these central cores formed fast and quenched rapidly. This happens in ETG and early type spirals, that in our sample  coincide with galaxies more massive than a few $\times$10$^{10}\,M_\odot$. Thus, the ETG and Sa-Sb in our sample have formed $> 80\%$ of their central core mass earlier than $z=2$, and with $\Sigma_{\rm SFR}$ significantly  above $10^3 M_\odot$\,pc$^{-2}\,$Gyr$^{-1}$ (up to $5 \times10^3 M_\odot$\,pc$^{-2}\,$Gyr$^{-1}$ in present-day massive ellipticals). These cores can be a relic of the red nuggets\footnote{ Very compact massive red galaxies that are $\sim$ 5 times smaller than equal-mass analogs.} detected at high and low redshift \citep{barro13,ferre-mateu17}.


\subsection{The rejuvenation in spirals}

Going back to Fig.\ \ref{fig:massfractionES}, but now focusing on the more recent past and examining the SFHs for all morphologies, one sees that all  E and S0 in our sample experience a further decrease of the star formation in the last at $\sim 4$ Gyr, while in spirals there is a new activation of the star formation over this same period (also seen as the increase in $\Sigma_{\rm SFR}$ and sSFR in Figs.\ \ref{fig:sSFR} and \ref{fig:SFRintensity}). This kind of rejuvenation has been also observed in the mass cumulative curve of low mass galaxies of the extended CALIFA sample (Garc\'\i a-Benito et al. 2017, in preparation), where the mass grows by more than 20\% in the last 2 Gyr.  This is also observed by \citet{leitner12} in low mass star forming galaxies.

This rejuvenation in spirals can be produced by infall of new gas or by the consumption of the residual gas already in the galaxy. This  phase is clearly less intense in early than in late type spirals, and is very significant in the outer regions of Sbc-Sc galaxies. It explains why Sbc-Sc-Sd galaxies, and in particular the disk regions, are  the major contributor to the present day star formation rate density of the Universe \citep{gonzalezdelgado16}. 

It is well known that the fraction of the mass in HI increases with both decreasing mass and later type
\citep{vandriel16}, so such galaxies have plenty of fuel for forming stars in their disks. The rejuvenation will be stronger in low mass galaxies ($\log \ M_\star < 10$), where $M_{\rm HI}/M_\star \sim$ 1.  Thus, they have enough gas if all is used to fuel the galaxy for a mass doubling time which is about a Hubble time (see Fig.\ \ref{fig:sSFR}).

However, it is difficult to know how this can happen. 
Gravitational torques are probably needed so it could be that this is related to minor mergers. This is in fact the case in M31, where a small interaction with M32 and M33 or minor mergers are the cause of the global enhancement of star formation in the disk during the last 2-4 Gyr, producing  $\sim$60$\%$ of the total mass formed in the disk during the last 5 Gyr  \citep{williams15}. Bars may also act to bring the HI gas from the outer disk and trigger the star formation. This is a plausible mechanism because bars are very frequent in late type galaxies   \citep{moles95, buta13, buta15}.


\subsection{Modes of galaxy growth}

Fig.\ \ref{fig:massfractionES} also suggests that galaxies can grow in two different modes. For the early evolution of Sa-Sbc spirals and the entire evolution of E and S0, the logarithm of their mass fraction ($\log \ m$) declines with $\log \ t$. For the later type (Sc-Sd) spirals and low mass galaxies,  $\log \ m$ is almost constant and independent of time. Thus, one mode is exponential and the other is scale free. This is a feature across all mass and morphological classes, although the balance between the two modes is different.

Perhaps the exponential mode represents the transition between the formation of a thick and a thin disk. The thick disk is a self-regulated mode, where strong outflows and turbulence drive the high intensity of the star formation rate occurring very early on the evolution \citep{lehnert15}. The thin disk is a scale free mode regulated by the secular processes; a phase driven by self-gravity, and the energy injection from the stellar population is not relevant for global regulation \citep{lehnert15}.

\section{Summary and conclusions}
\label{sec:Summary}

We have applied the fossil record method of the stellar populations to a sample of 436 galaxies observed by CALIFA at the 3.5m telescope in Calar Alto, to investigate their SFH in seven bins of morphology (E, S0, Sa, Sb, Sbc, Sc, Sd) and several stellar masses, in the range  $\sim10^9$ to $7\times10^{11} M_\odot$ (for a Salpeter IMF). A full spectral fitting analysis was performed using the \starlight\ code and a combination of composite stellar populations (CSP) spectra derived with the models of \citet{gonzalezdelgado05} and \citet{Vazdekis15}. This base comprises 18 logarithmically spaced  age bins centered at ages from 0.00245 to 11.50 Gyr, and 8 metallicities from $\log Z/Z_\odot = -2.28$ to $+0.40$.
The spectral fitting results are processed with our  \pycasso\ pipeline to derive the 2D ($R \times t$) map of the SFH for each galaxy, from which we obtain the spatial and temporal evolution information of the star formation rate (SFR), specific SFR (sSFR), and the intensity of the SFR ($\Sigma_{\rm SFR}$).

Our main results are:

\begin{enumerate}

\item
These nearby galaxies formed very fast. All of them have their peak of star formation at the earliest time ($z > 2$), independently of their stellar mass.  However, the subsequent SFH varies with  $M_\star$, with less massive galaxies showing a longer period of star formation. This is a manifestation of the ``downsizing'' effect observed in other surveys.

\item 
SFRs decline rapidly as the Universe evolves; at any epoch, the SFR is proportional to $M_\star$, and the most massive galaxies had the highest absolute SFR. The sSFR also decreases with time. The ${\rm sSFR}(t)$ curves vary systematically with $M_\star$ in the central regions, but not in the outer disk of spirals nor on the envelopes of ellipticals. At the present epoch, ${\rm sSFR} \geq 0.1$ Gyr$^{-1}$ for galaxies that are in the MSSF and in regions located in the disk of spirals of Hubble type later than Sa.  

\item 
The star formation intensity ($\Sigma_{\rm SFR}$) also declines rapidly as the Universe evolves. At the present epoch, the spatially averaged $\Sigma_{\rm SFR}$ is  similar in all spirals ($\sim 37 M_\odot\,$Gyr$^{-1}\,$pc$^{-2}$),  significantly higher than in ETG. In the past, however, $\Sigma_{\rm SFR}$ increases systematically from late to early Hubble types. The highest values are found among the progenitors of present day E's and S0's, with $\Sigma_{\rm SFR} \sim 10^3 M_\odot\,$Gyr$^{-1}\,$pc$^{-2}$ at $z > 2$. 
These values are similar to those reported for high redshift star forming galaxies.

\item
There is a remarkable similarity between the SFH of Sbc galaxies of $M_\star \sim 7\times 10^{10} M_\odot$ in our sample and the predictions for MW-like galaxies proposed by \citet{haywood15} and \citet{lehnert15} for the formation of the thick disk. In agreement with these models, we obtain that in the central ($R< 0.5$ HLR) $\Sigma_{\rm SFR}$ shows a significant drop from $600 M_\odot\,$Gyr$^{-1}\,$pc$^{-2}$ at $z > 2$ to $100 M_\odot\,$Gyr$^{-1}\,$pc$^{-2}$ at $z < 1$, while the global SFR decreases from $\sim 10$ to $2 \, M_\odot\,$yr$^{-1}$ over the same period. These comparisons suggest that the formation of a thick disk may be a common phase early on in the life of late type spirals.

\item
In regions located in the envelope of E and S0 (akin to $1.5 < R < 2$ HLR in Figs.\ \ref{fig:massfraction}--\ref{fig:SFRintensity}), the mass fraction, SFR, sSFR, and $\Sigma_{\rm SFR}$ decline with redshift 
($2 > z > 0.4$) more slowly than regions in the disk of early spirals. 
The central cores ($R< 0.5$ HLR) have grown their mass very early (70\% and 83\% at $z > 2$ in E's and S0's, respectively), and only increase their mass by a factor 1.2  between $0.4 < z< 2$, while the outer regions have increased their mass by a factor $\sim 1.6$. 
These results suggest that $z$ around 1 is an active epoch  of envelope assembly in E and S0, in line with the expectations of the two phase scenario proposed for the formation of ETG.

\item
A fast quenching is not observed in the SFH of most of the CALIFA galaxies. One exception occurs  in the most massive ($M_\star  > 10^{11.3} M_\odot$)  E galaxies, in which SFR and  $\Sigma_{\rm SFR}$ decline very rapidly; in particular, in the central region where most of the mass formed at $z > 2$, and the sSFR drops suddenly after $z= 1$. However, this fast quenching does not happen in galaxies of the same stellar mass but later type (S0 and Sa in our sample). These results suggest that the halo-quenching is not the main driver for the shut down of the star formation in these galaxies.

\item
The previous item notwithstanding, the SFH of most nuclei ($\sim$ 1 kpc) in these galaxies formed fast and quenched rapidly. They formed more than 80\% of their mass at $z >2$, when $\Sigma_{\rm SFR}>10^3$ M$_\odot$ Gyr$^{-1}$ pc$^{-2}$. They can be the relic of the ``red nuggets'' detected in high redshift galaxies.

\end{enumerate}

In summary, the SFH of  nearby spirals as represented by the CALIFA survey are compatible with a scenario of fast formation and a relatively long declining phase for the shut down of the star formation, that may be sustained by slow consumption of residual gas from the initial flow. A re-activation of the star formation happens in more recent times, the last 4 Gyr, with an intensity significantly below  the peak values, producing a rejuvenation of the disks that is significantly more relevant in low mass and late spiral types than in more massive and early type spirals. In ETG, the initial phases in the formation are similar to those in massive early type spirals, but E and S0 also have an active long phase of growth  between 
$0.4 <z<2$ that is relevant for the growth of their external envelope.

\begin{acknowledgements} 
CALIFA is the first legacy survey carried out at Calar Alto. The CALIFA collaboration would like to thank the IAA-CSIC and MPIA-MPG as major partners of the observatory, and CAHA itself, for the unique access to telescope time and support in manpower and infrastructures.  We also thank the CAHA staff for the dedication to this project.
Support from the Spanish Ministerio de Econom\'\i a y Competitividad, through projects AYA2016-77846-P, AYA2014-57490-P, AYA2010-15081, and Junta de Andaluc\'\i a FQ1580, AYA2010-22111-C03-03, AYA2010-10904E, AYA2013-42227P, RyC-2011-09461, AYA2013-47742-C4-3-P, EU SELGIFS exchange programme FP7-PEOPLE-2013-IRSES-612701, CONACYT-125180, DGAPA-IA100815, and PAPIIT-DGAPA-IA101217.
We also thank the Viabilidad, Dise\~no, Acceso y Mejora funding program, ICTS-2009-10, for funding the data acquisition of this project. 
ALdA, EADL and RCF thanks the hospitality of the IAA and the support of CNPq. RGD acknowledges the support of CNPq (Brazil) through Programa Ci\^encia sem Fronteiras (401452/2012-3). CJW acknowledges support through the Marie Curie Career Integration Grant 303912. We thank the support of the IAA Computing group.

\end{acknowledgements}



\bibliographystyle{aa}
\bibliography{Califa7_SFH}

\begin{thebibliography}{183}
\expandafter\ifx\csname natexlab\endcsname\relax\def\natexlab#1{#1}\fi

\bibitem[{{Abramson} {et~al.}(2016){Abramson}, {Gladders}, {Dressler},
  {Oemler}, {Poggianti}, \& {Vulcani}}]{abramson16}
{Abramson}, L.~E., {Gladders}, M.~D., {Dressler}, A., {et~al.} 2016, \apj, 832,
  7

\bibitem[{{Asari} {et~al.}(2007){Asari}, {Cid Fernandes}, {Stasi{\'n}ska},
  {Torres-Papaqui}, {Mateus}, {Sodr{\'e}}, {Schoenell}, \& {Gomes}}]{asari07}
{Asari}, N.~V., {Cid Fernandes}, R., {Stasi{\'n}ska}, G., {et~al.} 2007,
  \mnras, 381, 263

\bibitem[{{Baldry} {et~al.}(2004){Baldry}, {Glazebrook}, {Brinkmann},
  {Ivezi{\'c}}, {Lupton}, {Nichol}, \& {Szalay}}]{baldry04}
{Baldry}, I.~K., {Glazebrook}, K., {Brinkmann}, J., {et~al.} 2004, \apj, 600,
  681

\bibitem[{{Barden} {et~al.}(2005){Barden}, {Rix}, {Somerville}, {Bell},
  {H{\"a}u{\ss}ler}, {Peng}, {Borch}, {Beckwith}, {Caldwell}, {Heymans},
  {Jahnke}, {Jogee}, {McIntosh}, {Meisenheimer}, {S{\'a}nchez}, {Wisotzki}, \&
  {Wolf}}]{barden05}
{Barden}, M., {Rix}, H.-W., {Somerville}, R.~S., {et~al.} 2005, \apj, 635, 959

\bibitem[{{Barro} {et~al.}(2013){Barro}, {Faber}, {P{\'e}rez-Gonz{\'a}lez},
  {Koo}, {Williams}, {Kocevski}, {Trump}, {Mozena}, {McGrath}, {van der Wel},
  {Wuyts}, {Bell}, {Croton}, {Ceverino}, {Dekel}, {Ashby}, {Cheung},
  {Ferguson}, {Fontana}, {Fang}, {Giavalisco}, {Grogin}, {Guo}, {Hathi},
  {Hopkins}, {Huang}, {Koekemoer}, {Kartaltepe}, {Lee}, {Newman}, {Porter},
  {Primack}, {Ryan}, {Rosario}, {Somerville}, {Salvato}, \& {Hsu}}]{barro13}
{Barro}, G., {Faber}, S.~M., {P{\'e}rez-Gonz{\'a}lez}, P.~G., {et~al.} 2013,
  \apj, 765, 104

\bibitem[{{Behroozi} {et~al.}(2013){Behroozi}, {Wechsler}, \&
  {Conroy}}]{behroozi13}
{Behroozi}, P.~S., {Wechsler}, R.~H., \& {Conroy}, C. 2013, \apj, 770, 57

\bibitem[{{Bekerait{\.e}} {et~al.}(2016){Bekerait{\.e}}, {Walcher}, {Wisotzki},
  {Croton}, {Falc{\'o}n-Barroso}, {Lyubenova}, {Obreschkow}, {S{\'a}nchez},
  {Spekkens}, {Torrey}, {van de Ven}, {Zwaan}, {Ascasibar}, {Bland-Hawthorn},
  {Gonz{\'a}lez Delgado}, {Husemann}, {Marino}, {Vogelsberger}, \&
  {Ziegler}}]{bekeraite16}
{Bekerait{\.e}}, S., {Walcher}, C.~J., {Wisotzki}, L., {et~al.} 2016, \apjl,
  827, L36

\bibitem[{{Belfiore} {et~al.}(2017){Belfiore}, {Maiolino}, {Maraston},
  {Emsellem}, {Bershady}, {Masters}, {Bizyaev}, {Boquien}, {Brownstein},
  {Bundy}, {Diamond-Stanic}, {Drory}, {Heckman}, {Law}, {Malanushenko},
  {Oravetz}, {Pan}, {Roman-Lopes}, {Thomas}, {Weijmans}, {Westfall}, \&
  {Yan}}]{belfiore17}
{Belfiore}, F., {Maiolino}, R., {Maraston}, C., {et~al.} 2017, \mnras, 466,
  2570

\bibitem[{{Bell} {et~al.}(2004){Bell}, {McIntosh}, {Barden}, {Wolf},
  {Caldwell}, {Rix}, {Beckwith}, {Borch}, {H{\"a}ussler}, {Jahnke}, {Jogee},
  {Meisenheimer}, {Peng}, {Sanchez}, {Somerville}, \& {Wisotzki}}]{bell04}
{Bell}, E.~F., {McIntosh}, D.~H., {Barden}, M., {et~al.} 2004, \apjl, 600, L11

\bibitem[{{Bell} {et~al.}(2007){Bell}, {Zheng}, {Papovich}, {Borch}, {Wolf}, \&
  {Meisenheimer}}]{bell07}
{Bell}, E.~F., {Zheng}, X.~Z., {Papovich}, C., {et~al.} 2007, \apj, 663, 834

\bibitem[{{Bernard} {et~al.}(2015){Bernard}, {Ferguson}, {Richardson}, {Irwin},
  {Barker}, {Hidalgo}, {Aparicio}, {Chapman}, {Ibata}, {Lewis}, {McConnachie},
  \& {Tanvir}}]{bernard15}
{Bernard}, E.~J., {Ferguson}, A.~M.~N., {Richardson}, J.~C., {et~al.} 2015,
  \mnras, 446, 2789

\bibitem[{{Birnboim} {et~al.}(2007){Birnboim}, {Dekel}, \&
  {Neistein}}]{birnboim07}
{Birnboim}, Y., {Dekel}, A., \& {Neistein}, E. 2007, \mnras, 380, 339

\bibitem[{{Bland-Hawthorn} \& {Gerhard}(2016)}]{bland-hawthorn16}
{Bland-Hawthorn}, J. \& {Gerhard}, O. 2016, \araa, 54, 529

\bibitem[{{Blanton} {et~al.}(2003){Blanton}, {Hogg}, {Bahcall}, {Brinkmann},
  {Britton}, {Connolly}, {Csabai}, {Fukugita}, {Loveday}, {Meiksin}, {Munn},
  {Nichol}, {Okamura}, {Quinn}, {Schneider}, {Shimasaku}, {Strauss}, {Tegmark},
  {Vogeley}, \& {Weinberg}}]{blanton03}
{Blanton}, M.~R., {Hogg}, D.~W., {Bahcall}, N.~A., {et~al.} 2003, \apj, 592,
  819

\bibitem[{{Blanton} \& {Moustakas}(2009)}]{blanton09}
{Blanton}, M.~R. \& {Moustakas}, J. 2009, \araa, 47, 159

\bibitem[{{Bovy} {et~al.}(2012){Bovy}, {Rix}, \& {Hogg}}]{bovy12}
{Bovy}, J., {Rix}, H.-W., \& {Hogg}, D.~W. 2012, \apj, 751, 131

\bibitem[{{Brinchmann} {et~al.}(2004){Brinchmann}, {Charlot}, {White},
  {Tremonti}, {Kauffmann}, {Heckman}, \& {Brinkmann}}]{brinchmann04}
{Brinchmann}, J., {Charlot}, S., {White}, S.~D.~M., {et~al.} 2004, \mnras, 351,
  1151

\bibitem[{{Bruzual A.} \& {Kron}(1980)}]{bruzual80}
{Bruzual A.}, G. \& {Kron}, R.~G. 1980, \apj, 241, 25

\bibitem[{{Bryant} {et~al.}(2015){Bryant}, {Owers}, {Robotham}, {Croom},
  {Driver}, {Drinkwater}, {Lorente}, {Cortese}, {Scott}, {Colless}, {Schaefer},
  {Taylor}, {Konstantopoulos}, {Allen}, {Baldry}, {Barnes}, {Bauer},
  {Bland-Hawthorn}, {Bloom}, {Brooks}, {Brough}, {Cecil}, {Couch}, {Croton},
  {Davies}, {Ellis}, {Fogarty}, {Foster}, {Glazebrook}, {Goodwin}, {Green},
  {Gunawardhana}, {Hampton}, {Ho}, {Hopkins}, {Kewley}, {Lawrence},
  {Leon-Saval}, {Leslie}, {McElroy}, {Lewis}, {Liske}, {L{\'o}pez-S{\'a}nchez},
  {Mahajan}, {Medling}, {Metcalfe}, {Meyer}, {Mould}, {Obreschkow}, {O'Toole},
  {Pracy}, {Richards}, {Shanks}, {Sharp}, {Sweet}, {Thomas}, {Tonini}, \&
  {Walcher}}]{bryant15}
{Bryant}, J.~J., {Owers}, M.~S., {Robotham}, A.~S.~G., {et~al.} 2015, \mnras,
  447, 2857

\bibitem[{{Buitrago} {et~al.}(2008){Buitrago}, {Trujillo}, {Conselice},
  {Bouwens}, {Dickinson}, \& {Yan}}]{buitrago08}
{Buitrago}, F., {Trujillo}, I., {Conselice}, C.~J., {et~al.} 2008, \apjl, 687,
  L61

\bibitem[{{Buitrago} {et~al.}(2013){Buitrago}, {Trujillo}, {Conselice}, \&
  {H{\"a}u{\ss}ler}}]{buitrago13}
{Buitrago}, F., {Trujillo}, I., {Conselice}, C.~J., \& {H{\"a}u{\ss}ler}, B.
  2013, \mnras, 428, 1460

\bibitem[{{Bundy} {et~al.}(2015){Bundy}, {Bershady}, {Law}, {Yan}, {Drory},
  {MacDonald}, {Wake}, {Cherinka}, {S{\'a}nchez-Gallego}, {Weijmans}, {Thomas},
  {Tremonti}, {Masters}, {Coccato}, {Diamond-Stanic}, {Arag{\'o}n-Salamanca},
  {Avila-Reese}, {Badenes}, {Falc{\'o}n-Barroso}, {Belfiore}, {Bizyaev},
  {Blanc}, {Bland-Hawthorn}, {Blanton}, {Brownstein}, {Byler}, {Cappellari},
  {Conroy}, {Dutton}, {Emsellem}, {Etherington}, {Frinchaboy}, {Fu}, {Gunn},
  {Harding}, {Johnston}, {Kauffmann}, {Kinemuchi}, {Klaene}, {Knapen},
  {Leauthaud}, {Li}, {Lin}, {Maiolino}, {Malanushenko}, {Malanushenko}, {Mao},
  {Maraston}, {McDermid}, {Merrifield}, {Nichol}, {Oravetz}, {Pan}, {Parejko},
  {Sanchez}, {Schlegel}, {Simmons}, {Steele}, {Steinmetz}, {Thanjavur},
  {Thompson}, {Tinker}, {van den Bosch}, {Westfall}, {Wilkinson}, {Wright},
  {Xiao}, \& {Zhang}}]{bundy15}
{Bundy}, K., {Bershady}, M.~A., {Law}, D.~R., {et~al.} 2015, \apj, 798, 7

\bibitem[{{Buta}(2013)}]{buta13}
{Buta}, R.~J. 2013, {Galaxy Morphology}, ed. J.~{Falc{\'o}n-Barroso} \& J.~H.
  {Knapen}, 155

\bibitem[{{Buta} {et~al.}(2015){Buta}, {Sheth}, {Athanassoula}, {Bosma},
  {Knapen}, {Laurikainen}, {Salo}, {Elmegreen}, {Ho}, {Zaritsky}, {Courtois},
  {Hinz}, {Mu{\~n}oz-Mateos}, {Kim}, {Regan}, {Gadotti}, {Gil de Paz}, {Laine},
  {Men{\'e}ndez-Delmestre}, {Comer{\'o}n}, {Erroz Ferrer}, {Seibert},
  {Mizusawa}, {Holwerda}, \& {Madore}}]{buta15}
{Buta}, R.~J., {Sheth}, K., {Athanassoula}, E., {et~al.} 2015, \apjs, 217, 32

\bibitem[{{Cano-D{\'{\i}}az} {et~al.}(2016){Cano-D{\'{\i}}az}, {S{\'a}nchez},
  {Zibetti}, {Ascasibar}, {Bland-Hawthorn}, {Ziegler}, {Gonz{\'a}lez Delgado},
  {Walcher}, {Garc{\'{\i}}a-Benito}, {Mast}, {Mendoza-P{\'e}rez},
  {Falc{\'o}n-Barroso}, {Galbany}, {Husemann}, {Kehrig}, {Marino},
  {S{\'a}nchez-Bl{\'a}zquez}, {L{\'o}pez-Cob{\'a}}, {L{\'o}pez-S{\'a}nchez}, \&
  {Vilchez}}]{canodiaz16}
{Cano-D{\'{\i}}az}, M., {S{\'a}nchez}, S.~F., {Zibetti}, S., {et~al.} 2016,
  \apjl, 821, L26

\bibitem[{{Cappellari}(2013)}]{capellari13E}
{Cappellari}, M. 2013, \apjl, 778, L2

\bibitem[{{Cappellari}(2016)}]{cappellari16}
{Cappellari}, M. 2016, \araa, 54, 597

\bibitem[{{Cappellari} \& {Copin}(2003)}]{cappellari03}
{Cappellari}, M. \& {Copin}, Y. 2003, \mnras, 342, 345

\bibitem[{{Cappellari} {et~al.}(2011){Cappellari}, {Emsellem}, {Krajnovi{\'c}},
  {McDermid}, {Scott}, {Verdoes Kleijn}, {Young}, {Alatalo}, {Bacon}, {Blitz},
  {Bois}, {Bournaud}, {Bureau}, {Davies}, {Davis}, {de Zeeuw}, {Duc},
  {Khochfar}, {Kuntschner}, {Lablanche}, {Morganti}, {Naab}, {Oosterloo},
  {Sarzi}, {Serra}, \& {Weijmans}}]{cappellari11}
{Cappellari}, M., {Emsellem}, E., {Krajnovi{\'c}}, D., {et~al.} 2011, \mnras,
  413, 813

\bibitem[{{Cappellari} {et~al.}(2013){Cappellari}, {McDermid}, {Alatalo},
  {Blitz}, {Bois}, {Bournaud}, {Bureau}, {Crocker}, {Davies}, {Davis}, {de
  Zeeuw}, {Duc}, {Emsellem}, {Khochfar}, {Krajnovi{\'c}}, {Kuntschner},
  {Morganti}, {Naab}, {Oosterloo}, {Sarzi}, {Scott}, {Serra}, {Weijmans}, \&
  {Young}}]{cappellari13}
{Cappellari}, M., {McDermid}, R.~M., {Alatalo}, K., {et~al.} 2013, \mnras, 432,
  1862

\bibitem[{{Cardelli} {et~al.}(1989){Cardelli}, {Clayton}, \&
  {Mathis}}]{cardelli89}
{Cardelli}, J.~A., {Clayton}, G.~C., \& {Mathis}, J.~S. 1989, \apj, 345, 245

\bibitem[{{Carollo} {et~al.}(2013){Carollo}, {Bschorr}, {Renzini}, {Lilly},
  {Capak}, {Cibinel}, {Ilbert}, {Onodera}, {Scoville}, {Cameron}, {Mobasher},
  {Sanders}, \& {Taniguchi}}]{carollo13}
{Carollo}, C.~M., {Bschorr}, T.~J., {Renzini}, A., {et~al.} 2013, \apj, 773,
  112

\bibitem[{{Cassar{\`a}} {et~al.}(2016){Cassar{\`a}}, {Maccagni}, {Garilli},
  {Scodeggio}, {Thomas}, {Le F{\`e}vre}, {Zamorani}, {Schaerer}, {Lemaux},
  {Cassata}, {Le Brun}, {Pentericci}, {Tasca}, {Vanzella}, {Zucca},
  {Amor{\'{\i}}n}, {Bardelli}, {Castellano}, {Cimatti}, {Cucciati}, {Durkalec},
  {Fontana}, {Giavalisco}, {Grazian}, {Hathi}, {Ilbert}, {Paltani}, {Ribeiro},
  {Sommariva}, {Talia}, {Tresse}, {Vergani}, {Capak}, {Charlot}, {Contini}, {de
  la Torre}, {Dunlop}, {Fotopoulou}, {Guaita}, {Koekemoer},
  {L{\'o}pez-Sanjuan}, {Mellier}, {Pforr}, {Salvato}, {Scoville}, {Taniguchi},
  \& {Wang}}]{cassara16}
{Cassar{\`a}}, L.~P., {Maccagni}, D., {Garilli}, B., {et~al.} 2016, \aap, 593,
  A9

\bibitem[{{Catal{\'a}n-Torrecilla} {et~al.}(2015){Catal{\'a}n-Torrecilla}, {Gil
  de Paz}, {Castillo-Morales}, {Iglesias-P{\'a}ramo}, {S{\'a}nchez},
  {Kennicutt}, {P{\'e}rez-Gonz{\'a}lez}, {Marino}, {Walcher}, {Husemann},
  {Garc{\'{\i}}a-Benito}, {Mast}, {Gonz{\'a}lez Delgado}, {Mu{\~n}oz-Mateos},
  {Bland-Hawthorn}, {Bomans}, {Del Olmo}, {Galbany}, {Gomes}, {Kehrig},
  {L{\'o}pez-S{\'a}nchez}, {Mendoza}, {Monreal-Ibero}, {P{\'e}rez-Torres},
  {S{\'a}nchez-Bl{\'a}zquez}, {Vilchez}, \& {Califa Collaboration}}]{catalan15}
{Catal{\'a}n-Torrecilla}, C., {Gil de Paz}, A., {Castillo-Morales}, A.,
  {et~al.} 2015, \aap, 584, A87

\bibitem[{{Charbonnel} {et~al.}(1993){Charbonnel}, {Meynet}, {Maeder},
  {Schaller}, \& {Schaerer}}]{charbonnel93}
{Charbonnel}, C., {Meynet}, G., {Maeder}, A., {Schaller}, G., \& {Schaerer}, D.
  1993, \aaps, 101, 415

\bibitem[{{Cid Fernandes} {et~al.}(2005){Cid Fernandes}, {Mateus}, {Sodr{\'e}},
  {Stasi{\'n}ska}, \& {Gomes}}]{cidfernandes05}
{Cid Fernandes}, R., {Mateus}, A., {Sodr{\'e}}, L., {Stasi{\'n}ska}, G., \&
  {Gomes}, J.~M. 2005, \mnras, 358, 363

\bibitem[{{Cid Fernandes} {et~al.}(2013){Cid Fernandes}, {P{\'e}rez},
  {Garc{\'{\i}}a Benito}, {Gonz{\'a}lez Delgado}, {de Amorim}, {S{\'a}nchez},
  {Husemann}, {Falc{\'o}n Barroso}, {S{\'a}nchez-Bl{\'a}zquez}, {Walcher}, \&
  {Mast}}]{cidfernandes13}
{Cid Fernandes}, R., {P{\'e}rez}, E., {Garc{\'{\i}}a Benito}, R., {et~al.}
  2013, \aap, 557, A86

\bibitem[{{Cimatti} {et~al.}(2004){Cimatti}, {Daddi}, {Renzini}, {Cassata},
  {Vanzella}, {Pozzetti}, {Cristiani}, {Fontana}, {Rodighiero}, {Mignoli}, \&
  {Zamorani}}]{cimatti04}
{Cimatti}, A., {Daddi}, E., {Renzini}, A., {et~al.} 2004, \nat, 430, 184

\bibitem[{{Citro} {et~al.}(2016){Citro}, {Pozzetti}, {Moresco}, \&
  {Cimatti}}]{citro16}
{Citro}, A., {Pozzetti}, L., {Moresco}, M., \& {Cimatti}, A. 2016, \aap, 592,
  A19

\bibitem[{{Cordier} {et~al.}(2007){Cordier}, {Pietrinferni}, {Cassisi}, \&
  {Salaris}}]{cordier07}
{Cordier}, D., {Pietrinferni}, A., {Cassisi}, S., \& {Salaris}, M. 2007, \aj,
  133, 468

\bibitem[{{Croton} {et~al.}(2006){Croton}, {Springel}, {White}, {De Lucia},
  {Frenk}, {Gao}, {Jenkins}, {Kauffmann}, {Navarro}, \& {Yoshida}}]{croton06}
{Croton}, D.~J., {Springel}, V., {White}, S.~D.~M., {et~al.} 2006, \mnras, 365,
  11

\bibitem[{{Daddi} {et~al.}(2007){Daddi}, {Dickinson}, {Morrison}, {Chary},
  {Cimatti}, {Elbaz}, {Frayer}, {Renzini}, {Pope}, {Alexander}, {Bauer},
  {Giavalisco}, {Huynh}, {Kurk}, \& {Mignoli}}]{daddi07}
{Daddi}, E., {Dickinson}, M., {Morrison}, G., {et~al.} 2007, \apj, 670, 156

\bibitem[{{Dekel} \& {Birnboim}(2006)}]{dekel06}
{Dekel}, A. \& {Birnboim}, Y. 2006, \mnras, 368, 2

\bibitem[{{Dekel} {et~al.}(2009){Dekel}, {Birnboim}, {Engel}, {Freundlich},
  {Goerdt}, {Mumcuoglu}, {Neistein}, {Pichon}, {Teyssier}, \&
  {Zinger}}]{dekelbirnboim09}
{Dekel}, A., {Birnboim}, Y., {Engel}, G., {et~al.} 2009, \nat, 457, 451

\bibitem[{{Dekel} \& {Burkert}(2014)}]{dekel14}
{Dekel}, A. \& {Burkert}, A. 2014, \mnras, 438, 1870

\bibitem[{{Di Matteo} {et~al.}(2014){Di Matteo}, {Haywood}, {G{\'o}mez}, {van
  Damme}, {Combes}, {Hall{\'e}}, {Semelin}, {Lehnert}, \& {Katz}}]{diMatteo14}
{Di Matteo}, P., {Haywood}, M., {G{\'o}mez}, A., {et~al.} 2014, \aap, 567, A122

\bibitem[{{Di Matteo} {et~al.}(2005){Di Matteo}, {Springel}, \&
  {Hernquist}}]{diMatteoT05}
{Di Matteo}, T., {Springel}, V., \& {Hernquist}, L. 2005, \nat, 433, 604

\bibitem[{{Dopita} \& {Ryder}(1994)}]{dopita94}
{Dopita}, M.~A. \& {Ryder}, S.~D. 1994, \apj, 430, 163

\bibitem[{{Dressler} {et~al.}(2016){Dressler}, {Kelson}, {Abramson},
  {Gladders}, {Oemler}, {Poggianti}, {Mulchaey}, {Vulcani}, {Shectman},
  {Williams}, \& {McCarthy}}]{dressler16}
{Dressler}, A., {Kelson}, D.~D., {Abramson}, L.~E., {et~al.} 2016, \apj, 833,
  251

\bibitem[{{Dutton} {et~al.}(2010){Dutton}, {van den Bosch}, \&
  {Dekel}}]{dutton10}
{Dutton}, A.~A., {van den Bosch}, F.~C., \& {Dekel}, A. 2010, \mnras, 405, 1690

\bibitem[{{Elbaz} {et~al.}(2007){Elbaz}, {Daddi}, {Le Borgne}, {Dickinson},
  {Alexander}, {Chary}, {Starck}, {Brandt}, {Kitzbichler}, {MacDonald},
  {Nonino}, {Popesso}, {Stern}, \& {Vanzella}}]{elbaz07}
{Elbaz}, D., {Daddi}, E., {Le Borgne}, D., {et~al.} 2007, \aap, 468, 33

\bibitem[{{Elbaz} {et~al.}(2011){Elbaz}, {Dickinson}, {Hwang},
  {D{\'{\i}}az-Santos}, {Magdis}, {Magnelli}, {Le Borgne}, {Galliano},
  {Pannella}, {Chanial}, {Armus}, {Charmandaris}, {Daddi}, {Aussel}, {Popesso},
  {Kartaltepe}, {Altieri}, {Valtchanov}, {Coia}, {Dannerbauer}, {Dasyra},
  {Leiton}, {Mazzarella}, {Alexander}, {Buat}, {Burgarella}, {Chary}, {Gilli},
  {Ivison}, {Juneau}, {Le Floc'h}, {Lutz}, {Morrison}, {Mullaney}, {Murphy},
  {Pope}, {Scott}, {Brodwin}, {Calzetti}, {Cesarsky}, {Charlot}, {Dole},
  {Eisenhardt}, {Ferguson}, {F{\"o}rster Schreiber}, {Frayer}, {Giavalisco},
  {Huynh}, {Koekemoer}, {Papovich}, {Reddy}, {Surace}, {Teplitz}, {Yun}, \&
  {Wilson}}]{elbaz11}
{Elbaz}, D., {Dickinson}, M., {Hwang}, H.~S., {et~al.} 2011, \aap, 533, A119

\bibitem[{{Elmegreen} {et~al.}(2008){Elmegreen}, {Bournaud}, \&
  {Elmegreen}}]{elmegreen08}
{Elmegreen}, B.~G., {Bournaud}, F., \& {Elmegreen}, D.~M. 2008, \apj, 688, 67

\bibitem[{{Elmegreen} \& {Elmegreen}(2006)}]{elmegreen06}
{Elmegreen}, B.~G. \& {Elmegreen}, D.~M. 2006, \apj, 650, 644

\bibitem[{{Faber} {et~al.}(2007){Faber}, {Willmer}, {Wolf}, {Koo}, {Weiner},
  {Newman}, {Im}, {Coil}, {Conroy}, {Cooper}, {Davis}, {Finkbeiner}, {Gerke},
  {Gebhardt}, {Groth}, {Guhathakurta}, {Harker}, {Kaiser}, {Kassin},
  {Kleinheinrich}, {Konidaris}, {Kron}, {Lin}, {Luppino}, {Madgwick},
  {Meisenheimer}, {Noeske}, {Phillips}, {Sarajedini}, {Schiavon}, {Simard},
  {Szalay}, {Vogt}, \& {Yan}}]{Faber07}
{Faber}, S.~M., {Willmer}, C.~N.~A., {Wolf}, C., {et~al.} 2007, \apj, 665, 265

\bibitem[{{Fardal} {et~al.}(2007){Fardal}, {Katz}, {Weinberg}, \&
  {Dav{\'e}}}]{fardal07}
{Fardal}, M.~A., {Katz}, N., {Weinberg}, D.~H., \& {Dav{\'e}}, R. 2007, \mnras,
  379, 985

\bibitem[{{Ferr{\'e}-Mateu} {et~al.}(2017){Ferr{\'e}-Mateu}, {Trujillo},
  {Mart{\'{\i}}n-Navarro}, {Vazdekis}, {Mezcua}, {Balcells}, \&
  {Dom{\'{\i}}nguez}}]{ferre-mateu17}
{Ferr{\'e}-Mateu}, A., {Trujillo}, I., {Mart{\'{\i}}n-Navarro}, I., {et~al.}
  2017, ArXiv e-prints [\eprint[arXiv]{1701.05197}]

\bibitem[{{Feulner} {et~al.}(2005){Feulner}, {Gabasch}, {Salvato}, {Drory},
  {Hopp}, \& {Bender}}]{feulner05}
{Feulner}, G., {Gabasch}, A., {Salvato}, M., {et~al.} 2005, \apjl, 633, L9

\bibitem[{{F{\"o}rster Schreiber} {et~al.}(2011){F{\"o}rster Schreiber},
  {Shapley}, {Erb}, {Genzel}, {Steidel}, {Bouch{\'e}}, {Cresci}, \&
  {Davies}}]{forsterscheiber11}
{F{\"o}rster Schreiber}, N.~M., {Shapley}, A.~E., {Erb}, D.~K., {et~al.} 2011,
  \apj, 731, 65

\bibitem[{{Gallagher} {et~al.}(1984){Gallagher}, {Hunter}, \&
  {Tutukov}}]{gallagher84}
{Gallagher}, III, J.~S., {Hunter}, D.~A., \& {Tutukov}, A.~V. 1984, \apj, 284,
  544

\bibitem[{{Gallazzi} {et~al.}(2014){Gallazzi}, {Bell}, {Zibetti}, {Brinchmann},
  \& {Kelson}}]{gallazzi14}
{Gallazzi}, A., {Bell}, E.~F., {Zibetti}, S., {Brinchmann}, J., \& {Kelson},
  D.~D. 2014, \apj, 788, 72

\bibitem[{{Gallazzi} {et~al.}(2005){Gallazzi}, {Charlot}, {Brinchmann},
  {White}, \& {Tremonti}}]{gallazzi05}
{Gallazzi}, A., {Charlot}, S., {Brinchmann}, J., {White}, S.~D.~M., \&
  {Tremonti}, C.~A. 2005, \mnras, 362, 41

\bibitem[{{Garc{\'{\i}}a-Benito} {et~al.}(2015){Garc{\'{\i}}a-Benito},
  {Zibetti}, {S{\'a}nchez}, {Husemann}, {de Amorim}, {Castillo-Morales}, {Cid
  Fernandes}, {Ellis}, {Falc{\'o}n-Barroso}, {Galbany}, {Gil de Paz},
  {Gonz{\'a}lez Delgado}, {Lacerda}, {L{\'o}pez-Fernandez}, {de
  Lorenzo-C{\'a}ceres}, {Lyubenova}, {Marino}, {Mast}, {Mendoza}, {P{\'e}rez},
  {Vale Asari}, {Aguerri}, {Ascasibar}, {Bekerait*error*{\.e}},
  {Bland-Hawthorn}, {Barrera-Ballesteros}, {Bomans}, {Cano-D{\'{\i}}az},
  {Catal{\'a}n-Torrecilla}, {Cortijo}, {Delgado-Inglada}, {Demleitner},
  {Dettmar}, {D{\'{\i}}az}, {Florido}, {Gallazzi}, {Garc{\'{\i}}a-Lorenzo},
  {Gomes}, {Holmes}, {Iglesias-P{\'a}ramo}, {Jahnke}, {Kalinova}, {Kehrig},
  {Kennicutt}, {L{\'o}pez-S{\'a}nchez}, {M{\'a}rquez}, {Masegosa}, {Meidt},
  {Mendez-Abreu}, {Moll{\'a}}, {Monreal-Ibero}, {Morisset}, {del Olmo},
  {Papaderos}, {P{\'e}rez}, {Quirrenbach}, {Rosales-Ortega}, {Roth},
  {Ruiz-Lara}, {S{\'a}nchez-Bl{\'a}zquez}, {S{\'a}nchez-Menguiano}, {Singh},
  {Spekkens}, {Stanishev}, {Torres-Papaqui}, {van de Ven}, {Vilchez},
  {Walcher}, {Wild}, {Wisotzki}, {Ziegler}, {Alves}, {Barrado}, {Quintana}, \&
  {Aceituno}}]{garciabenito15}
{Garc{\'{\i}}a-Benito}, R., {Zibetti}, S., {S{\'a}nchez}, S.~F., {et~al.} 2015,
  \aap, 576, A135

\bibitem[{{Genzel} {et~al.}(2014){Genzel}, {F{\"o}rster Schreiber}, {Lang},
  {Tacchella}, {Tacconi}, {Wuyts}, {Bandara}, {Burkert}, {Buschkamp},
  {Carollo}, {Cresci}, {Davies}, {Eisenhauer}, {Hicks}, {Kurk}, {Lilly},
  {Lutz}, {Mancini}, {Naab}, {Newman}, {Peng}, {Renzini}, {Shapiro Griffin},
  {Sternberg}, {Vergani}, {Wisnioski}, {Wuyts}, \& {Zamorani}}]{genzel14}
{Genzel}, R., {F{\"o}rster Schreiber}, N.~M., {Lang}, P., {et~al.} 2014, \apj,
  785, 75

\bibitem[{{Girardi} {et~al.}(2000){Girardi}, {Bressan}, {Bertelli}, \&
  {Chiosi}}]{girardi00}
{Girardi}, L., {Bressan}, A., {Bertelli}, G., \& {Chiosi}, C. 2000, \aaps, 141,
  371

\bibitem[{{Gladders} {et~al.}(2013){Gladders}, {Oemler}, {Dressler},
  {Poggianti}, {Vulcani}, \& {Abramson}}]{gladders13}
{Gladders}, M.~D., {Oemler}, A., {Dressler}, A., {et~al.} 2013, \apj, 770, 64

\bibitem[{{Goddard} {et~al.}(2016){Goddard}, {Thomas}, {Maraston}, {Westfall},
  {Etherington}, {Riffel}, {Mallmann}, {Zheng}, {Argudo-Fernandez}, {Lian},
  {Bershady}, {Bundy}, {Drory}, {Law}, {Yan}, {Wake}, {Weijmans}, {Bizyaev},
  {Brownstein}, {Lane}, {Maiolino}, {Masters}, {Merrifield}, {Nitschelm},
  {Pan}, {Roman-Lopes}, {Storchi-Bergmann}, \& {Schneider}}]{goddard16}
{Goddard}, D., {Thomas}, D., {Maraston}, C., {et~al.} 2016, ArXiv e-prints
  [\eprint[arXiv]{1612.01546}]

\bibitem[{{Gonz{\'a}lez Delgado} {et~al.}(2005){Gonz{\'a}lez Delgado},
  {Cervi{\~n}o}, {Martins}, {Leitherer}, \& {Hauschildt}}]{gonzalezdelgado05}
{Gonz{\'a}lez Delgado}, R.~M., {Cervi{\~n}o}, M., {Martins}, L.~P.,
  {Leitherer}, C., \& {Hauschildt}, P.~H. 2005, \mnras, 357, 945

\bibitem[{{Gonz{\'a}lez Delgado} {et~al.}(2014{\natexlab{a}}){Gonz{\'a}lez
  Delgado}, {Cid Fernandes}, {Garc{\'{\i}}a-Benito}, {P{\'e}rez}, {de Amorim},
  {Cortijo-Ferrero}, {Lacerda}, {L{\'o}pez Fern{\'a}ndez}, {S{\'a}nchez}, {Vale
  Asari}, {Alves}, {Bland-Hawthorn}, {Galbany}, {Gallazzi}, {Husemann},
  {Bekeraite}, {Jungwiert}, {L{\'o}pez-S{\'a}nchez}, {de Lorenzo-C{\'a}ceres},
  {Marino}, {Mast}, {Moll{\'a}}, {del Olmo}, {S{\'a}nchez-Bl{\'a}zquez}, {van
  de Ven}, {V{\'{\i}}lchez}, {Walcher}, {Wisotzki}, {Ziegler}, \&
  {Collaboration920}}]{gonzalezdelgado14b}
{Gonz{\'a}lez Delgado}, R.~M., {Cid Fernandes}, R., {Garc{\'{\i}}a-Benito}, R.,
  {et~al.} 2014{\natexlab{a}}, \apjl, 791, L16

\bibitem[{{Gonz{\'a}lez Delgado} {et~al.}(2016){Gonz{\'a}lez Delgado}, {Cid
  Fernandes}, {P{\'e}rez}, {Garc{\'{\i}}a-Benito}, {L{\'o}pez Fern{\'a}ndez},
  {Lacerda}, {Cortijo-Ferrero}, {de Amorim}, {Vale Asari}, {S{\'a}nchez},
  {Walcher}, {Wisotzki}, {Mast}, {Alves}, {Ascasibar}, {Bland-Hawthorn},
  {Galbany}, {Kennicutt}, {M{\'a}rquez}, {Masegosa}, {Moll{\'a}},
  {S{\'a}nchez-Bl{\'a}zquez}, \& {V{\'{\i}}lchez}}]{gonzalezdelgado16}
{Gonz{\'a}lez Delgado}, R.~M., {Cid Fernandes}, R., {P{\'e}rez}, E., {et~al.}
  2016, \aap, 590, A44

\bibitem[{{Gonz{\'a}lez Delgado} {et~al.}(2015){Gonz{\'a}lez Delgado},
  {Garc{\'{\i}}a-Benito}, {P{\'e}rez}, {Cid Fernandes}, {de Amorim},
  {Cortijo-Ferrero}, {Lacerda}, {L{\'o}pez Fern{\'a}ndez}, {Vale-Asari},
  {S{\'a}nchez}, {Moll{\'a}}, {Ruiz-Lara}, {S{\'a}nchez-Bl{\'a}zquez},
  {Walcher}, {Alves}, {Aguerri}, {Bekerait{\'e}}, {Bland-Hawthorn}, {Galbany},
  {Gallazzi}, {Husemann}, {Iglesias-P{\'a}ramo}, {Kalinova},
  {L{\'o}pez-S{\'a}nchez}, {Marino}, {M{\'a}rquez}, {Masegosa}, {Mast},
  {M{\'e}ndez-Abreu}, {Mendoza}, {del Olmo}, {P{\'e}rez}, {Quirrenbach}, \&
  {Zibetti}}]{gonzalezdelgado15}
{Gonz{\'a}lez Delgado}, R.~M., {Garc{\'{\i}}a-Benito}, R., {P{\'e}rez}, E.,
  {et~al.} 2015, \aap, 581, A103

\bibitem[{{Gonz{\'a}lez Delgado} {et~al.}(2014{\natexlab{b}}){Gonz{\'a}lez
  Delgado}, {P{\'e}rez}, {Cid Fernandes}, {Garc{\'{\i}}a-Benito}, {de Amorim},
  {S{\'a}nchez}, {Husemann}, {Cortijo-Ferrero}, {L{\'o}pez Fern{\'a}ndez},
  {S{\'a}nchez-Bl{\'a}zquez}, {Bekeraite}, {Walcher}, {Falc{\'o}n-Barroso},
  {Gallazzi}, {van de Ven}, {Alves}, {Bland-Hawthorn}, {Kennicutt}, {Kupko},
  {Lyubenova}, {Mast}, {Moll{\'a}}, {Marino}, {Quirrenbach}, {V{\'{\i}}lchez},
  \& {Wisotzki}}]{gonzalezdelgado14a}
{Gonz{\'a}lez Delgado}, R.~M., {P{\'e}rez}, E., {Cid Fernandes}, R., {et~al.}
  2014{\natexlab{b}}, \aap, 562, A47

\bibitem[{{Hammer} {et~al.}(2005){Hammer}, {Flores}, {Elbaz}, {Zheng}, {Liang},
  \& {Cesarsky}}]{hammer05}
{Hammer}, F., {Flores}, H., {Elbaz}, D., {et~al.} 2005, \aap, 430, 115

\bibitem[{{Haywood} {et~al.}(2013){Haywood}, {Di Matteo}, {Lehnert}, {Katz}, \&
  {G{\'o}mez}}]{haywood13}
{Haywood}, M., {Di Matteo}, P., {Lehnert}, M.~D., {Katz}, D., \& {G{\'o}mez},
  A. 2013, \aap, 560, A109

\bibitem[{{Haywood} {et~al.}(2015){Haywood}, {Di Matteo}, {Snaith}, \&
  {Lehnert}}]{haywood15}
{Haywood}, M., {Di Matteo}, P., {Snaith}, O., \& {Lehnert}, M.~D. 2015, \aap,
  579, A5

\bibitem[{{Haywood} {et~al.}(2016){Haywood}, {Lehnert}, {Di Matteo}, {Snaith},
  {Schultheis}, {Katz}, \& {G{\'o}mez}}]{haywood16}
{Haywood}, M., {Lehnert}, M.~D., {Di Matteo}, P., {et~al.} 2016, \aap, 589, A66

\bibitem[{{Heavens} {et~al.}(2004){Heavens}, {Panter}, {Jimenez}, \&
  {Dunlop}}]{heavens04}
{Heavens}, A., {Panter}, B., {Jimenez}, R., \& {Dunlop}, J. 2004, \nat, 428,
  625

\bibitem[{{Hopkins} \& {Beacom}(2006)}]{hopkinsbeacom06}
{Hopkins}, A.~M. \& {Beacom}, J.~F. 2006, \apj, 651, 142

\bibitem[{{Huang} {et~al.}(2013){Huang}, {Kauffmann}, {Chen}, {Moran},
  {Heckman}, {Dav{\'e}}, \& {Johansson}}]{huang13}
{Huang}, M.-L., {Kauffmann}, G., {Chen}, Y.-M., {et~al.} 2013, \mnras, 431,
  2622

\bibitem[{{Huertas-Company} {et~al.}(2015){Huertas-Company},
  {P{\'e}rez-Gonz{\'a}lez}, {Mei}, {Shankar}, {Bernardi}, {Daddi}, {Barro},
  {Cabrera-Vives}, {Cattaneo}, {Dimauro}, \& {Gravet}}]{huertas15}
{Huertas-Company}, M., {P{\'e}rez-Gonz{\'a}lez}, P.~G., {Mei}, S., {et~al.}
  2015, \apj, 809, 95

\bibitem[{{Husemann} {et~al.}(2013){Husemann}, {Jahnke}, {S{\'a}nchez},
  {Barrado}, {Bekerait*error*{\.e}}, {Bomans}, {Castillo-Morales},
  {Catal{\'a}n-Torrecilla}, {Cid Fernandes}, {Falc{\'o}n-Barroso},
  {Garc{\'{\i}}a-Benito}, {Gonz{\'a}lez Delgado}, {Iglesias-P{\'a}ramo},
  {Johnson}, {Kupko}, {L{\'o}pez-Fernandez}, {Lyubenova}, {Marino}, {Mast},
  {Miskolczi}, {Monreal-Ibero}, {Gil de Paz}, {P{\'e}rez}, {P{\'e}rez},
  {Rosales-Ortega}, {Ruiz-Lara}, {Schilling}, {van de Ven}, {Walcher}, {Alves},
  {de Amorim}, {Backsmann}, {Barrera-Ballesteros}, {Bland-Hawthorn}, {Cortijo},
  {Dettmar}, {Demleitner}, {D{\'{\i}}az}, {Enke}, {Florido}, {Flores},
  {Galbany}, {Gallazzi}, {Garc{\'{\i}}a-Lorenzo}, {Gomes}, {Gruel}, {Haines},
  {Holmes}, {Jungwiert}, {Kalinova}, {Kehrig}, {Kennicutt}, {Klar}, {Lehnert},
  {L{\'o}pez-S{\'a}nchez}, {de Lorenzo-C{\'a}ceres}, {M{\'a}rmol-Queralt{\'o}},
  {M{\'a}rquez}, {Mendez-Abreu}, {Moll{\'a}}, {del Olmo}, {Meidt}, {Papaderos},
  {Puschnig}, {Quirrenbach}, {Roth}, {S{\'a}nchez-Bl{\'a}zquez}, {Spekkens},
  {Singh}, {Stanishev}, {Trager}, {Vilchez}, {Wild}, {Wisotzki}, {Zibetti}, \&
  {Ziegler}}]{husemann13}
{Husemann}, B., {Jahnke}, K., {S{\'a}nchez}, S.~F., {et~al.} 2013, \aap, 549,
  A87

\bibitem[{{Ibarra-Medel} {et~al.}(2016){Ibarra-Medel}, {S{\'a}nchez},
  {Avila-Reese}, {Hern{\'a}ndez-Toledo}, {Gonz{\'a}lez}, {Drory}, {Bundy},
  {Bizyaev}, {Cano-D{\'{\i}}az}, {Malanushenko}, {Pan}, {Roman-Lopes}, \&
  {Thomas}}]{ibarra16}
{Ibarra-Medel}, H.~J., {S{\'a}nchez}, S.~F., {Avila-Reese}, V., {et~al.} 2016,
  \mnras, 463, 2799

\bibitem[{{Karim} {et~al.}(2011){Karim}, {Schinnerer},
  {Mart{\'{\i}}nez-Sansigre}, {Sargent}, {van der Wel}, {Rix}, {Ilbert},
  {Smol{\v c}i{\'c}}, {Carilli}, {Pannella}, {Koekemoer}, {Bell}, \&
  {Salvato}}]{karim11}
{Karim}, A., {Schinnerer}, E., {Mart{\'{\i}}nez-Sansigre}, A., {et~al.} 2011,
  \apj, 730, 61

\bibitem[{{Kauffmann} {et~al.}(2003){Kauffmann}, {Heckman}, {White}, {Charlot},
  {Tremonti}, {Peng}, {Seibert}, {Brinkmann}, {Nichol}, {SubbaRao}, \&
  {York}}]{kauffmann03}
{Kauffmann}, G., {Heckman}, T.~M., {White}, S.~D.~M., {et~al.} 2003, \mnras,
  341, 54

\bibitem[{{Kaviraj} {et~al.}(2015){Kaviraj}, {Devriendt}, {Dubois}, {Slyz},
  {Welker}, {Pichon}, {Peirani}, \& {Le Borgne}}]{kaviraj15}
{Kaviraj}, S., {Devriendt}, J., {Dubois}, Y., {et~al.} 2015, \mnras, 452, 2845

\bibitem[{{Kelz} {et~al.}(2006){Kelz}, {Verheijen}, {Roth}, {Bauer}, {Becker},
  {Paschke}, {Popow}, {S{\'a}nchez}, \& {Laux}}]{kelz06}
{Kelz}, A., {Verheijen}, M.~A.~W., {Roth}, M.~M., {et~al.} 2006, \pasp, 118,
  129

\bibitem[{{Kennicutt}(1998)}]{kennicutt98}
{Kennicutt}, Jr., R.~C. 1998, \araa, 36, 189

\bibitem[{{Kere{\v s}} {et~al.}(2005){Kere{\v s}}, {Katz}, {Weinberg}, \&
  {Dav{\'e}}}]{kere05}
{Kere{\v s}}, D., {Katz}, N., {Weinberg}, D.~H., \& {Dav{\'e}}, R. 2005,
  \mnras, 363, 2

\bibitem[{{Koleva} {et~al.}(2009){Koleva}, {Prugniel}, {Bouchard}, \&
  {Wu}}]{koleva09}
{Koleva}, M., {Prugniel}, P., {Bouchard}, A., \& {Wu}, Y. 2009, \aap, 501, 1269

\bibitem[{{Koleva} {et~al.}(2011){Koleva}, {Prugniel}, {de Rijcke}, \&
  {Zeilinger}}]{koleva11}
{Koleva}, M., {Prugniel}, P., {de Rijcke}, S., \& {Zeilinger}, W.~W. 2011,
  \mnras, 417, 1643

\bibitem[{{Law} {et~al.}(2015){Law}, {Yan}, {Bershady}, {Bundy}, {Cherinka},
  {Drory}, {MacDonald}, {S{\'a}nchez-Gallego}, {Wake}, {Weijmans}, {Blanton},
  {Klaene}, {Moran}, {Sanchez}, \& {Zhang}}]{law15}
{Law}, D.~R., {Yan}, R., {Bershady}, M.~A., {et~al.} 2015, \aj, 150, 19

\bibitem[{{Lehnert} {et~al.}(2014){Lehnert}, {Di Matteo}, {Haywood}, \&
  {Snaith}}]{lehnert14}
{Lehnert}, M.~D., {Di Matteo}, P., {Haywood}, M., \& {Snaith}, O.~N. 2014,
  \apjl, 789, L30

\bibitem[{{Lehnert} {et~al.}(2013){Lehnert}, {Le Tiran}, {Nesvadba}, {van
  Driel}, {Boulanger}, \& {Di Matteo}}]{lehnert13}
{Lehnert}, M.~D., {Le Tiran}, L., {Nesvadba}, N.~P.~H., {et~al.} 2013, \aap,
  555, A72

\bibitem[{{Lehnert} {et~al.}(2015){Lehnert}, {van Driel}, {Le Tiran}, {Di
  Matteo}, \& {Haywood}}]{lehnert15}
{Lehnert}, M.~D., {van Driel}, W., {Le Tiran}, L., {Di Matteo}, P., \&
  {Haywood}, M. 2015, \aap, 577, A112

\bibitem[{{Leitner}(2012)}]{leitner12}
{Leitner}, S.~N. 2012, \apj, 745, 149

\bibitem[{{Licquia} \& {Newman}(2015)}]{licquia15}
{Licquia}, T.~C. \& {Newman}, J.~A. 2015, \apj, 806, 96

\bibitem[{{Lilly} {et~al.}(2013){Lilly}, {Carollo}, {Pipino}, {Renzini}, \&
  {Peng}}]{lilly13}
{Lilly}, S.~J., {Carollo}, C.~M., {Pipino}, A., {Renzini}, A., \& {Peng}, Y.
  2013, \apj, 772, 119

\bibitem[{{Lilly} {et~al.}(1996){Lilly}, {Le Fevre}, {Hammer}, \&
  {Crampton}}]{lilly96}
{Lilly}, S.~J., {Le Fevre}, O., {Hammer}, F., \& {Crampton}, D. 1996, \apjl,
  460, L1

\bibitem[{{Madau} \& {Dickinson}(2014)}]{madau14}
{Madau}, P. \& {Dickinson}, M. 2014, \araa, 52, 415

\bibitem[{{Madau} {et~al.}(1998){Madau}, {Pozzetti}, \& {Dickinson}}]{madau98}
{Madau}, P., {Pozzetti}, L., \& {Dickinson}, M. 1998, \apj, 498, 106

\bibitem[{{Magdis} {et~al.}(2010){Magdis}, {Elbaz}, {Daddi}, {Morrison},
  {Dickinson}, {Rigopoulou}, {Gobat}, \& {Hwang}}]{magdis10}
{Magdis}, G.~E., {Elbaz}, D., {Daddi}, E., {et~al.} 2010, \apj, 714, 1740

\bibitem[{{Man} {et~al.}(2012){Man}, {Toft}, {Zirm}, {Wuyts}, \& {van der
  Wel}}]{man12}
{Man}, A.~W.~S., {Toft}, S., {Zirm}, A.~W., {Wuyts}, S., \& {van der Wel}, A.
  2012, \apj, 744, 85

\bibitem[{{Maragkoudakis} {et~al.}(2016){Maragkoudakis}, {Zezas}, {Ashby}, \&
  {Willner}}]{maragkoudakis16}
{Maragkoudakis}, A., {Zezas}, A., {Ashby}, M.~L.~N., \& {Willner}, S.~P. 2016,
  ArXiv e-prints [\eprint[arXiv]{1611.10085}]

\bibitem[{{Maraston} {et~al.}(2013){Maraston}, {Pforr}, {Henriques}, {Thomas},
  {Wake}, {Brownstein}, {Capozzi}, {Tinker}, {Bundy}, {Skibba}, {Beifiori},
  {Nichol}, {Edmondson}, {Schneider}, {Chen}, {Masters}, {Steele}, {Bolton},
  {York}, {Weaver}, {Higgs}, {Bizyaev}, {Brewington}, {Malanushenko},
  {Malanushenko}, {Snedden}, {Oravetz}, {Pan}, {Shelden}, \&
  {Simmons}}]{maraston13}
{Maraston}, C., {Pforr}, J., {Henriques}, B.~M., {et~al.} 2013, \mnras, 435,
  2764

\bibitem[{{Maraston} {et~al.}(2010){Maraston}, {Pforr}, {Renzini}, {Daddi},
  {Dickinson}, {Cimatti}, \& {Tonini}}]{maraston10}
{Maraston}, C., {Pforr}, J., {Renzini}, A., {et~al.} 2010, \mnras, 407, 830

\bibitem[{{Martig} {et~al.}(2009){Martig}, {Bournaud}, {Teyssier}, \&
  {Dekel}}]{martig09}
{Martig}, M., {Bournaud}, F., {Teyssier}, R., \& {Dekel}, A. 2009, \apj, 707,
  250

\bibitem[{{Mateus} {et~al.}(2006){Mateus}, {Sodr{\'e}}, {Cid Fernandes},
  {Stasi{\'n}ska}, {Schoenell}, \& {Gomes}}]{mateus06}
{Mateus}, A., {Sodr{\'e}}, L., {Cid Fernandes}, R., {et~al.} 2006, \mnras, 370,
  721

\bibitem[{{McCarthy} {et~al.}(2004){McCarthy}, {Le Borgne}, {Crampton}, {Chen},
  {Abraham}, {Glazebrook}, {Savaglio}, {Carlberg}, {Marzke}, {Roth},
  {J{\o}rgensen}, {Hook}, {Murowinski}, \& {Juneau}}]{mcCarthy04}
{McCarthy}, P.~J., {Le Borgne}, D., {Crampton}, D., {et~al.} 2004, \apjl, 614,
  L9

\bibitem[{{McDermid} {et~al.}(2015){McDermid}, {Alatalo}, {Blitz}, {Bournaud},
  {Bureau}, {Cappellari}, {Crocker}, {Davies}, {Davis}, {de Zeeuw}, {Duc},
  {Emsellem}, {Khochfar}, {Krajnovi{\'c}}, {Kuntschner}, {Morganti}, {Naab},
  {Oosterloo}, {Sarzi}, {Scott}, {Serra}, {Weijmans}, \& {Young}}]{mcdermid15}
{McDermid}, R.~M., {Alatalo}, K., {Blitz}, L., {et~al.} 2015, \mnras, 448, 3484

\bibitem[{{Moles} {et~al.}(1995){Moles}, {Marquez}, \& {Perez}}]{moles95}
{Moles}, M., {Marquez}, I., \& {Perez}, E. 1995, \apj, 438, 604

\bibitem[{{Mosleh} {et~al.}(2012){Mosleh}, {Williams}, {Franx}, {Gonzalez},
  {Bouwens}, {Oesch}, {Labbe}, {Illingworth}, \& {Trenti}}]{mosleh12}
{Mosleh}, M., {Williams}, R.~J., {Franx}, M., {et~al.} 2012, \apjl, 756, L12

\bibitem[{{Naab} {et~al.}(2009){Naab}, {Johansson}, \& {Ostriker}}]{naab09}
{Naab}, T., {Johansson}, P.~H., \& {Ostriker}, J.~P. 2009, \apjl, 699, L178

\bibitem[{{Naab} \& {Ostriker}(2016)}]{naab16}
{Naab}, T. \& {Ostriker}, J.~P. 2016, ArXiv e-prints
  [\eprint[arXiv]{1612.06891}]

\bibitem[{{Neistein} \& {Dekel}(2008)}]{neistein08}
{Neistein}, E. \& {Dekel}, A. 2008, \mnras, 388, 1792

\bibitem[{{Neistein} {et~al.}(2006){Neistein}, {van den Bosch}, \&
  {Dekel}}]{neistein06}
{Neistein}, E., {van den Bosch}, F.~C., \& {Dekel}, A. 2006, \mnras, 372, 933

\bibitem[{{Newman} {et~al.}(2012){Newman}, {Ellis}, {Bundy}, \&
  {Treu}}]{newman12}
{Newman}, A.~B., {Ellis}, R.~S., {Bundy}, K., \& {Treu}, T. 2012, \apj, 746,
  162

\bibitem[{{Noeske} {et~al.}(2007){Noeske}, {Faber}, {Weiner}, {Koo}, {Primack},
  {Dekel}, {Papovich}, {Conselice}, {Le Floc'h}, {Rieke}, {Coil}, {Lotz},
  {Somerville}, \& {Bundy}}]{noeske07}
{Noeske}, K.~G., {Faber}, S.~M., {Weiner}, B.~J., {et~al.} 2007, \apjl, 660,
  L47

\bibitem[{{Ocvirk} {et~al.}(2006){Ocvirk}, {Pichon}, {Lan{\c c}on}, \&
  {Thi{\'e}baut}}]{ocvirk06}
{Ocvirk}, P., {Pichon}, C., {Lan{\c c}on}, A., \& {Thi{\'e}baut}, E. 2006,
  \mnras, 365, 74

\bibitem[{{Ocvirk} {et~al.}(2008){Ocvirk}, {Pichon}, \& {Teyssier}}]{ocvirk08}
{Ocvirk}, P., {Pichon}, C., \& {Teyssier}, R. 2008, \mnras, 390, 1326

\bibitem[{{Oemler} {et~al.}(2013){Oemler}, {Dressler}, {Gladders}, {Fritz},
  {Poggianti}, {Vulcani}, \& {Abramson}}]{oemler13}
{Oemler}, Jr., A., {Dressler}, A., {Gladders}, M.~G., {et~al.} 2013, \apj, 770,
  63

\bibitem[{{Oliver} {et~al.}(2010){Oliver}, {Frost}, {Farrah},
  {Gonzalez-Solares}, {Shupe}, {Henriques}, {Roseboom}, {Alfonso-Luis},
  {Babbedge}, {Frayer}, {Lencz}, {Lonsdale}, {Masci}, {Padgett}, {Polletta},
  {Rowan-Robinson}, {Siana}, {Smith}, {Surace}, \& {Vaccari}}]{oliver10}
{Oliver}, S., {Frost}, M., {Farrah}, D., {et~al.} 2010, \mnras, 405, 2279

\bibitem[{{Pacifici} {et~al.}(2016){Pacifici}, {Kassin}, {Weiner}, {Holden},
  {Gardner}, {Faber}, {Ferguson}, {Koo}, {Primack}, {Bell}, {Dekel}, {Gawiser},
  {Giavalisco}, {Rafelski}, {Simons}, {Barro}, {Croton}, {Dav{\'e}}, {Fontana},
  {Grogin}, {Koekemoer}, {Lee}, {Salmon}, {Somerville}, \&
  {Behroozi}}]{pacifici16}
{Pacifici}, C., {Kassin}, S.~A., {Weiner}, B.~J., {et~al.} 2016, \apj, 832, 79

\bibitem[{{Panter} {et~al.}(2003){Panter}, {Heavens}, \& {Jimenez}}]{panter03}
{Panter}, B., {Heavens}, A.~F., \& {Jimenez}, R. 2003, \mnras, 343, 1145

\bibitem[{{Panter} {et~al.}(2008){Panter}, {Jimenez}, {Heavens}, \&
  {Charlot}}]{panter08}
{Panter}, B., {Jimenez}, R., {Heavens}, A.~F., \& {Charlot}, S. 2008, \mnras,
  391, 1117

\bibitem[{{Papovich} {et~al.}(2011){Papovich}, {Finkelstein}, {Ferguson},
  {Lotz}, \& {Giavalisco}}]{papovich11}
{Papovich}, C., {Finkelstein}, S.~L., {Ferguson}, H.~C., {Lotz}, J.~M., \&
  {Giavalisco}, M. 2011, \mnras, 412, 1123

\bibitem[{{P{\'e}rez} {et~al.}(2013){P{\'e}rez}, {Cid Fernandes}, {Gonz{\'a}lez
  Delgado}, {Garc{\'{\i}}a-Benito}, {S{\'a}nchez}, {Husemann}, {Mast},
  {Rod{\'o}n}, {Kupko}, {Backsmann}, {de Amorim}, {van de Ven}, {Walcher},
  {Wisotzki}, {Cortijo-Ferrero}, \& {CALIFA Collaboration}}]{perez13}
{P{\'e}rez}, E., {Cid Fernandes}, R., {Gonz{\'a}lez Delgado}, R.~M., {et~al.}
  2013, \apjl, 764, L1

\bibitem[{{Pietrinferni} {et~al.}(2004){Pietrinferni}, {Cassisi}, {Salaris}, \&
  {Castelli}}]{pietrinferni04}
{Pietrinferni}, A., {Cassisi}, S., {Salaris}, M., \& {Castelli}, F. 2004, \apj,
  612, 168

\bibitem[{{Pietrinferni} {et~al.}(2006){Pietrinferni}, {Cassisi}, {Salaris}, \&
  {Castelli}}]{pietrinferni06}
{Pietrinferni}, A., {Cassisi}, S., {Salaris}, M., \& {Castelli}, F. 2006, \apj,
  642, 797

\bibitem[{{Pietrinferni} {et~al.}(2013){Pietrinferni}, {Cassisi}, {Salaris}, \&
  {Hidalgo}}]{pietrinferni13}
{Pietrinferni}, A., {Cassisi}, S., {Salaris}, M., \& {Hidalgo}, S. 2013, \aap,
  558, A46

\bibitem[{{Pietrinferni} {et~al.}(2009){Pietrinferni}, {Cassisi}, {Salaris},
  {Percival}, \& {Ferguson}}]{pietrinferni09}
{Pietrinferni}, A., {Cassisi}, S., {Salaris}, M., {Percival}, S., \&
  {Ferguson}, J.~W. 2009, \apj, 697, 275

\bibitem[{{Renzini} \& {Peng}(2015)}]{renzinipeng15}
{Renzini}, A. \& {Peng}, Y.-j. 2015, \apjl, 801, L29

\bibitem[{{Roberts} \& {Haynes}(1994)}]{roberts94}
{Roberts}, M.~S. \& {Haynes}, M.~P. 1994, \araa, 32, 115

\bibitem[{{Rodighiero} {et~al.}(2010){Rodighiero}, {Vaccari}, {Franceschini},
  {Tresse}, {Le Fevre}, {Le Brun}, {Mancini}, {Matute}, {Cimatti}, {Marchetti},
  {Ilbert}, {Arnouts}, {Bolzonella}, {Zucca}, {Bardelli}, {Lonsdale}, {Shupe},
  {Surace}, {Rowan-Robinson}, {Garilli}, {Zamorani}, {Pozzetti}, {Bondi}, {de
  la Torre}, {Vergani}, {Santini}, {Grazian}, \& {Fontana}}]{rodighiero10}
{Rodighiero}, G., {Vaccari}, M., {Franceschini}, A., {et~al.} 2010, \aap, 515,
  A8

\bibitem[{{Roth} {et~al.}(2005){Roth}, {Kelz}, {Fechner}, {Hahn}, {Bauer},
  {Becker}, {B{\"o}hm}, {Christensen}, {Dionies}, {Paschke}, {Popow}, {Wolter},
  {Schmoll}, {Laux}, \& {Altmann}}]{Roth05}
{Roth}, M.~M., {Kelz}, A., {Fechner}, T., {et~al.} 2005, \pasp, 117, 620

\bibitem[{{Salim} {et~al.}(2007){Salim}, {Rich}, {Charlot}, {Brinchmann},
  {Johnson}, {Schiminovich}, {Seibert}, {Mallery}, {Heckman}, {Forster},
  {Friedman}, {Martin}, {Morrissey}, {Neff}, {Small}, {Wyder}, {Bianchi},
  {Donas}, {Lee}, {Madore}, {Milliard}, {Szalay}, {Welsh}, \& {Yi}}]{salim07}
{Salim}, S., {Rich}, R.~M., {Charlot}, S., {et~al.} 2007, \apjs, 173, 267

\bibitem[{{S{\'a}nchez} {et~al.}(2016){S{\'a}nchez}, {Garc{\'{\i}}a-Benito},
  {Zibetti}, {Walcher}, {Husemann}, {Mendoza}, {Galbany}, {Falc{\'o}n-Barroso},
  {Mast}, {Aceituno}, {Aguerri}, {Alves}, {Amorim}, {Ascasibar},
  {Barrado-Navascues}, {Barrera-Ballesteros}, {Bekerait{\`e}},
  {Bland-Hawthorn}, {Cano D{\'{\i}}az}, {Cid Fernandes}, {Cavichia}, {Cortijo},
  {Dannerbauer}, {Demleitner}, {D{\'{\i}}az}, {Dettmar}, {de
  Lorenzo-C{\'a}ceres}, {del Olmo}, {Galazzi}, {Garc{\'{\i}}a-Lorenzo}, {Gil de
  Paz}, {Gonz{\'a}lez Delgado}, {Holmes}, {Igl{\'e}sias-P{\'a}ramo}, {Kehrig},
  {Kelz}, {Kennicutt}, {Kleemann}, {Lacerda}, {L{\'o}pez Fern{\'a}ndez},
  {L{\'o}pez S{\'a}nchez}, {Lyubenova}, {Marino}, {M{\'a}rquez},
  {Mendez-Abreu}, {Moll{\'a}}, {Monreal-Ibero}, {Ortega Minakata},
  {Torres-Papaqui}, {P{\'e}rez}, {Rosales-Ortega}, {Roth},
  {S{\'a}nchez-Bl{\'a}zquez}, {Schilling}, {Spekkens}, {Vale Asari}, {van den
  Bosch}, {van de Ven}, {Vilchez}, {Wild}, {Wisotzki}, {Y{\i}ld{\i}r{\i}m}, \&
  {Ziegler}}]{sanchez16}
{S{\'a}nchez}, S.~F., {Garc{\'{\i}}a-Benito}, R., {Zibetti}, S., {et~al.} 2016,
  \aap, 594, A36

\bibitem[{{S{\'a}nchez} {et~al.}(2012){S{\'a}nchez}, {Kennicutt}, {Gil de Paz},
  {van de Ven}, {V{\'{\i}}lchez}, {Wisotzki}, {Walcher}, {Mast}, {Aguerri},
  {Albiol-P{\'e}rez}, {Alonso-Herrero}, {Alves}, {Bakos}, {Bart{\'a}kov{\'a}},
  {Bland-Hawthorn}, {Boselli}, {Bomans}, {Castillo-Morales}, {Cortijo-Ferrero},
  {de Lorenzo-C{\'a}ceres}, {Del Olmo}, {Dettmar}, {D{\'{\i}}az}, {Ellis},
  {Falc{\'o}n-Barroso}, {Flores}, {Gallazzi}, {Garc{\'{\i}}a-Lorenzo},
  {Gonz{\'a}lez Delgado}, {Gruel}, {Haines}, {Hao}, {Husemann},
  {Igl{\'e}sias-P{\'a}ramo}, {Jahnke}, {Johnson}, {Jungwiert}, {Kalinova},
  {Kehrig}, {Kupko}, {L{\'o}pez-S{\'a}nchez}, {Lyubenova}, {Marino},
  {M{\'a}rmol-Queralt{\'o}}, {M{\'a}rquez}, {Masegosa}, {Meidt},
  {Mendez-Abreu}, {Monreal-Ibero}, {Montijo}, {Mour{\~a}o}, {Palacios-Navarro},
  {Papaderos}, {Pasquali}, {Peletier}, {P{\'e}rez}, {P{\'e}rez}, {Quirrenbach},
  {Rela{\~n}o}, {Rosales-Ortega}, {Roth}, {Ruiz-Lara},
  {S{\'a}nchez-Bl{\'a}zquez}, {Sengupta}, {Singh}, {Stanishev}, {Trager},
  {Vazdekis}, {Viironen}, {Wild}, {Zibetti}, \& {Ziegler}}]{sanchez12}
{S{\'a}nchez}, S.~F., {Kennicutt}, R.~C., {Gil de Paz}, A., {et~al.} 2012,
  \aap, 538, A8

\bibitem[{{S{\'a}nchez Almeida} {et~al.}(2014){S{\'a}nchez Almeida},
  {Elmegreen}, {Mu{\~n}oz-Tu{\~n}{\'o}n}, \& {Elmegreen}}]{jsa14}
{S{\'a}nchez Almeida}, J., {Elmegreen}, B.~G., {Mu{\~n}oz-Tu{\~n}{\'o}n}, C.,
  \& {Elmegreen}, D.~M. 2014, \aapr, 22, 71

\bibitem[{{S{\'a}nchez-Bl{\'a}zquez} {et~al.}(2006){S{\'a}nchez-Bl{\'a}zquez},
  {Peletier}, {Jim{\'e}nez-Vicente}, {Cardiel}, {Cenarro},
  {Falc{\'o}n-Barroso}, {Gorgas}, {Selam}, \& {Vazdekis}}]{sanchez-blazquez06}
{S{\'a}nchez-Bl{\'a}zquez}, P., {Peletier}, R.~F., {Jim{\'e}nez-Vicente}, J.,
  {et~al.} 2006, \mnras, 371, 703

\bibitem[{{S{\'a}nchez-Bl{\'a}zquez} {et~al.}(2014){S{\'a}nchez-Bl{\'a}zquez},
  {Rosales-Ortega}, {M{\'e}ndez-Abreu}, {P{\'e}rez}, {S{\'a}nchez}, {Zibetti},
  {Aguerri}, {Bland-Hawthorn}, {Catal{\'a}n-Torrecilla}, {Cid Fernandes}, {de
  Amorim}, {de Lorenzo-Caceres}, {Falc{\'o}n-Barroso}, {Galazzi},
  {Garc{\'{\i}}a Benito}, {Gil de Paz}, {Gonz{\'a}lez Delgado}, {Husemann},
  {Iglesias-P{\'a}ramo}, {Jungwiert}, {Marino}, {M{\'a}rquez}, {Mast},
  {Mendoza}, {Moll{\'a}}, {Papaderos}, {Ruiz-Lara}, {van de Ven}, {Walcher}, \&
  {Wisotzki}}]{sanchez-blazquez14}
{S{\'a}nchez-Bl{\'a}zquez}, P., {Rosales-Ortega}, F.~F., {M{\'e}ndez-Abreu},
  J., {et~al.} 2014, \aap, 570, A6

\bibitem[{{Sandage}(1986)}]{sandage86}
{Sandage}, A. 1986, \aap, 161, 89

\bibitem[{{Scalo}(1986)}]{scalo86}
{Scalo}, J.~M. 1986, \fcp, 11, 1

\bibitem[{{Schaerer} {et~al.}(1993){Schaerer}, {Charbonnel}, {Meynet},
  {Maeder}, \& {Schaller}}]{schaerer93}
{Schaerer}, D., {Charbonnel}, C., {Meynet}, G., {Maeder}, A., \& {Schaller}, G.
  1993, \aaps, 102, 339

\bibitem[{{Schaller} {et~al.}(1992){Schaller}, {Schaerer}, {Meynet}, \&
  {Maeder}}]{schaller92}
{Schaller}, G., {Schaerer}, D., {Meynet}, G., \& {Maeder}, A. 1992, \aaps, 96,
  269

\bibitem[{{Schiminovich} {et~al.}(2007){Schiminovich}, {Wyder}, {Martin},
  {Johnson}, {Salim}, {Seibert}, {Treyer}, {Budav{\'a}ri}, {Hoopes},
  {Zamojski}, {Barlow}, {Forster}, {Friedman}, {Morrissey}, {Neff}, {Small},
  {Bianchi}, {Donas}, {Heckman}, {Lee}, {Madore}, {Milliard}, {Rich}, {Szalay},
  {Welsh}, \& {Yi}}]{schiminovich07}
{Schiminovich}, D., {Wyder}, T.~K., {Martin}, D.~C., {et~al.} 2007, \apjs, 173,
  315

\bibitem[{{Searle} {et~al.}(1973){Searle}, {Sargent}, \& {Bagnuolo}}]{searle73}
{Searle}, L., {Sargent}, W.~L.~W., \& {Bagnuolo}, W.~G. 1973, \apj, 179, 427

\bibitem[{{Shi} {et~al.}(2011){Shi}, {Helou}, {Yan}, {Armus}, {Wu}, {Papovich},
  \& {Stierwalt}}]{shi11}
{Shi}, Y., {Helou}, G., {Yan}, L., {et~al.} 2011, \apj, 733, 87

\bibitem[{{Snaith} {et~al.}(2015){Snaith}, {Haywood}, {Di Matteo}, {Lehnert},
  {Combes}, {Katz}, \& {G{\'o}mez}}]{snaith15}
{Snaith}, O., {Haywood}, M., {Di Matteo}, P., {et~al.} 2015, \aap, 578, A87

\bibitem[{{Snaith} {et~al.}(2014){Snaith}, {Haywood}, {Di Matteo}, {Lehnert},
  {Combes}, {Katz}, \& {G{\'o}mez}}]{snaith14}
{Snaith}, O.~N., {Haywood}, M., {Di Matteo}, P., {et~al.} 2014, \apjl, 781, L31

\bibitem[{{Speagle} {et~al.}(2014){Speagle}, {Steinhardt}, {Capak}, \&
  {Silverman}}]{speagle14}
{Speagle}, J.~S., {Steinhardt}, C.~L., {Capak}, P.~L., \& {Silverman}, J.~D.
  2014, \apjs, 214, 15

\bibitem[{{Stark} {et~al.}(2013){Stark}, {Schenker}, {Ellis}, {Robertson},
  {McLure}, \& {Dunlop}}]{stark13}
{Stark}, D.~P., {Schenker}, M.~A., {Ellis}, R., {et~al.} 2013, \apj, 763, 129

\bibitem[{{Tacchella} {et~al.}(2015){Tacchella}, {Carollo}, {Renzini},
  {Schreiber}, {Lang}, {Wuyts}, {Cresci}, {Dekel}, {Genzel}, {Lilly},
  {Mancini}, {Newman}, {Onodera}, {Shapley}, {Tacconi}, {Woo}, \&
  {Zamorani}}]{tacchella15}
{Tacchella}, S., {Carollo}, C.~M., {Renzini}, A., {et~al.} 2015, Science, 348,
  314

\bibitem[{{Tacchella} {et~al.}(2016){Tacchella}, {Dekel}, {Carollo},
  {Ceverino}, {DeGraf}, {Lapiner}, {Mandelker}, \& {Primack}}]{tacchella16}
{Tacchella}, S., {Dekel}, A., {Carollo}, C.~M., {et~al.} 2016, \mnras, 458, 242

\bibitem[{{Tacconi} {et~al.}(2013){Tacconi}, {Neri}, {Genzel}, {Combes},
  {Bolatto}, {Cooper}, {Wuyts}, {Bournaud}, {Burkert}, {Comerford}, {Cox},
  {Davis}, {F{\"o}rster Schreiber}, {Garc{\'{\i}}a-Burillo}, {Gracia-Carpio},
  {Lutz}, {Naab}, {Newman}, {Omont}, {Saintonge}, {Shapiro Griffin}, {Shapley},
  {Sternberg}, \& {Weiner}}]{tacconi13}
{Tacconi}, L.~J., {Neri}, R., {Genzel}, R., {et~al.} 2013, \apj, 768, 74

\bibitem[{{Tinsley}(1968)}]{tinsley68}
{Tinsley}, B.~M. 1968, \apj, 151, 547

\bibitem[{{Tinsley}(1972)}]{tinsley72}
{Tinsley}, B.~M. 1972, \aap, 20, 383

\bibitem[{{Tojeiro} {et~al.}(2016){Tojeiro}, {Eardley}, {Peacock}, {Norberg},
  {Alpaslan}, {Driver}, {Henriques}, {Hopkins}, {Kafle}, {Robotham}, {Thomas},
  {Tonini}, \& {Wild}}]{tojeiro16}
{Tojeiro}, R., {Eardley}, E., {Peacock}, J.~A., {et~al.} 2016, ArXiv e-prints
  [\eprint[arXiv]{1612.08595}]

\bibitem[{{Tojeiro} {et~al.}(2007){Tojeiro}, {Heavens}, {Jimenez}, \&
  {Panter}}]{tojeiro07}
{Tojeiro}, R., {Heavens}, A.~F., {Jimenez}, R., \& {Panter}, B. 2007, \mnras,
  381, 1252

\bibitem[{{Tojeiro} {et~al.}(2011){Tojeiro}, {Percival}, {Heavens}, \&
  {Jimenez}}]{tojeiro11}
{Tojeiro}, R., {Percival}, W.~J., {Heavens}, A.~F., \& {Jimenez}, R. 2011,
  \mnras, 413, 434

\bibitem[{{Tojeiro} {et~al.}(2009){Tojeiro}, {Wilkins}, {Heavens}, {Panter}, \&
  {Jimenez}}]{tojeiro09}
{Tojeiro}, R., {Wilkins}, S., {Heavens}, A.~F., {Panter}, B., \& {Jimenez}, R.
  2009, \apjs, 185, 1

\bibitem[{{Trujillo} {et~al.}(2007){Trujillo}, {Conselice}, {Bundy}, {Cooper},
  {Eisenhardt}, \& {Ellis}}]{trujillo07}
{Trujillo}, I., {Conselice}, C.~J., {Bundy}, K., {et~al.} 2007, \mnras, 382,
  109

\bibitem[{{Trujillo} {et~al.}(2006){Trujillo}, {F{\"o}rster Schreiber},
  {Rudnick}, {Barden}, {Franx}, {Rix}, {Caldwell}, {McIntosh}, {Toft},
  {H{\"a}ussler}, {Zirm}, {van Dokkum}, {Labb{\'e}}, {Moorwood},
  {R{\"o}ttgering}, {van der Wel}, {van der Werf}, \& {van
  Starkenburg}}]{trujillo06}
{Trujillo}, I., {F{\"o}rster Schreiber}, N.~M., {Rudnick}, G., {et~al.} 2006,
  \apj, 650, 18

\bibitem[{{van der Wel} {et~al.}(2014){van der Wel}, {Franx}, {van Dokkum},
  {Skelton}, {Momcheva}, {Whitaker}, {Brammer}, {Bell}, {Rix}, {Wuyts},
  {Ferguson}, {Holden}, {Barro}, {Koekemoer}, {Chang}, {McGrath},
  {H{\"a}ussler}, {Dekel}, {Behroozi}, {Fumagalli}, {Leja}, {Lundgren},
  {Maseda}, {Nelson}, {Wake}, {Patel}, {Labb{\'e}}, {Faber}, {Grogin}, \&
  {Kocevski}}]{vanderwel14}
{van der Wel}, A., {Franx}, M., {van Dokkum}, P.~G., {et~al.} 2014, \apj, 788,
  28

\bibitem[{{van Dokkum} {et~al.}(2013){van Dokkum}, {Leja}, {Nelson}, {Patel},
  {Skelton}, {Momcheva}, {Brammer}, {Whitaker}, {Lundgren}, {Fumagalli},
  {Conroy}, {F{\"o}rster Schreiber}, {Franx}, {Kriek}, {Labb{\'e}},
  {Marchesini}, {Rix}, {van der Wel}, \& {Wuyts}}]{vandokkum13}
{van Dokkum}, P.~G., {Leja}, J., {Nelson}, E.~J., {et~al.} 2013, \apjl, 771,
  L35

\bibitem[{{van Dokkum} {et~al.}(2015){van Dokkum}, {Nelson}, {Franx}, {Oesch},
  {Momcheva}, {Brammer}, {F{\"o}rster Schreiber}, {Skelton}, {Whitaker}, {van
  der Wel}, {Bezanson}, {Fumagalli}, {Illingworth}, {Kriek}, {Leja}, \&
  {Wuyts}}]{vandokkum15}
{van Dokkum}, P.~G., {Nelson}, E.~J., {Franx}, M., {et~al.} 2015, \apj, 813, 23

\bibitem[{{van Dokkum} {et~al.}(2010){van Dokkum}, {Whitaker}, {Brammer},
  {Franx}, {Kriek}, {Labb{\'e}}, {Marchesini}, {Quadri}, {Bezanson},
  {Illingworth}, {Muzzin}, {Rudnick}, {Tal}, \& {Wake}}]{vandokkum10}
{van Dokkum}, P.~G., {Whitaker}, K.~E., {Brammer}, G., {et~al.} 2010, \apj,
  709, 1018

\bibitem[{{van Driel} {et~al.}(2016){van Driel}, {Butcher}, {Schneider},
  {Lehnert}, {Minchin}, {Blyth}, {Chemin}, {Hallet}, {Joseph}, {Kotze},
  {Kraan-Korteweg}, {Olofsson}, \& {Ramatsoku}}]{vandriel16}
{van Driel}, W., {Butcher}, Z., {Schneider}, S., {et~al.} 2016, \aap, 595, A118

\bibitem[{{Vazdekis} {et~al.}(2015){Vazdekis}, {Coelho}, {Cassisi},
  {Ricciardelli}, {Falc{\'o}n-Barroso}, {S{\'a}nchez-Bl{\'a}zquez}, {Barbera},
  {Beasley}, \& {Pietrinferni}}]{Vazdekis15}
{Vazdekis}, A., {Coelho}, P., {Cassisi}, S., {et~al.} 2015, \mnras, 449, 1177

\bibitem[{{Vazdekis} {et~al.}(2010){Vazdekis}, {S{\'a}nchez-Bl{\'a}zquez},
  {Falc{\'o}n-Barroso}, {Cenarro}, {Beasley}, {Cardiel}, {Gorgas}, \&
  {Peletier}}]{Vazdekis10}
{Vazdekis}, A., {S{\'a}nchez-Bl{\'a}zquez}, P., {Falc{\'o}n-Barroso}, J.,
  {et~al.} 2010, \mnras, 404, 1639

\bibitem[{{Verheijen} {et~al.}(2004){Verheijen}, {Bershady}, {Andersen},
  {Swaters}, {Westfall}, {Kelz}, \& {Roth}}]{verheijen04}
{Verheijen}, M.~A.~W., {Bershady}, M.~A., {Andersen}, D.~R., {et~al.} 2004,
  Astronomische Nachrichten, 325, 151

\bibitem[{{Walcher} {et~al.}(2008){Walcher}, {Lamareille}, {Vergani},
  {Arnouts}, {Buat}, {Charlot}, {Tresse}, {Le F{\`e}vre}, {Bolzonella},
  {Brinchmann}, {Pozzetti}, {Zamorani}, {Bottini}, {Garilli}, {Le Brun},
  {Maccagni}, {Milliard}, {Scaramella}, {Scodeggio}, {Vettolani}, {Zanichelli},
  {Adami}, {Bardelli}, {Cappi}, {Ciliegi}, {Contini}, {Franzetti}, {Foucaud},
  {Gavignaud}, {Guzzo}, {Ilbert}, {Iovino}, {McCracken}, {Marano}, {Marinoni},
  {Mazure}, {Meneux}, {Merighi}, {Paltani}, {Pell{\`o}}, {Pollo}, {Radovich},
  {Zucca}, {Lonsdale}, \& {Martin}}]{walcher08}
{Walcher}, C.~J., {Lamareille}, F., {Vergani}, D., {et~al.} 2008, \aap, 491,
  713

\bibitem[{{Walcher} {et~al.}(2014){Walcher}, {Wisotzki}, {Bekerait{\'e}},
  {Husemann}, {Iglesias-P{\'a}ramo}, {Backsmann}, {Barrera Ballesteros},
  {Catal{\'a}n-Torrecilla}, {Cortijo}, {del Olmo}, {Garcia Lorenzo},
  {Falc{\'o}n-Barroso}, {Jilkova}, {Kalinova}, {Mast}, {Marino},
  {M{\'e}ndez-Abreu}, {Pasquali}, {S{\'a}nchez}, {Trager}, {Zibetti},
  {Aguerri}, {Alves}, {Bland-Hawthorn}, {Boselli}, {Castillo Morales}, {Cid
  Fernandes}, {Flores}, {Galbany}, {Gallazzi}, {Garc{\'{\i}}a-Benito}, {Gil de
  Paz}, {Gonz{\'a}lez-Delgado}, {Jahnke}, {Jungwiert}, {Kehrig}, {Lyubenova},
  {M{\'a}rquez Perez}, {Masegosa}, {Monreal Ibero}, {P{\'e}rez}, {Quirrenbach},
  {Rosales-Ortega}, {Roth}, {Sanchez-Blazquez}, {Spekkens}, {Tundo}, {van de
  Ven}, {Verheijen}, {Vilchez}, \& {Ziegler}}]{walcher14}
{Walcher}, C.~J., {Wisotzki}, L., {Bekerait{\'e}}, S., {et~al.} 2014, \aap,
  569, A1

\bibitem[{{Whitaker} {et~al.}(2014){Whitaker}, {Franx}, {Leja}, {van Dokkum},
  {Henry}, {Skelton}, {Fumagalli}, {Momcheva}, {Brammer}, {Labb{\'e}},
  {Nelson}, \& {Rigby}}]{whitaker14}
{Whitaker}, K.~E., {Franx}, M., {Leja}, J., {et~al.} 2014, \apj, 795, 104

\bibitem[{{Whitaker} {et~al.}(2012){Whitaker}, {van Dokkum}, {Brammer}, \&
  {Franx}}]{whitaker12}
{Whitaker}, K.~E., {van Dokkum}, P.~G., {Brammer}, G., \& {Franx}, M. 2012,
  \apjl, 754, L29

\bibitem[{{Whitaker} {et~al.}(2013){Whitaker}, {van Dokkum}, {Brammer},
  {Momcheva}, {Skelton}, {Franx}, {Kriek}, {Labb{\'e}}, {Fumagalli},
  {Lundgren}, {Nelson}, {Patel}, \& {Rix}}]{whitaker13}
{Whitaker}, K.~E., {van Dokkum}, P.~G., {Brammer}, G., {et~al.} 2013, \apjl,
  770, L39

\bibitem[{{Williams} {et~al.}(2015){Williams}, {Dalcanton}, {Dolphin}, {Weisz},
  {Lewis}, {Lang}, {Bell}, {Boyer}, {Fouesneau}, {Gilbert}, {Monachesi}, \&
  {Skillman}}]{williams15}
{Williams}, B.~F., {Dalcanton}, J.~J., {Dolphin}, A.~E., {et~al.} 2015, \apj,
  806, 48

\bibitem[{{Williams} {et~al.}(2011){Williams}, {Quadri}, \&
  {Franx}}]{williams11}
{Williams}, R.~J., {Quadri}, R.~F., \& {Franx}, M. 2011, \apjl, 738, L25

\bibitem[{{Worthey}(1994)}]{worthey94}
{Worthey}, G. 1994, \apjs, 95, 107

\bibitem[{{Wuyts} {et~al.}(2013){Wuyts}, {F{\"o}rster Schreiber}, {Nelson},
  {van Dokkum}, {Brammer}, {Chang}, {Faber}, {Ferguson}, {Franx}, {Fumagalli},
  {Genzel}, {Grogin}, {Kocevski}, {Koekemoer}, {Lundgren}, {Lutz}, {McGrath},
  {Momcheva}, {Rosario}, {Skelton}, {Tacconi}, {van der Wel}, \&
  {Whitaker}}]{wuyts13}
{Wuyts}, S., {F{\"o}rster Schreiber}, N.~M., {Nelson}, E.~J., {et~al.} 2013,
  \apj, 779, 135

\bibitem[{{Wuyts} {et~al.}(2011){Wuyts}, {F{\"o}rster Schreiber}, {van der
  Wel}, {Magnelli}, {Guo}, {Genzel}, {Lutz}, {Aussel}, {Barro}, {Berta},
  {Cava}, {Graci{\'a}-Carpio}, {Hathi}, {Huang}, {Kocevski}, {Koekemoer},
  {Lee}, {Le Floc'h}, {McGrath}, {Nordon}, {Popesso}, {Pozzi}, {Riguccini},
  {Rodighiero}, {Saintonge}, \& {Tacconi}}]{wuyts11}
{Wuyts}, S., {F{\"o}rster Schreiber}, N.~M., {van der Wel}, A., {et~al.} 2011,
  \apj, 742, 96

\bibitem[{{Yuma} {et~al.}(2011){Yuma}, {Ohta}, {Yabe}, {Kajisawa}, \&
  {Ichikawa}}]{yuma11}
{Yuma}, S., {Ohta}, K., {Yabe}, K., {Kajisawa}, M., \& {Ichikawa}, T. 2011,
  \apj, 736, 92

\bibitem[{{Zheng} {et~al.}(2017){Zheng}, {Wang}, {Ge}, {Mao}, {Li}, {Li}, {Mo},
  {Goddard}, {Bundy}, {Li}, {Nair}, {Lin}, {Long}, {Riffel}, {Thomas},
  {Masters}, {Bizyaev}, {Brownstein}, {Zhang}, {Law}, {Drory}, {Roman Lopes},
  \& {Malanushenko}}]{zheng17}
{Zheng}, Z., {Wang}, H., {Ge}, J., {et~al.} 2017, \mnras, 465, 4572

\bibitem[{{Zibetti} {et~al.}(2017){Zibetti}, {Gallazzi}, {Ascasibar},
  {Charlot}, {Galbany}, {Garcia Benito}, {Kehrig}, {de Lorenzo-Caceres},
  {Lyubenova}, {Marino}, {Marquez}, {Sanchez}, {van de Ven}, {Walcher}, \&
  {Wisotzki}}]{zibetti17}
{Zibetti}, S., {Gallazzi}, A.~R., {Ascasibar}, Y., {et~al.} 2017, ArXiv
  e-prints [\eprint[arXiv]{1701.06570}]

\end{thebibliography}

\appendix
\section{Dependence of  the stellar population properties on CSP models}

\label{sec:CSP}

\begin{figure*}
\includegraphics[width=\textwidth]{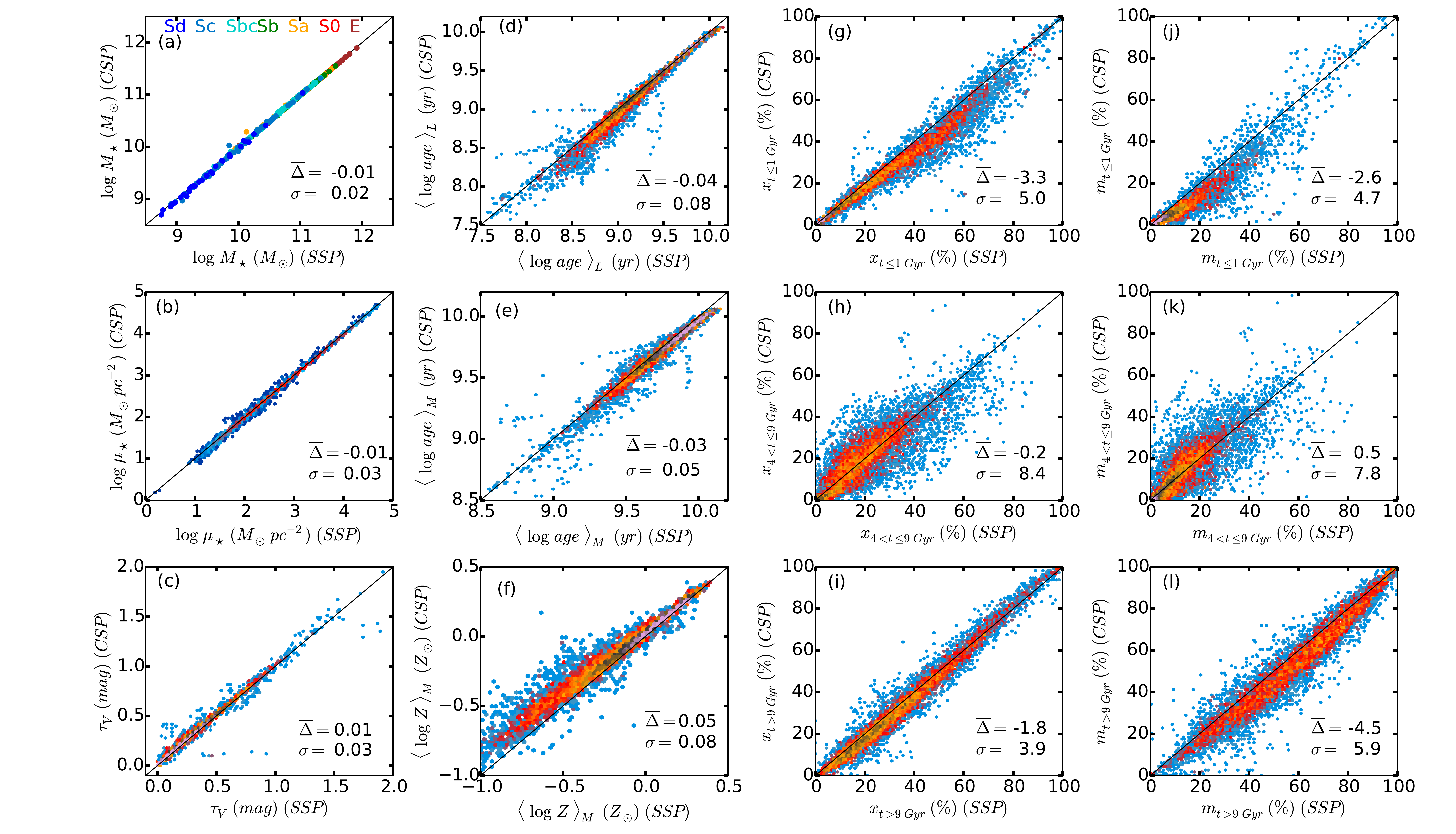}
\caption{
Comparison of several stellar population properties as obtained with the CSP (y-axis) and SSP (x-axis) templates. The average difference between the property in the y and x-axis is labeled as $\overline{\Delta}$ in each panel, and its standard deviation as $\sigma$.  Panel (a) shows the galaxy mass, with galaxies colored by their Hubble type. In the other panels, the values of the property measured  every 0.1 HLR are compared, and the color indicates the density of points in a logarithmic scale (from 0 to 2).
 } 
\label{fig:CSP_SSP}
\end{figure*}

To evaluate to what extent our results depend on the choice of the CSPs instead of the SSPs, we compare the properties derived with two sets of bases: a) CSPs, the one used in the main text and described in \ref{sec:Method}; b) SSPs, built in a similar way as the base $GMe$ (e.g. \citealt{gonzalezdelgado14a}), but now using  \citet{Vazdekis15} instead of \citet{Vazdekis10}. This new SSP base is a combination of 254 SSPs, with  8 metallicities from $\log Z/Z_\odot = -2.28$ to $+0.40$, the same as our CSP base. The age is sampled by 37 SSPs per metallicity covering from 1 Myr to 14 Gyr. Like our CSPs, these SSPs, with $t > 63$ Myr, are based on models that match the Galactic abundance pattern imprinted in the MILES stars \citep{sanchez-blazquez06}. In both bases, the IMF is Salpeter, and the dust effects are modeled as a foreground screen with a \citet{cardelli89} reddening law. More details of this base are in Garc\'\i a-Benito et al. (2017, submitted).

Using our pipeline \pycasso\ we obtained the radial distribution of the stellar population properties for each galaxy with spatial sampling of 0.1 HLR. Fig.\ \ref{fig:CSP_SSP}, panels (b) to (l), compares the results obtained with the CSPs (y-axis) and with the SSPs (x-axis) for a total of 8720 points corresponding to a maximum of 20 radial points (from the nucleus to 2 HLR), for each of the 436 galaxies analyzed in this work. Each panel quotes the mean  $\overline{\Delta}$ = property(CSP) - property(SPS), and the standard deviation ($\sigma$). The color indicates the density of points in a logarithmic scale.
Panel (a) in Fig.\ \ref{fig:CSP_SSP} compares M$_\star$ with the two bases, with galaxies colored by their Hubble type.

On average, both methods provide similar results. There are no significant differences between the two sets of results for M$_\star$ and for $\mu_\star$, with  $\overline{\Delta} \sim$ 0 and $\sigma \sim$ 0.03 dex. Other properties, such as $\tau_V$, light weighted age  $\langle  \log \ t \rangle_L$, mass weighted age $\langle  \log \ t \rangle_M$, and mass weighted metallicity $\langle  \log \ Z \rangle_M$, are also very similar, with  $\overline{\Delta}$  ($\sigma$) = 0.01 (0.03), -0.04 (0.08), -0.03 (0.05), and 0.05 (0.08),  respectively. A slight age-metallicity degeneracy is seen here; CSP models tend to obtain slightly younger ages and higher metallicities. Thus, the spectral fits with CSPs and SSPs provide equal average stellar population properties in galaxies. 
 
To find out the  differences in the SFH, we plot in Fig.\ \ref{fig:CSP_SSP}  the light ($x$, panels in the third column) and mass ($m$, panels in the fourth column) fractions compressed in three ($t \leq$ 1 Gyr, 4 $< t \leq$ 9 Gyr, and $t >$ 9 Gyr) of the four representative lookback time ranges analyzed in the main text, and discussed in  Fig.\ \ref{sec:LighAndMassFractions}. The results are very similar, with the largest difference occurring for the light fraction of stars younger than 1 Gyr ($\overline{\Delta}$ = -3.3$\%$ and $\sigma$ = 5$\%$), and the mass fraction of stars older than 9 Gyr ($\overline{\Delta}$ = -4.5$\%$ and $\sigma$ = 5.9$\%$). These differences,  explain why the ages obtained with CSP are slightly younger than with SSPs. However, they are quite small, and they are smaller than the differences found  when comparing results obtained with $GMe$ and $CBe$ bases, as seen in Figure B.2 in \citet{gonzalezdelgado15} and Figure A.1 in \citet{gonzalezdelgado16}.

\end{document}